\documentclass[fleqn,twoside]{elsart}

\usepackage[letterpaper]{geometry}

\usepackage{graphicx}
\usepackage{amssymb}
\usepackage{amsmath}
\usepackage{epsfig}
\usepackage{bbm}
\usepackage{color} % cjbf

\newcommand{\nuC}{\nu_{\rm c}}

\newcommand{\aaa}{a}
\newcommand{\bbb}{b}
\newcommand{\EF}{E_{\rm F}}
\begin{document}

\begin{frontmatter}

% Title, authors and addresses

% use the thanksref command within \title, \author or \address for footnotes;
% use the corauthref command within \author for corresponding author footnotes;
% use the ead command for the email address,
% and the form \ead[url] for the home page:
% \title{Title\thanksref{label1}}
% \thanks[label1]{}
% \author{Name\corauthref{cor1}\thanksref{label2}}
% \ead{email address}
% \ead[url]{home page}
% \thanks[label2]{}
% \corauth[cor1]{}
% \address{Address\thanksref{label3}}
% \thanks[label3]{}

\title{Electron interactions in an antidot in the integer quantum Hall regime}

% use optional labels to link authors explicitly to addresses:
% \author[label1,label2]{}
% \address[label1]{}
% \address[label2]{}

\author{
H.-S. Sim\corauthref{cor} } \corauth[cor]{Corresponding author.}
\ead{hssim@kaist.ac.kr}

\address{
Department of Physics, Korea Advanced Institute of Science and
Technology, 373-1 Guseong-dong, Yuseong-gu, Daejeon 305-701,
Korea}

\author{M. Kataoka and C. J. B. Ford}

\address{Cavendish Laboratory,
J J Thomson Avenue, Cambridge CB3 0HE, United Kingdom}

\begin{abstract}

A quantum antidot, a submicron depletion region in a two-dimensional
electron system, has been actively studied in the past two decades,
providing a powerful tool for understanding quantum Hall systems. In
a perpendicular magnetic field, electrons form bound states around
the antidot. Aharonov-Bohm resonances through such bound states
have been experimentally studied, showing interesting
phenomena such as Coulomb charging, $h/2e$ oscillations, spectator
modes, signatures of electron interactions in the line shape, Kondo
effect, etc. None of them can be explained by a simple
noninteracting electron approach. Theoretical models for the
above observations have been developed recently, such as a
capacitive-interaction model for explaining the $h/2e$ oscillations
and the Kondo effect, numerical prediction of a hole
maximum-density-droplet antidot ground state, and
spin density-functional theory for investigating the compressibility
of antidot edges.  In this review, we summarize such experimental
and theoretical works on electron interactions in antidots.
\end{abstract}

\begin{keyword}
Antidot \sep Quantum Hall effects \sep Edge states \sep
Aharonov-Bohm effect \sep Electron-electron interactions \sep
Electron transport \sep Mesoscopic systems
% keywords here, in the form: keyword \sep keyword

% PACS codes here, in the form: \PACS code \sep code
\PACS
73.43.-f, %Quantum Hall effects
73.23.Hk, %Coulomb blockade; single-electron tunneling
73.23.-b, %Electronic transport in mesoscopic systems
72.15.Qm  %Scattering mechanisms and Kondo effect
\end{keyword}
\end{frontmatter}

\tableofcontents

\section{Introduction}
\label{INTRO}

An antidot is a potential hill in a two-dimensional electron gas
(2DEG) formed in a GaAs/AlGaAs heterostructure. It can be created by
a negative voltage on a surface gate (for example, see Refs.
\cite{Ford,Sachrajda}) or by an etched pit in the wafer surface
(Refs. \cite{SWHwang,Goldman}). It is often regarded as an
artificial repulsive impurity and thus considered to be the
counterpart of a quantum dot \cite{ReviewDot,ReviewDot2}. The idea
of an antidot in a quantum Hall system was first introduced by
Jain and Kivelson \cite{Kivelson}
in order to explain
resistance peaks observed in a narrow Hall bar \cite{Timp}.  The
first experimental observation of Aharonov-Bohm oscillations in a
gate-controllable antidot was made by Smith and his coworkers in
1989 \cite{Smith}.

\begin{figure}[th]
\centerline{\psfig{file=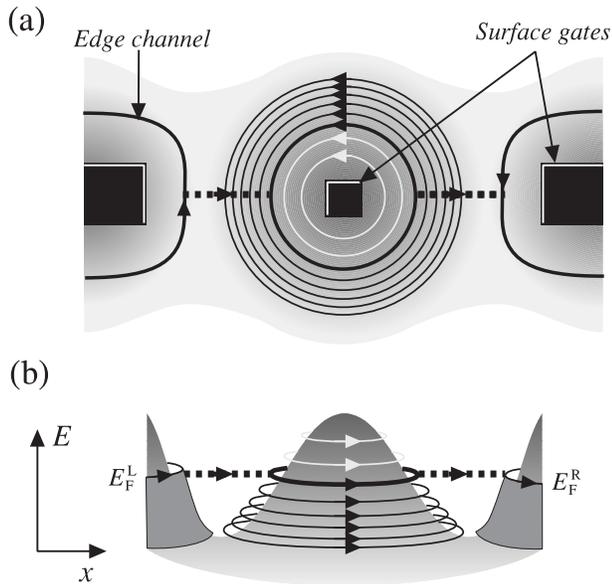,width=8cm}} \vspace*{8pt}
\caption{Single-particle states formed around an antidot in the
integer quantum Hall regime (ignoring self-consistent effects). (a)
Top view: The antidot potential is created by the voltage applied on
the central surface gate, while the constriction width between the
antidot states and the extended quantum Hall edge channels can be
controlled by tuning the side gates. The central and side surface
gates are shown as filled boxes. Filled and empty single-particle
states encircling the antidot are shown in black and light-grey
lines, respectively. The dotted lines represent the tunneling
between the antidot state (shown as bold) and the extended
edge channels along the left and right edges (with the Fermi levels
$E^{\rm L}_{\rm F}$ and $E^{\rm R}_{\rm F}$, respectively). (b)
Side view: the potential profile created by the voltages on the
antidot and side gates, depicting the continuum of states along the
extended edges, and the discrete states around the antidot.}
\label{antidot}
\end{figure}

By applying magnetic fields perpendicular to the 2DEG, antidots
have been extensively and intensively investigated to understand
coherent electron scattering, localized antidot states, the
quantum Hall effect, and many-body interaction effects. In the
regime of weak magnetic fields, an antidot behaves as a simple
scattering center. In this regime, antidot superlattices provide a
good tool for studying semiclassical motion of electrons under
periodic scattering centers, for example, commensurate Weiss
oscillations \cite{Weiss,Kang,Smet} and chaotic motions
\cite{Fleischmann,Weiss2}. On the other hand, under strong
magnetic fields of the order of a Tesla, an antidot provides a
local depletion region inside quantum Hall systems
\cite{Chakraborty}. In this regime, quantum Hall edge states
\cite{Halperin} can form closed chiral orbits around the antidot
(see Fig.~\ref{antidot}). The localized orbits around a single
antidot
\cite{Ford,Sachrajda,Kirczenow,Mace,Barnes96,Kirczenow2,Maasilta,Kataoka_charging,Kataoka_double,Karakurt,Kataoka_Kondo,Kataoka_Spin}
and around an antidot molecule
\cite{Gould_AMD2,Gould,Kirczenow_AMD} have been studied in the
integer quantum Hall regime by measuring Aharonov-Bohm
oscillations of conductance, which occur when the localized orbits
couple to extended edge channels along the boundary of the 2DEG.
The antidot has also been investigated in the fractional quantum
Hall regime experimentally
\cite{Goldman,Kang,Franklin,Franklin2,Maasilta_FQHE,Maasilta_molecule,Goldman01,Goldman_FQHE}
and theoretically
\cite{Geller,Geller2,Braggio,Bonderson,Stern,DasSarma} to
understand quasiparticle tunneling, charge and statistics,
composite fermions, chiral Luttinger liquids, and non-Abelian
statistics of the fractional quantum Hall 5/2 state.

In a simple description of the integer quantum Hall effect,
electron-electron interactions are often neglected. Thus, one
might expect that single-particle localized edge states are
enough to describe the antidot states in the integer quantum Hall
regime. However, contrary to this naive expectation, there has
recently been a series of interesting experimental observations in
the regimes of antidot local filling factor one or two, which
include $h/2e$ Aharonov-Bohm conductance oscillations
\cite{Ford,Sachrajda,Kataoka_double,Kataoka_Spin}, the signature of
electron interactions shown in the line shape of conductance peaks
\cite{Maasilta}, the detection of antidot charging effect
\cite{Kataoka_charging}, Kondo effect \cite{Kataoka_Kondo}, and
spectator modes in an antidot molecule \cite{Gould}. None
of these observations can be understood within single-particle
models \cite{Mace,Kirczenow2,Takagaki,Kirczenow_th}. Note that the
importance of Coulomb interactions has also been reported in other
experimental works on integer quantum Hall systems such as in
scanning probe studies of quantum Hall localized states
\cite{Zhitenev,Finkelstein,Ilani} and in the realization of
electronic quantum Hall Mach-Zehnder interferometers
\cite{Ji,Neder}.

The starting point of understanding such experimental observations
in the integer quantum Hall antidots is {\em excess charge}
\cite{Ford}. The excess charge can be formed around an antidot due
to the magnetic flux quantization. When the magnetic field $B$
(perpendicular to the surface of the sample) changes adiabatically,
each single-particle state encircling the antidot moves with respect
to the antidot potential, adjusting the enclosed area $S$ in order
to keep the flux $BS$ constant (see Sec.\/ \ref{THexcess} for
details). The phase change of the wavefunction around the antidot
increases by $2\pi$ for each unit of flux $h/e$ enclosed, through
the Aharonov-Bohm effect. If the occupation of these states does not
change, the electron displacement results in a local charge
imbalance around the antidot, i.e., local accumulation of excess
charges. Excess charges can provide a source of electron-electron
interactions in the antidot.

Recently, a phenomenological model for integer quantum Hall antidots
has been proposed \cite{Sim} to describe the capacitive interaction
of excess charges. This model provides a good explanation of the
experimental observations of the $h/2e$ Aharonov-Bohm oscillations,
the antidot charging effect, and the antidot Kondo effect,
indicating that the capacitive interaction of excess charges is a
good starting point for understanding electron interactions in
antidots. The model is reminiscent of the constant-interaction model
for quantum dots \cite{ReviewDot,ReviewDot2,Glazman_CI}. To see the
microscopic nature of electron interactions and of many-body antidot
states, numerical Hartree-Fock calculations have been performed
\cite{Sim_HF,Hwang,Yang_Antidot}. It has been found \cite{Hwang}
that the Hartree-Fock results for a large antidot can be well
described by the capacitive interaction model.

The formation of compressible regions around an antidot is also an
interesting issue. In the quantum Hall regime, there is no charge
screening in incompressible bulk regions because of the quantum Hall
energy gap at the Fermi level, while screening effects are important
along compressible extended edge regions \cite{Beenakker,Chang}
where the local filling factor decreases from the bulk value to
zero. The combined effects of the energy gap and the screening
result in the formation of alternate compressible and incompressible
strips along the extended edge regions of the 2DEG. The widths of
the compressible strips can be determined in a self-consistent
treatment of Coulomb interactions \cite{Chklovskii1,Chklovskii2}. On
the other hand, an antidot provides a closed finite edge of a
quantum Hall system. Due to finite-size effects (discrete energy
levels), the properties of the compressible regions formed around an
antidot are expected to deviate from those formed along
the extended edge regions, as the antidot size decreases. There has been
controversy \cite{Kataoka_double,Karakurt,Kataoka_comment,Goldman_reply,Michael}
about whether
compressible regions can be formed around a small-size antidot, or
in what range of fields. Recent numerical work on an antidot, based
on spin density functional theory, has shown \cite{Ihnatsenka} that
compressible regions around an antidot can be narrower than those
along the extended edges because of weaker screening effects. It was
also shown that exchange interactions play an important role in
determining the width of compressible regions around an antidot.

In this article, we review recent experimental and theoretical works
devoted to electron interactions in antidots in the integer quantum
Hall regime. In the experimental parts, we describe the direct
evidence of the antidot charging effect \cite{Kataoka_charging}, the
$h/2e$ Aharonov-Bohm oscillations of conductance
\cite{Ford,Sachrajda,Kataoka_double,Kataoka_Spin}, the spectator
mode in an antidot molecule \cite{Gould}, the signature of electron
interactions shown in the line shape of Aharonov-Bohm peaks
\cite{Maasilta}, the experimental search for compressible regions
\cite{Kataoka_double,Karakurt,Kataoka_comment,Goldman_reply,Michael},
and the antidot Kondo effect
\cite{Kataoka_Kondo}. In the theoretical parts, we explain the
origin of excess charges on an antidot and introduce the
capacitive interaction of excess charges, which
may give rise to the antidot charging effect. Then, we
show that the capacitive interaction model can give a good
explanation \cite{Sim} for the $h/2e$ oscillations and the
antidot Kondo effect. We also include the prediction \cite{Hwang} of
a hole maximum-density droplet of antidot states based on a
numerical Hartree-Fock approach and the numerical
spin density-functional search for compressible regions of
an antidot \cite{Ihnatsenka}.
Despite the above experimental and theoretical efforts, there
are still many open problems in antidots in the integer quantum Hall
regime.
We suggest some possible open questions in the
concluding part of the review.
We note that antidots in the
fractional quantum Hall regime are not addressed in this review.

This paper is organized as follows. In Sec.\/ \ref{SP}, the
single-particle description of an integer quantum Hall antidot is
briefly introduced. Single-particle energy levels (Sec.
\ref{SPENERGY}), forward and backward scattering of extended edge
channels by antidots, a noninteracting electron description of the
Aharonov-Bohm oscillations, and the two-terminal conductance of an
antidot (Sec.\/ \ref{SPTUNNELING}) are discussed. In Sec.\/
\ref{EXP}, experimental works on electron interactions are
mentioned. This section includes typical experimental setups (Sec.
\ref{EXPsetup}) and experimental data (Sec.\/ \ref{LOWB_AB}), the
detection of the antidot charging effect (Sec.\/ \ref{EXPcharging}),
the $h/2e$ Aharonov-Bohm oscillations (Sec.\/ \ref{EXPdouble}), the
spectator modes in an antidot molecule (Sec.\/ \ref{EXPmolecule}),
the line shape of Aharonov-Bohm peaks (Sec.\/ \ref{EXPline}), the
search for compressible regions around an antidot (Sec.\/
\ref{EXPcompressible}), and the antidot Kondo effect (Sec.\/
\ref{EXPKondo}). In Sec.\/ \ref{TH}, we describe theoretical models
dealing with electron interactions in an antidot. This section
provides the total Hamiltonian of an antidot system (Sec.
\ref{THHamiltonian}), the origin of excess charges (Sec.\/
\ref{THexcess}), the capacitive interaction model for an antidot
(Sec.\/ \ref{THcapacitive}), the explanation of the $h/2e$
oscillations and the antidot Kondo effect (Sec.\/ \ref{THKondo}),
the prediction of a hole maximum-density-droplet antidot ground
state (Sec.\/ \ref{THHF}), and the numerical search for compressible
regions (Sec.\/ \ref{THcompressible}). Finally, open problems and
perspectives of the integer quantum Hall antidot research will be
given in Sec.\/ \ref{SUMMARY}.

\section{Noninteracting description of an integer quantum Hall antidot}
\label{SP}

In this section, we sketch a single-particle model of an integer
quantum Hall antidot. In Sec.\/ \ref{SPENERGY}, we describe
single-particle localized states around an isolated antidot. The
properties of the single-particle states are governed by the
Aharonov-Bohm flux enclosed by them. In Sec.\/ \ref{SPTUNNELING}, we
discuss resonant scattering when electrons can tunnel between the
extended edges of the 2DEG and the antidot. The resonant scattering
can result in the flux-dependent Aharonov-Bohm oscillations of
conductance, the line shape of which depends on the detailed
coupling nature of the antidot to the extended edge channels.

\subsection{Single-particle energy levels}
\label{SPENERGY}

We first consider an isolated antidot where the antidot states are
decoupled from the extended edge channels of the 2DEG. In a
magnetic field $B$ perpendicular to the 2DEG, classical skipping
orbits appear around the antidot due to $\vec{E} \times \vec{B}$
drift with electric field $\vec{E}$ created by the antidot
potential. The skipping orbits are quantized (see
Fig.~\ref{antidot}) so that their enclosing area $S_m$ satisfies
the Aharonov-Bohm condition
\begin{equation}
BS_m \sim m \phi_0,
\end{equation}
where $m$ is the orbital quantum number, $\phi_0 = h/e$ is
the magnetic flux quantum, $h$ is Planck's constant, and $e$ is the
magnitude of the electronic charge. Due to the sloping shape of the
antidot potential, the quantized states form a ladder in energy
around the antidot [Fig.~\ref{antidot}(b)]. At zero temperature
these states are filled up to the Fermi level $E_{\rm F}$ when
electron-electron interactions are neglected. For simplicity,
hereafter, we assume that the antidot orbits have a circular shape.

\begin{figure}[th]
\centerline{\psfig{file=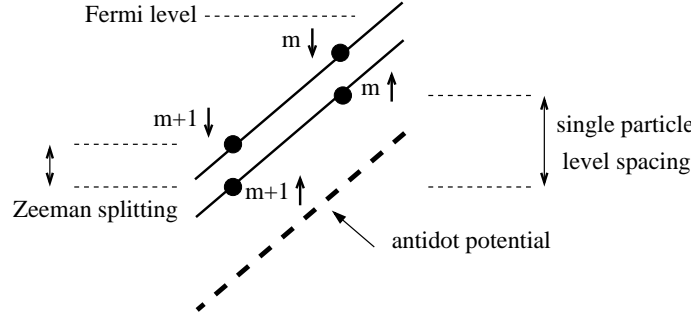,width=9cm}}
\vspace*{8pt} \caption{Schematic energy diagram of single-particle
antidot states. For simplicity, only the spin-split branches of the
lowest Landau level are drawn. The single-particle energy spacing
$\delta \epsilon_m$ between the $m$-th and ($m+1$)-th localized
states and Zeeman energy splitting of spin-up (up arrows) and down
(down arrows) states are shown. Spin-up states are assumed to have
smaller Zeeman energy than spin-down states. The dashed
line indicates the antidot potential.}\label{desp}
\end{figure}

The properties of the single-particle antidot states can be
understood from the Landau levels. Using the symmetric gauge, one
can have, for example, the $m$-th orbital state $\phi_m$ in the
lowest Landau level \cite{Chakraborty},
\begin{eqnarray}
\phi_m (\vec{r}) = \frac{1}{\sqrt{2^{m} \pi (m-1)!}~ l_B}
(\frac{z}{l_B})^{m-1} \exp (\frac{-|z|^2}{4 l_B^2}),
\label{LLsymmetricgauge}
\end{eqnarray}
where $z = x + iy$ is the complex coordinate of the two-dimensional
plane, $m \ge 1$ is the orbital quantum number used above, and $l_B
= \sqrt{\hbar/(eB)} = 25.6 / \sqrt{B[{\rm T}]} \textrm{[nm]}$ is the
magnetic length.
The state $\phi_m$ encloses $m$ units $h/e$ of magnetic flux,
{\it i.e.}, the magnetic flux enclosed within $r_m$ (given
by $r_m^2 \equiv \langle \phi_m | r^2 | \phi_m \rangle = 2 m
l_B^2$) is quantized.
The state is confined radially within the magnetic length scale $l_B$
due to the perpendicular magnetic field. In the presence of the
antidot potential $V_{\rm AD}(r)$, when $V_{\rm AD}(r)$ varies
sufficiently slowly on the scale of $l_B$, one can use the states
$\phi_m$ as the single-particle states of the isolated antidot with
single-particle energy $\epsilon_m$,
\begin{eqnarray}
\epsilon_{m \sigma} \simeq \frac{1}{2} \hbar \omega_c + \bar{V}_{\rm
AD}(m) + \epsilon^{\rm Z}_\sigma, \label{LLenergy}
\end{eqnarray}
where $\omega_c = e B / m^*$ is the cyclotron frequency, $m^*$ is
the electron effective mass ($m^* = 0.067 m_{\rm e}$ for
GaAs), $\bar{V}_{\rm AD}(m) = \langle \phi_m | V_{\rm AD} | \phi_m
\rangle$ is the mean antidot potential energy, $\epsilon^{\rm
Z}_\sigma = g \mu_{\rm B} B \sigma / 2$ is the Zeeman
energy of a spin-$\sigma$ electron, $\mu_{\rm B}$ is the
Boltzmann factor, $g$ is the Land\'{e} $g$ factor ($g = -0.44$ for
GaAs), and $\sigma = 1$ ($-1$) for spin-up (down) electrons. Here,
we consider the antidot states coming from the lowest Landau level
only. The spatial separation $\Delta r_m \equiv r_{m+1} - r_m$
between two adjacent states depends on $B$. For large $m \gg 1$, one
can show that
\begin{equation}
\Delta r_m \simeq \frac{\phi_0}{2\pi r_m B}.
\end{equation}
In the case where the antidot potential $V_{\rm AD}(r)$ is linear
over a width much greater than the separation of the states around
the radius $r_m$, for $m \gg 1$ (see Fig.~\ref{desp}), the
single-particle energy gap $\delta \epsilon_m$ between two
neighboring states is proportional to $1/B$,
\begin{equation}
\delta \epsilon_m \equiv \epsilon_{(m-1) \sigma} - \epsilon_{m
\sigma} \simeq -\Delta r_m \frac{dV_{\rm AD}(r)}{dr}|_{r=r_m} =
\frac{\phi_0}{2\pi r_m B} \left|\frac{dV_{\rm AD}(r)}{dr}|_{r=r_m}
\right|. \label{SPlevelspacing}
\end{equation}

The single-particle antidot states can be experimentally controlled
by tuning either the antidot gate voltage or the magnetic field $B$.
The dependence of their energies on the antidot gate voltage or on
$B$ is governed by the {\em quantization of enclosed magnetic flux}.
For example, as $B$ increases, the antidot states move towards the
center of the antidot, keeping the average magnetic flux enclosed
constant. As a consequence, the states rise up in energy, and the
highest occupied states become empty one by one as they pass through
the Fermi level $E_{\rm{F}}$. This depopulation process is periodic
in $B$ with period $\Delta B$,
\begin{equation}
\Delta B \simeq \frac{\phi_0}{S},
\label{ABperiod}
\end{equation}
where $S = \pi r^2$ is the effective antidot area enclosed by the
state at the Fermi level; the finite width of the spatial
distribution of single-particle antidot states can slightly modify
this depopulation condition because of the spatial variation
of the antidot potential within the finite width.
The same processes happen when the antidot gate voltage varies,
changing the potential at each radius relative to $E_{\rm
F}$. We emphasize that the $B$-dependent depopulation of the
antidot states is reminiscent of a gate-voltage controlling the
states of quantum dots \cite{ReviewDot,ReviewDot2}. Note that when
electron-electron interactions play a role, magnetic flux
quantization provides still a crucial restriction on antidot states.
We will return to this issue in Sec.\/ \ref{THexcess} where excess
charges around an antidot will be discussed.

The depopulation of single-particle states near the Fermi level can
result in Aharonov-Bohm resonance oscillations in antidot
conductance when one allows electron tunneling between the antidot
states and the current-carrying extended edge channels of the 2DEG
by adjusting the side gate voltages shown in Fig.~\ref{antidot}. The
conductance oscillations are the subject of the next section
(\ref{SPTUNNELING}).

\subsection{Aharonov-Bohm oscillations}
\label{SPTUNNELING}

\begin{figure}[th]
\centerline{\psfig{file=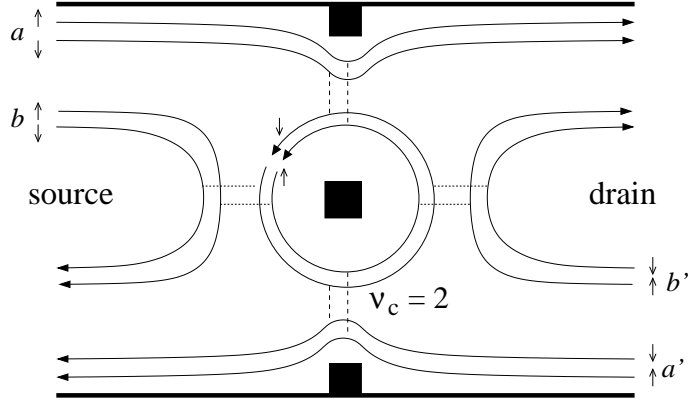,width=9cm}} \vspace*{8pt}
\caption{A schematic diagram of a two-terminal Hall bar with an
antidot. The black square at the center of the Hall bar is the
antidot gate, while those attached to the upper and lower edges
(thick lines) of the Hall bar represent the side gates, which
control the effective widths of the constrictions between the
antidot and the edges. In this diagram, the filling factor in each
constriction $\nuC = 2$, so the antidot states (circles) and the
extended edge channels (channels $\aaa \sigma$ and $\aaa'
\sigma$ with spin $\sigma$) in the constrictions come from the
spin-up and down lowest Landau levels. The bulk filling factor is
chosen to be $\nu_{\rm bulk} = 4$, so that there are two additional
extended edge channels (channels $\bbb \sigma$ and $\bbb'
\sigma$), which come from the second lowest Landau level
and are perfectly reflected at the constrictions. The left side of
the Hall bar is chosen to be the current source while the right is
the drain. The couplings of antidot states to the channels $\aaa$
and $\aaa'$ are represented by dashed lines (intra-Landau-level
coupling), and those to the channels $\bbb$ and $\bbb'$ by
dotted lines (inter-Landau-level coupling). A magnetic field $B$ is
applied perpendicular to the 2DEG plane. This diagram may
equivalently describe the cases of $\nu_{\rm bulk}
> 4$ because the inter-Landau-level coupling between the antidot
states and the extended edge channels coming from Landau
levels higher than the second is negligible when $B$ is strong
enough.} \label{geometry}
\end{figure}

The localized antidot states discussed in the previous section can
be observed in conductance measurements when they couple to the
current-carrying extended edge channels of the 2DEG. In
Fig.~\ref{geometry}, we depict schematically an experimental
two-terminal setup for the measurement of electron current through
an antidot in the quantum Hall regime. By tuning the side gates, one
can modify the effective widths of the constrictions between the
antidot and the extended edges of the 2DEG. As the widths decrease,
the electron density in the constrictions becomes smaller,
decreasing the constriction filling factor $\nuC$ from the filling
factor $\nu_{\rm bulk} = 2 \pi l_B^2 n_{\rm bulk}$ of the bulk 2DEG,
where $n_{\rm bulk}$ is the electron density in the bulk region. At
the same time, the overlap between the antidot state at the Fermi
level and the extended edge channels becomes larger, so that
resonant electron tunneling becomes allowed between them. In this
way, the antidot states have been experimentally studied by
measuring the two-terminal conductance. In this section, we first
introduce the expression \cite{Kivelson,Mace,Geller2,Takagaki,Kirczenow_th}
for the two-terminal conductance for a noninteracting antidot in the
limit of zero temperature and zero bias, and discuss Aharonov-Bohm
resonance oscillations in the conductance as well as the backward
and forward scatterings of the extended edge channels by the
antidot. Finally, we derive an expression for the two-terminal
conductance in terms of the antidot Green's function, which is
applicable to the cases \cite{Sim} where electron interactions are
important.

We consider a typical two-terminal setup shown in
Fig.~\ref{geometry}. For simplicity, the local antidot filling
factor is chosen as $\nuC =2$, while the bulk filling factor of the
2DEG is $\nu_{\rm bulk} = 4$. The following discussion can be easily
generalized to the cases of other integer $\nuC$ and $\nu_{\rm
bulk}$. In the case of $\nuC = 2$ and $\nu_{\rm bulk} =4 $, there
are two types of extended edge channels, one from the lowest Landau
level (see the channels $\aaa$ and $\aaa'$ in Fig.~\ref{geometry})
and the other from the second lowest Landau level (the channels
$\bbb$ and $\bbb'$). The channels $\aaa$ and $\aaa'$ pass through
the antidot constriction, while $\bbb$ and $\bbb'$ are perfectly
reflected at the constriction. These two different types of edge
channels make electron transport through the antidot a mixture of
forward and backward scattering.

We first discuss a simple case where the antidot states couple only
to the channels $\aaa$ and $\aaa'$ and are completely decoupled from $\bbb$
and $\bbb'$. In this case, electrons in the channel $\aaa \sigma$ can be
scattered into $\aaa' \sigma$ via a resonant antidot state with spin
$\sigma$, thus resonant {\em back}-scattering occurs as shown in the
following expression \cite{Kivelson,Takagaki,Kirczenow_th} for the
total probability $T_\sigma(E)$ of electron transmission from the
source to the drain,
\begin{equation}
T_\sigma (E) =  1 - \frac{p_{\aaa\sigma} p_{\aaa'\sigma}}{1 - 2 \sqrt{q_{\aaa
\sigma} q_{\aaa' \sigma}} \cos (\Phi_{{\rm AD} \sigma} (E)) + q_{\aaa
\sigma} q_{\aaa' \sigma}}. \label{BWDsc1}
\end{equation}
Here, $E$ is the energy of incident channel, $p_{i \sigma}$ is the
scattering probability of spin-$\sigma$ electron between the channel
$i \in \{\aaa, \aaa'\}$ and the antidot state, $q_{i \sigma} = 1 - p_{i
\sigma}$, and $\Phi_{\rm{AD} \sigma} (E)$ is the phase accumulated
along the complete orbit of the antidot state, which includes the
phase shifts acquired in the scattering processes between the
channels and the antidot. Note that the energy dependence of $p_{i
\sigma}$ is ignored. A localized antidot state comes on to resonance
when the phase accumulation is an integer multiple of $2 \pi$, {\it
i.e.} $\Phi_{\rm{AD} \sigma} = 2 \pi l$, causing a minimum in
$T_{\sigma}$. We note that the transmission (\ref{BWDsc1}) has the
inverse Breit-Wigner resonance line shape around the resonance
energy $E_{\rm res}$ in the limit of $\Gamma$ smaller than the
single-particle level spacing since
\begin{equation}
T_{\sigma} (E) \simeq 1- \frac{\Gamma_\sigma^\aaa
\Gamma_\sigma^{\aaa'}}{(E - E_{\rm res})^2 + (\Gamma_\sigma^\aaa +
\Gamma_\sigma^{\aaa'})^2/4}. \label{BreitWigner}
\end{equation}
Here, $\Gamma_\sigma^{i} = p_{i \sigma} (dE / d \Phi_{{\rm AD}
\sigma})$ is the broadening width of the resonance due to the
coupling to the channel $i \in \{ \aaa, \aaa' \}$, and we have used the
approximation that $1 - \cos (\Phi_{{\rm AD} \sigma} (E)) \simeq (d
\Phi_{{\rm AD} \sigma} / dE)^2 (E - E_{\rm res})^2 / 2$ around $E =
E_{\rm res}$.

There is also a simple case of resonant {\em forward} scattering,
where the antidot states couple only to the channels $\bbb$ and $\bbb'$
and are completely decoupled from $\aaa$ and $\aaa'$. In this case, the
total transmission probability $T_{\sigma}$ from the source to the
drain can be obtained as
\begin{equation}
T_{\sigma} (E) = 1 + \frac{p_{\bbb\sigma} p_{\bbb'\sigma}}{1 - 2
\sqrt{q_{\bbb \sigma} q_{\bbb' \sigma}} \cos (\Phi_{{\rm AD} \sigma} (E))
+ q_{\bbb \sigma} q_{\bbb' \sigma}}. \label{FWDsc1}
\end{equation}
Here, $p_{i \sigma}$ is the scattering probability of spin-$\sigma$
electrons between channel $i \in \{\bbb, \bbb' \}$ and the antidot state,
and $q_{i \sigma} = 1 - p_{i \sigma}$. The first term of Eq.
(\ref{FWDsc1}) comes from the perfect transmission of the channels
$\aaa$ and $\aaa'$. The resonances occur with the maximum value of
$T_{\sigma}$ at the same condition of $\Phi_{\rm{AD} \sigma} = 2 \pi
l$ as the above back-scattering case.

In the general case where the antidot couples to both types of
channel ($\aaa$,$\aaa'$) and ($\bbb$,$\bbb'$), the total
transmission from the source to the drain is a combination
of the forward and backward scattering given by \cite{Kirczenow_th}
\begin{equation}
T_{\sigma} (E)  = 1 + \frac{q_{\aaa \sigma} + q_{\aaa' \sigma} + q_{\aaa
\sigma} q_{\aaa' \sigma} ( q_{\bbb \sigma} q_{\bbb' \sigma} - q_{\bbb \sigma} -
q_{\bbb' \sigma}) - 1}{1 - 2 \sqrt{q_{\aaa \sigma} q_{\aaa' \sigma} q_{\bbb
\sigma} q_{\bbb' \sigma}} \cos (\Phi_{{\rm AD} \sigma} (E)) + q_{\aaa
\sigma} q_{\aaa' \sigma} q_{\bbb \sigma} q_{\bbb' \sigma}}. \label{FWDBWDsc1}
\end{equation}
On resonance, $T_{\sigma}$ can be either larger or smaller than $1$,
determined by the competition between the antidot coupling strengths
$\Gamma^i_\sigma$ to the extended edge channels $i \in
\{\aaa,\aaa',\bbb,\bbb'\}$. The phase accumulation $\Phi_{\rm{AD}
\sigma}$ depends on the magnetic field $B$ as $\Phi_{\rm{AD} \sigma}
= 2 \pi B S / \phi_0 + \Phi_{\rm{AD} \sigma, 0}$, where
$\Phi_{\rm{AD} \sigma,0}$ is a phase independent of $B$. Thus as a
function of $B$ the transmission shows Aharonov-Bohm oscillations
with period $\Delta B$ given by Eq. (\ref{ABperiod}) and line shape
given by Eq. (\ref{FWDBWDsc1}). This noninteracting description of
the Aharonov-Bohm oscillations will be used to analyze experimental
data in Sec.\/ \ref{EXP}. However, this noninteracting model fails
to explain all the data, so we need an expression for the
transmission that can be used for interacting electrons; this will
be derived based on a Green's-function method in the
remaining parts of this section.

Below, for the antidot system shown in Fig.~\ref{geometry}, we will
derive the total transmission from the source to the drain in terms
of the Green's function of the antidot in the limit of zero
temperature and zero bias. Following Ref. \cite{Ng}, we start with
the S-matrix of the extended edge channels $i \in \{
\aaa,\aaa',\bbb,\bbb' \}$. Because there is no spin-flip process at
each single-electron tunneling event (in the absence, for example,
of spin-orbit coupling), the S-matrix does not connect channels of
different spin, and it has the following form,
\begin{eqnarray}
\vec{A}_{\rm out, \sigma} & = & \tilde{S}_\sigma \vec{A}_{\rm
in, \sigma}, \nonumber \\
\vec{A}_{\rm in, \sigma}^{\rm T} & = & (u_{\aaa\sigma},
u_{\bbb\sigma}, u_{\aaa' \sigma}, u_{\bbb' \sigma}),
\nonumber \\
\vec{A}_{\rm out, \sigma}^{\rm T} & = & (v_{\aaa\sigma},
v_{\bbb\sigma}, v_{\aaa' \sigma}, v_{\bbb' \sigma}), \nonumber
\end{eqnarray}
where $\vec{A}_{{\rm in}, \sigma}$ is the incoming spin-$\sigma$
state vector to the antidot, $\vec{A}_{{\rm out}, \sigma}$ is the
outgoing state vector, and $\vec{A}^{\rm T}_{{\rm in/out}, \sigma}$
is the transpose of $\vec{A}_{{\rm in/out}, \sigma}$. The detailed
form of the S-matrix can be found as
\begin{eqnarray}
\tilde{S}_\sigma = \tilde{I} - 2\pi i N_\sigma {\cal G}_\sigma (E +
i \delta) \left(
\begin{array}{cccc} t_{\aaa \sigma}^2 & t_{\bbb \sigma} t_{\aaa \sigma}
& t_{\aaa' \sigma} t_{\aaa \sigma} &
t_{\bbb' \sigma} t_{\aaa \sigma} \\
t_{\aaa \sigma} t_{\bbb \sigma} & t_{\bbb \sigma}^2 & t_{\aaa' \sigma} t_{\bbb
\sigma} &
t_{\bbb' \sigma} t_{\bbb \sigma} \\
t_{\aaa \sigma} t_{\aaa' \sigma} & t_{\bbb \sigma} t_{\aaa' \sigma} & t_{\aaa'
\sigma}^2 &
t_{\bbb' \sigma} t_{\aaa' \sigma} \\
t_{\aaa \sigma} t_{\bbb' \sigma} & t_{\bbb \sigma} t_{\bbb' \sigma} & t_{\aaa'
\sigma} t_{\bbb' \sigma} & t_{\bbb' \sigma}^2
\end{array} \right),
\label{Smatrix}
\end{eqnarray}
where $\tilde{I}$ is the identity matrix, $N_\sigma(E)$ is the
density of states of noninteracting spin-$\sigma$ electrons of each
extended edge channel (the channel-dependence of $N_\sigma(E)$ is
ignored), and ${\cal G}_\sigma(E + i\delta)$ is the full Green's
function of the antidot. Also, $t_{i \sigma}$ is the hopping energy
between an antidot state with spin $\sigma$ and the edge channels $i
\sigma$, $i \in \{\aaa,\aaa',\bbb,\bbb' \}$. From $\tilde{S}_\sigma$, one can
define the transmission matrix $\tilde{T}_\sigma$:
\begin{eqnarray}
(v_{\aaa \sigma}, v_{\bbb' \sigma})^{\rm T} & = & \tilde{T}_\sigma
(u_{\aaa \sigma}, u_{\bbb \sigma})^{\rm T}, \nonumber \\
\tilde{T}_\sigma & = & \left( \begin{array}{cc} 1 - 2\pi i t_{\aaa
\sigma}^2 N_\sigma {\cal G}_\sigma &
- 2\pi i t_{\aaa \sigma} t_{\bbb \sigma} N_\sigma {\cal G}_\sigma \\
- 2\pi i t_{\bbb' \sigma} t_{\aaa \sigma} N_\sigma {\cal G}_\sigma & -
2\pi i t_{\bbb' \sigma} t_{\bbb \sigma} N_\sigma {\cal G}_\sigma
\end{array} \right).
\label{TransmissionMatrix}
\end{eqnarray}
The total transmission $T_\sigma$ of the channels $\aaa\sigma$ and
$\bbb\sigma$ incoming from the source to the channels $\aaa\sigma$ and $\bbb'
\sigma$ outgoing to the drain can be found as $T_\sigma =
\textrm{Tr} (\tilde{T}_\sigma^\dagger \tilde{T}_\sigma)$.

To derive $T_\sigma$ further, we use the imaginary part of the
self-energy of ${\cal G}_\sigma(\EF)$,
\begin{eqnarray}
\Im \Sigma_\sigma (\EF) = -\pi (t_{\aaa \sigma}^2 +
t_{\bbb \sigma}^2 + t_{\aaa' \sigma}^2 + t_{\bbb' \sigma}^2)
N_\sigma (\EF),
\end{eqnarray}
and the occupation number $\langle n_\sigma \rangle$ of the antidot,
\begin{eqnarray}
\langle n_\sigma \rangle = \pi^{-1} \Im ( \ln {\cal G}_\sigma
(\EF + i \delta)).
\end{eqnarray}
After some trivial algebra, one can find $T_\sigma$ in terms of
$\theta_\sigma = \pi \langle n_\sigma \rangle$ as
\begin{eqnarray}
T_\sigma & = & 1 + \frac{4(t_{\bbb \sigma}^2 t_{\bbb' \sigma}^2 - t_{\aaa
\sigma}^2 t_{\aaa' \sigma}^2) \sin^2 \theta_\sigma}{ (t_{\aaa \sigma}^2 +
t_{\bbb \sigma}^2 + t_{\aaa' \sigma}^2 + t_{\bbb' \sigma}^2)^2} \nonumber \\
& = & 1 + \frac{4(\Gamma^\bbb_\sigma \Gamma^{\bbb'}_\sigma -
\Gamma^\aaa_\sigma \Gamma^{\aaa'}_\sigma) \sin^2 \theta_\sigma}{
(\Gamma^\aaa_\sigma + \Gamma^\bbb_\sigma + \Gamma^{\aaa'}_\sigma +
\Gamma^{\bbb'}_\sigma)^2}. \label{transmission}
\end{eqnarray}
Then, in the linear-response regime at zero temperature,
the two-terminal conductance $G_{\rm ad}$ of the antidot system
shown in Fig.~\ref{geometry} is immediately written as
\begin{eqnarray}
G_{\rm ad} = \sum_\sigma G_\sigma = \sum_\sigma \frac{e^2}{h} T_\sigma.
\label{conductance}
\end{eqnarray}

Here, we provide a simple application of the transmission
(\ref{transmission}). In the weak coupling regime, the antidot
Green's function and $\sin^2 \theta_\sigma$ can be written around a
resonance energy $E_{res}$ as
\begin{eqnarray}
{\cal G}(E) & = & \frac{1}{E - E_{res} - i \Im \Sigma_\sigma},
\nonumber \\
\sin^2 \theta_\sigma & = & \frac{ (\Im \Sigma_\sigma )^2}{ (E -
E_{res})^2 + (\Im \Sigma_\sigma)^2}.
\label{SELF2}
\end{eqnarray}
If one assumes that the antidot is decoupled from the edge channels
$\bbb$ and $\bbb'$, {\it i.e.}, $\Gamma^\bbb_\sigma = \Gamma^{\bbb'}_\sigma =
0$, the expressions (\ref{transmission}) and (\ref{SELF2}) lead to
the inverse Breit-Wigner line shape equivalent to Eq.
(\ref{BreitWigner}), which describes the backscattering. On the
other hand, if one assumes $\Gamma^\aaa_\sigma = \Gamma^{\aaa'}_\sigma =
0$, one can find the expression for the forward scattering as well.

The expression (\ref{transmission}) for the total transmission is
applicable for interacting electrons. However, in order to use it,
one has to know the antidot Green's function or $\theta_\sigma$. For
example, $\theta_\sigma$ has been calculated \cite{Sim} by using the
numerical renormalization group method (see Sec.\/ \ref{THKondo}),
and the resulting line shape has been found to reproduce the
experimental Kondo features \cite{Kataoka_Kondo}.

\section{Experimental signatures of electron interactions
in antidots}
\label{EXP}

In the previous section, the single-particle antidot energy levels
and the Aharonov-Bohm forward and backward scattering by an antidot
have been discussed within a noninteracting model.
%In real experiments, to have the Aharonov-Bohm effects, the
%phase coherence length should be longer than the circumference of
%the antidot. The 2DEG formed in GaAs heterostructures satisfies
%\cite{Timp_AB,Ford_AB}
%this condition at enough low temperatures less than 100 mK,
%which is accessible in current measurement technique.
In experiments, Aharonov-Bohm oscillations with an antidot were
first observed by Smith and his coworkers \cite{Smith}. Since then,
the Aharonov-Bohm effect has been intensively investigated
experimentally
\cite{Ford,Sachrajda,Kirczenow,Mace,Barnes96,Kirczenow2,Maasilta,
Kataoka_charging,Kataoka_double,Karakurt,Kataoka_Kondo,Kataoka_Spin}
in the integer quantum Hall regime and revealed some interesting
features in the period, the line shape, and the temperature or bias
dependence of the Aharonov-Bohm conductance oscillations that cannot
be explained within the noninteracting model.

In this section, we describe the experimental work devoted to
electron interactions in an antidot in the integer quantum Hall
regime and compare the observations with the noninteracting model.
We discuss typical experimental setups (Sec.\/ \ref{EXPsetup}),
simple Aharonov-Bohm oscillations, including the
detection of the oscillation of the Fermi energy \cite{Barnes96}
(Sec.\/ \ref{LOWB_AB}), direct
evidence of the antidot charging effect \cite{Kataoka_charging}
found using a noninvasive voltage probe (Sec.\/ \ref{EXPcharging}),
the $h/2e$ Aharonov-Bohm oscillations in $\nuC = 2$ antidots
\cite{Ford,Sachrajda,Kataoka_double,Kataoka_Spin} (Sec.
\ref{EXPdouble}), the spectator behavior of the molecular modes in
antidot molecules \cite{Gould} (Sec.\/ \ref{EXPmolecule}), the
signature of electron interactions found in the line shape
\cite{Maasilta} (Sec.\/ \ref{EXPline}), an experimental search for
the compressible regions around an antidot
\cite{Kataoka_double,Karakurt,Kataoka_comment,Goldman_reply,Michael}
(Sec. \ref{EXPcompressible}), and the antidot Kondo effect
\cite{Kataoka_Kondo} (Sec.\/ \ref{EXPKondo}).

\subsection{Typical experimental setup}
\label{EXPsetup}

The typical experimental setup of an antidot is as follows. The
antidot potential and the constrictions (see Fig.~\ref{geometry})
can be formed by using Schottky gates and/or shallow etching on the
surface of a high-mobility GaAs-Al$_x$Ga$_{1-x}$As heterostructure.
The size or the energy levels of the antidot and the
electrical widths of the constrictions can be tuned by
applying negative voltages to the gates. The typical carrier
concentration and mobility are $(1-4) \times 10^{15} \, \rm{m}^{-2}$
and $>100 \, \rm{m}^2/\rm{Vs}$, respectively. The typical
lithographic width of the antidot gate (or etched pit) is $0.2-0.3
\, \mu{\rm m}$, and the lithographic width of the side
constrictions is $0.4-0.8 \, \mu{\rm m}$. A negative gate voltage of
the order of $-1$ V creates an antidot potential of effective radius
$r_{\rm AD} \sim 0.3-0.4 \, \mu {\rm m}$ at the Fermi level. We note
that the Aharonov-Bohm period in magnetic field for adding one more
flux quantum $\phi_0$ to a loop of radius $r_{\rm AD} \sim 0.3 \,
\mu {\rm m}$ is $\sim 15 \, \rm{mT}$. The conductance of the antidot
geometry has been measured at temperatures lower than $100 \,
\rm{mK}$ using standard low-bias ac lock-in techniques. The
constriction filling factors are usually adjusted to $\nuC = 1$ or
$2$, while the bulk filling factor $\nu_{\rm bulk}$ is larger than
$\nuC$.

\begin{figure}[th]
\centerline{\psfig{file=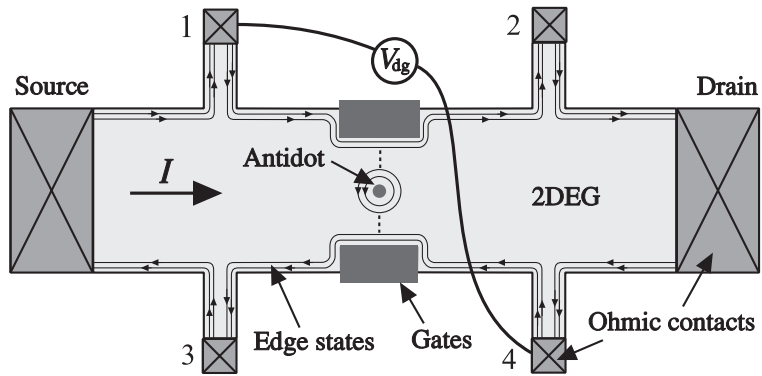,width=10cm}} \vspace*{8pt}
\caption{ A typical experimental setup for antidot conductance
measurements.} \label{HallBar}
\end{figure}

The antidot conductance can be measured in various ways,
two-terminally or four-terminally.  Following the
Landauer-B\"{u}ttiker formula \cite{Buttiker}, the two-terminal
antidot conductance in the absence of resonances with small
source-drain bias $V$ and measured current $I$ gives
\begin{eqnarray}
  G_{\rm ad-2T} = \frac{I}{V} = \nuC \frac{e^2}{h}.
\end{eqnarray}
The two-terminal conductance is useful, and intuitive, because this
background conductance is quantized and reveals the filling factor
in whichever antidot constriction is wider. Resonances through the
antidot states appear as peaks or dips on the background
conductance, depending on the nature of the scattering, following
Eq.~(\ref{conductance}). However, actual two-terminal measurements
include series resistance, which prevents the exact quantization of
the background conductance.  On the other hand, four-terminal
measurements of longitudinal conductance (the derivation of which is
provided in the next paragraph)
\begin{eqnarray}
G_{\rm ad-4T} = \frac{I}{V_{\rm L}} =
(\frac{1}{\nuC}-\frac{1}{\nu_{\rm bulk}})^{-1} \frac{e^2}{h},
\end{eqnarray}
where $V_{\rm L}$ is the longitudinal voltage drop, do not suffer
the effect of series resistance.  However, the background
conductance is not quantized and the bulk filling factor needs to be
taken into account in extracting the information on the antidot
filling factor.  The effect of the bulk filling factor can be
removed by adding in the Hall voltage with the correct sign, by
measuring the diagonal voltage drop $V_{\rm dg}$, as shown in
Fig.~\ref{HallBar}:
\begin{eqnarray}
  G_{\rm ad-dg} = \frac{I}{V_{\rm dg}} = \nuC \frac{e^2}{h}.
\end{eqnarray}
This is effectively equal to the ideal two-terminal conductance
(without the effect of series resistance). This feature comes from
the chiral and adiabatic transport of edge channels, which does not
produce any voltage drop between the voltage-probe reservoir 1 (4)
and the source (drain).

Before closing this subsection, we derive $G_{\rm ad-4T}$, based on
the Landauer-B\"{u}ttiker formalism \cite{Buttiker}. In the
formalism, the current at a reservoir $l$ is written as
\begin{eqnarray}
I_{l} = \frac{e}{h} \sum_{l_1 \ne l}( T_{l \to l_1} \mu_l - T_{l_1
\to l} \mu_{l_1}), \nonumber
\end{eqnarray}
where $T_{l \to l_1}$ is the total transmission from reservoir $l$
to $l_1$ and $\mu_{l(l_1)}$ is the electrochemical potential of
reservoir $l$ ($l_1$). The first term in the above relation shows
the current outgoing from the reservoir $l$ to $l_1$ while the
second means the contribution incoming from $l_1$ to $l$. This
relation can be applied for the antidot setup in Fig.~\ref{HallBar},
taking into account the chiral and adiabatic nature of edge-channel
transport, as
\begin{eqnarray}
I_{\rm source} & = & I = \frac{e}{h} (\nu_{\rm bulk} \mu_{\rm
source} - \nu_{\rm bulk} \mu_{\rm 3}), \nonumber \\
I_{1} & = & 0 = \frac{e}{h} (\nu_{\rm bulk} \mu_{1} - \nu_{\rm bulk}
\mu_{\rm source}), \nonumber \\
I_{2} & = & 0 = \frac{e}{h} [\nu_{\rm bulk} \mu_{2} - \nuC \mu_{1} -
(\nu_{\rm bulk} - \nuC) \mu_4], \nonumber \\
I_{3} & = & 0 = \frac{e}{h} [\nu_{\rm bulk} \mu_{3} - \nuC \mu_{4} -
(\nu_{\rm bulk} - \nuC) \mu_1], \nonumber \\
I_{4} & = & 0 = \frac{e}{h} (\nu_{\rm bulk} \mu_{4} - \nu_{\rm bulk}
\mu_{\rm drain}), \nonumber \\
I_{\rm drain} & = & -I  = \frac{e}{h} (\nu_{\rm bulk} \mu_{\rm
drain} - \nu_{\rm bulk} \mu_2), \nonumber
\end{eqnarray}
where $\mu_{\rm source}$, $\mu_{\rm drain}$, and $\mu_{l}$ are the
electrochemical potentials of the source, drain and reservoir $l =
1,2,3,4$, respectively. Here, $I_{l = 1,2,3,4} = 0$ since the
reservoirs $l$ are voltage probes. From the above set of current
relations, one can arrive at $\mu_2 - \mu_1 = (\nu_{\rm bulk}^{-1} -
\nuC^{-1})(h/e) I$, from which $G_{\rm ad-4T} = (\nuC^{-1} -
\nu_{\rm bulk}^{-1})^{-1} (e^2/h)$. $G_{\rm{ad-2T}}$ and
$G_{\rm{ad-dg}}$ can be derived in a similar manner.

\subsection{Simple Aharonov-Bohm oscillations}
\label{LOWB_AB}

\begin{figure}[th]
\centerline{\psfig{file=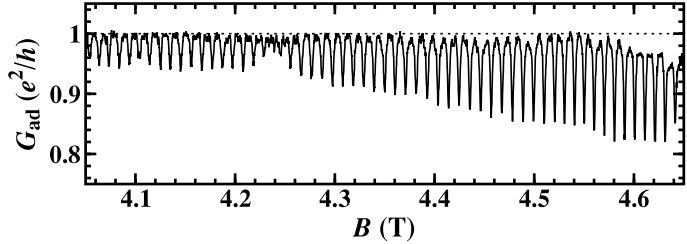,width=9cm}} \vspace*{8pt}
\caption{ Typical Aharonov-Bohm oscillations of conductance $G_{\rm
ad}$ in an antidot with $\nuC = 1$. Conductance dips appear from the
$\nuC = 1$ plateau (dotted line), indicating that back-scattering is
dominant. From \cite{Kataoka_unpubMace}.}\label{AB4T}
\end{figure}

Figure~\ref{AB4T} shows typical resonance structure in the
conductance $G_{\rm ad}$ for an antidot with $\nuC = 1$, as a
function of magnetic field. Back-scattering causes periodic
Aharonov-Bohm resonance dips from the $\nuC=1$ plateau.  From the
periodicity of 10.5~mT, the effective antidot radius is estimated to
be $0.35 \, \mu{\rm m}$ [see Eq. (\ref{ABperiod})].

\begin{figure}[th]
\centerline{\psfig{file=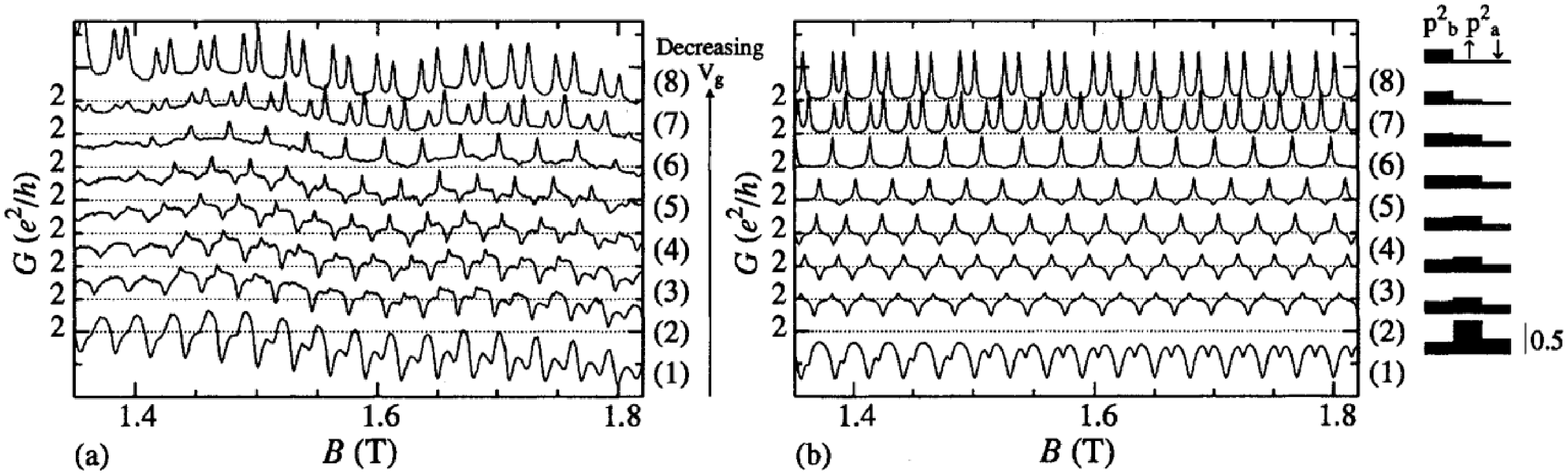,width=15cm}} \vspace*{8pt}
\caption{ (a) Aharonov-Bohm oscillations of antidot conductance $G$
with $\nuC = 2$, which show the transition from conductance peaks to
dips as the antidot gate voltage is increased negatively down the
curves. The curves are offset vertically for clarity. (b)
Corresponding model conductance curves and scattering probabilities.
From \cite{Mace}.} \label{LowB}
\end{figure}

The antidot states and their coupling to the extended edge channels
can be studied from the periodic nature of the Aharonov-Bohm
oscillations, the line shapes of the peaks or the dips, and their
heights.  For example, one can study the coupling of the antidot
states to the extended edge channels by varying the antidot gate
voltage. As the magnitude of the gate voltage decreases, the size of
the antidot is reduced and the coupling between the antidot states
and the extended edge channel $\aaa$ passing through the
constrictions (see Fig.~\ref{geometry}) becomes weaker.
As we have seen
in Sec.\/ \ref{SP}, the result is that the forward scattering is
favored more as the gate voltage becomes less negative.
Such competition [see Eq. (\ref{FWDBWDsc1})] between forward and
backward scattering was first studied experimentally in a
systematic way by Mace and coworkers \cite{Mace}.
%MODIF %at fairly low magnetic fields

\begin{figure}[th]
\centerline{\psfig{file=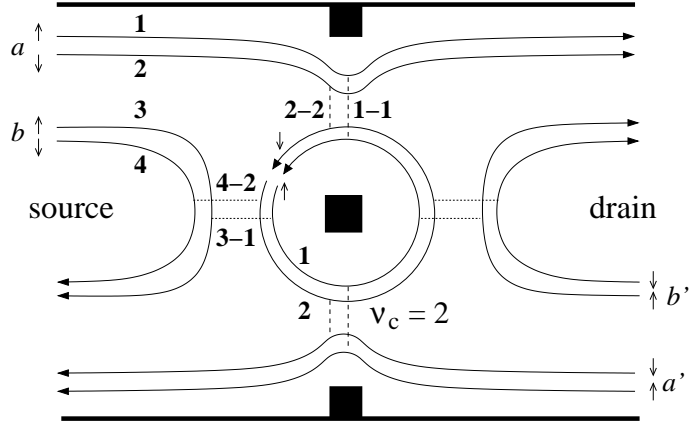,width=9cm}} \vspace*{8pt}
\caption{ The same schematic diagram of an antidot with $\nuC
= 2$ as in Fig. \ref{geometry}, except for introducing the following
notation for the Landau levels. The states coming from the spin-up
and spin-down states of the lowest Landau level are denoted as 1 and
2, respectively, and those from the spin-up and spin-down
states of the next Landau level as 3 and 4, respectively. This
notation is convenient to describe the scattering between
the antidot states and the extended edge channels. The
inter-Landau-level coupling between the antidot spin-up (down) state
and the channel $\bbb \uparrow$ ($\bbb \downarrow$) is represented
as 3--1 (4--2) scattering. Similarly, the intra-Landau-level
coupling between the antidot spin-up (down) state and the channel
$\aaa \uparrow$ ($\aaa \downarrow$) is represented as 1--1 (2--2)
scattering.} \label{geometry1234}
\end{figure}

In Fig.~\ref{LowB}, the antidot gate voltage is made more negative
for the lower traces.  In the topmost curve, paired peaks appear
from the $\nuC=2$ plateau. These peaks are caused by resonant
transmission through the two spin states of the lowest Landau level
(we call these edge states 1 and 2, see
Fig.~\ref{geometry1234}); the left-hand peak of the pairs is via
spin-down states (higher Zeeman energy, edge state 2), and the
right-hand peak is via spin-up states (lower Zeeman energy, edge
state 1). Let us call the spin-up and spin-down edge states of the
next Landau level 3 and 4, respectively. We denote inter-edge-state
scattering from edge state 4 to 2 as 4--2 and similarly for 3--1.
These processes scatter between Landau levels, preserving the spin.
The two peaks have similar amplitudes because the tunneling
distances of 4--2 and 3--1 scattering are almost the same (see
Fig.~\ref{geometry1234}). As the antidot voltage is increased
negatively, increasing the antidot size, these resonances shift
towards smaller magnetic field, as each state moves up in energy to
keep the enclosed flux constant. As the antidot voltage is increased
further, the amplitude of each left-hand peak starts to decrease,
and eventually the peaks turn into dips.  The right-hand peak
follows the same trend at more negative antidot gate voltages. This
behavior can be fully understood by the noninteracting model. As the
size of the antidot increases, the constriction width becomes
narrower, so the spin-down state starts to couple with the
corresponding extended edge channel transmitted through the
constrictions, of the same Landau level and spin. We denote this
tunneling as 2--2, and similarly 1--1 for the other spin. This
back-scattering is through exactly the same resonant state that
gives a left-hand resonant transmission (4--2) peak. In the absence
of 4--2 tunneling, this would give rise to a dip. With both 4--2 and
2--2 present, the dip and peak have the same linewidth, as that is
determined by the state's lifetime. The dip and peak compete,
reducing the peak height, and eventually the peak turns into a dip
as 2--2 becomes dominant. The back-scattering of the spin-up state
(1--1) starts at a later stage, when the constrictions are narrower,
since the tunneling distance for 1--1 is
longer than for 2--2 (see Fig.~\ref{geometry1234}).

\begin{figure}[th]
\centerline{\psfig{file=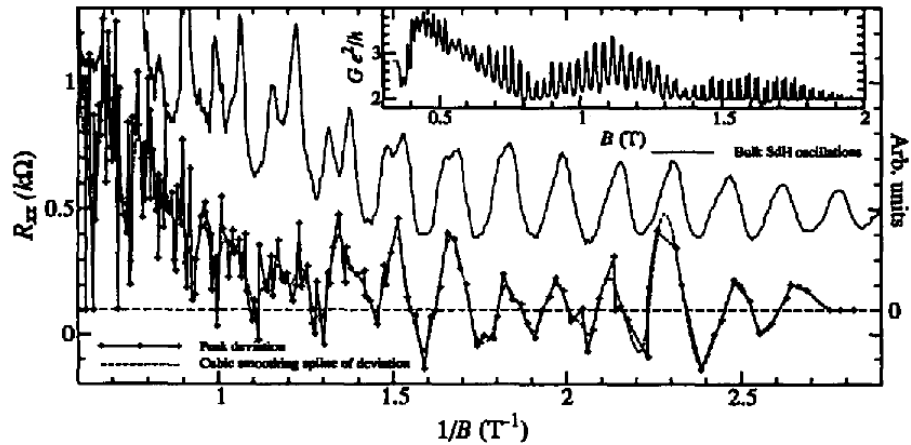,width=12cm}} \vspace*{8pt}
\caption{Inset: A typical conductance trace in which an aperiodicity
in the Aharonov-Bohm oscillations is clearly seen near 0.6 T. Upper
solid line: resistance $R_{xx}$ of the 2D region with that antidot
gate voltage set to zero. (At high field, the antidot is defined at
zero gate voltage, accounting for the absence of zeros.) Lower solid
line with points (right-hand axis): deviation of the peak position
against peak index, derived by removing a smoothed background; four
different antidot radii were used to obtain more points. Dashed
line: cubic smoothing spline performed on the solid line with
points. From \cite{Barnes96}.} \label{BarnesF1}\end{figure}

We note that
Aharonov-Bohm oscillations of antidot conductance can be used
to detect the oscillation of the Fermi energy in
the 2D system surrounding it \cite{Barnes96}. The deviation of the
period of the Aharonov-Bohm oscillations from the average
periodicity was found to oscillate in synchrony with the bulk
Shubnikov--de-Haas oscillations (see Fig.~\ref{BarnesF1}).
Numerical calculations were used to
show that the result was consistent with the oscillation of the
Fermi energy and the concomitant oscillation of the density of
states there.

As in these low-field examples, much can be learned about the behavior
and properties of antidot states in each field range by studying the
variation of the resonances as they are tuned with gate voltages.
Some experimental observations \cite{Kirczenow,Mace} can be
understood within the noninteracting electron model in Sec.\/
\ref{SP}, while others require an explanation taking electron
interactions into account. In the following sections, we describe
the experimental data that cannot be explained by a simple
noninteracting model, starting with the observation of the antidot
charging effect.

\subsection{Detection of antidot charging}
\label{EXPcharging}

The charging effect is a ubiquitous feature of quantum dots, almost
isolated small-size potential wells weakly coupled to reservoirs.
The large capacitive energy required to add an electronic charge to
the dot, and electron-electron interactions inside the quantum dots,
can prevent electron tunneling between the dot and the reservoirs,
resulting in Coulomb-blockade phenomena
\cite{ReviewDot,ReviewDot2}.

An antidot is an open system and does not electrostatically confine
any electrons. However, a sufficiently strong perpendicular magnetic
field applied to the antidot system can confine electrons around the
antidot and result in the formation of localized antidot states as
discussed in Sec.\/ \ref{SP}. Therefore, one may expect that
the Coulomb interactions between the localized antidot states cannot
be ignored and could play some role. Indeed, experimental
findings of $h/2e$ Aharonov-Bohm oscillations \cite{Ford,Sachrajda}
with a single antidot and unexpected behavior of molecular states in two
coupled antidots (antidot molecule) \cite{Gould} enhanced
this expectation, which will be reviewed in the next two
sections (Secs. \ref{EXPdouble} and \ref{EXPmolecule}).  Here, we
will focus on the observation of the direct evidence of the charging
effect in the antidot.

\begin{figure}[th]
\centerline{\psfig{file=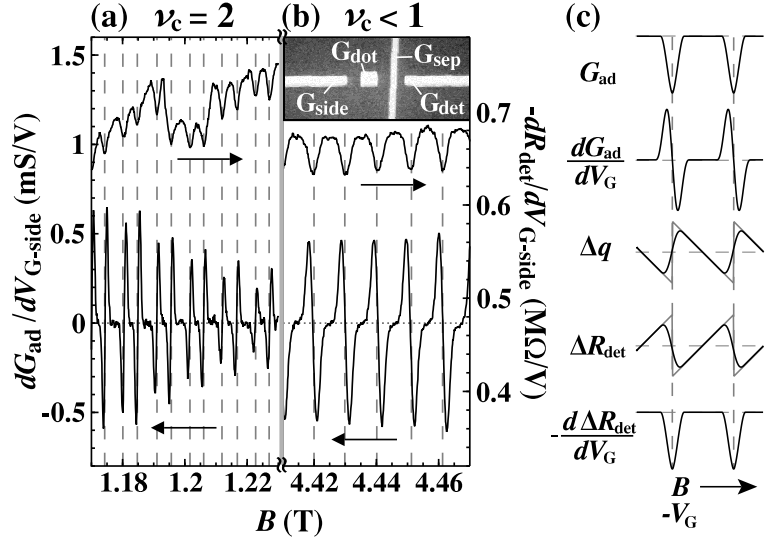,width=10cm}}
\vspace*{8pt} \caption{ An antidot setup with a noninvasive voltage
detector and the simultaneous measurement of the conductance $G_{\rm
ad}$ of the antidot system and the detector resistance $R_{\rm
det}$, as a function of magnetic field $B$. Left panel: $dG_{\rm
ad}/dV_{\rm G-side}$ and $-dR_{\rm det}/dV_{\rm G-side}$ are shown
in the two different regimes of antidot filling factors (a) $\nuC =
2$ and $\nu_{\rm bulk} = 7$ and (b) $\nuC < 1$ and $\nu_{\rm bulk} =
2$. Here, $V_{\rm G-side}$ is the side gate voltage. Vertical dashed
lines show that the dips of $-dR_{\rm det}/dV_{\rm G-side}$ coincide
with the zeros of $dG_{\rm ad}/dV_{\rm G-side}$, showing that the
Aharonov-Bohm oscillation of $G_{\rm ad}$ is accompanied by the
relaxation and accumulation of the excess charges $\Delta q$ around
the antidot. Inset: Scanning electron micrograph of the device.
Right panel (c): Illustration of the relation between the line
shapes of $G_{\rm ad}$, $dG_{\rm ad}/dV_{\rm G-side}$, $\Delta q$,
$R_{\rm det}$, and $dG_{\rm ad}/dV_{\rm G-side}$. From Ref.
\protect\cite{Kataoka_charging}. } \label{NoninvasiveDetection}
\end{figure}

Direct evidence of the antidot charging effect was found in Ref.
\cite{Kataoka_charging} using a noninvasive voltage probe
\cite{Field}. The noninvasive voltage detection was made (see
Fig.~\ref{NoninvasiveDetection}) by fabricating
a detector constriction
next to an antidot system, the lithographic dimensions of which are
similar to those described in Sec.\/ \ref{EXPsetup}. The detector
constriction is separated from the antidot system by the separation
gate $\rm{G}_{\rm sep}$ of width $0.1 \mu \rm{m}$, and the
resistance $R_{\rm det}$ of the constriction is adjusted to be very
sensitive to the variations of charges nearby by tuning the detector
gate to nearly pinch off that channel. The variation of the total
antidot charge is visible in the measurement of $R_{\rm det}$. Note
that a similar noninvasive voltage probe was first used by Field and
his coworkers to detect charge oscillations in a quantum dot
\cite{Field}.

Here, in order to enhance the detector sensitivity,
%MODIF %and to have a
%wider range of gate voltages for one setting of the detector,
the detector transresistance $-dR_{\rm det}/dV_{\rm G-side}$ was
measured by modulating the voltage on $\rm{G_{side}}$ at a low
frequency as shown in Figs.~\ref{NoninvasiveDetection}(a) and (b)
measured at two different antidot filling factors. Dips in
\sloppy$-dR_{\rm det}/dV_{\rm G-side}$ appear in phase with the
oscillations in the antidot transconductance $dG_{\rm ad}/dV_{\rm
G-side}$, which was measured simultaneously.  The relations between
the transresistance/transconductance line shapes and the
resistance/conductance line shapes are shown in
Fig.~\ref{NoninvasiveDetection}(c).  These show that the observed
dips in the detector transresistance correspond to sawtooth
oscillations in the detector resistance, which are the evidence of
the steady accumulation and sudden relaxation (presumably in units
of $e$) of some excess charge $\Delta q$ near the detector
constriction. The discrete steps in the charge oscillations coincide
with the resonances in the antidot conductance, as expected for
Coulomb blockade oscillations. Therefore it is concluded that the
source of the excess charge is the antidot.

\begin{figure}[th]
\centerline{\psfig{file=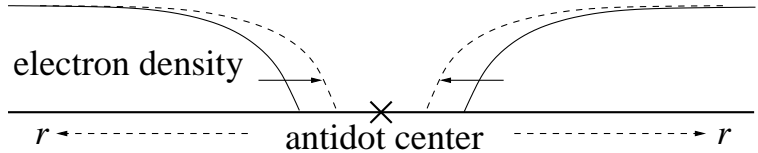,width=10cm}} \vspace*{8pt}
\caption{ Schematic diagram of the accumulation of excess charge
$\Delta q$ around the antidot as the magnetic field increases. The
electron density shift results from the enclosed magnetic flux
movement of the discrete single-particle states to keep the enclosed
magnetic flux constant. From Ref. \cite{Sim}.} \label{EXCESS}
\end{figure}

The above result confirms the following interpretation
\cite{Ford,Kataoka_charging,Sim}. As $B$ increases, all the states
encircling the antidot move inwards to keep the enclosed flux $BS_m$
constant, while the density of the positive background charges
(which come from ionized donors and gate voltages and preserve the
total charge neutrality of the whole system) is fixed (see
Fig.~\ref{EXCESS}). If the occupation of the antidot states does not
vary during their movement, excess (negative) charge $\Delta q$
accumulates around the antidot. The excess charge cannot be screened
by the electrons in completely filled (incompressible) states around
the antidot. The antidot therefore behaves like a quantum dot, with a
capacitive energy of $\Delta q^2 / (2C)$, where $C$ is an effective
capacitance of the antidot. The accumulated excess charge cannot
relax because of an energy cost for removing an electron,
until $\Delta q$ reaches $-e/2$, at which point
the total energy of the system
is the same as for an excess charge of $\pm e/2$, and
an electron can leave, making $\Delta q=e/2$;
here, the effect of the single-particle energy is ignored
for simplicity.
The process repeats,
with electrons tunneling from one lead to the other via this
resonant state with no energy cost, giving rise to a peak or dip in
$G_{\rm ad}$. Further increase of $B$ again gives rise to a positive
energy for adding or removing an electron, and so the system goes
off resonance. Resonant tunneling repeats with the periodicity of
the movement of the single-particle states, as each moves up to the
position formerly occupied by its neighbor. This corresponds to
adding $h/e$ of flux to the area enclosed by a state at the Fermi
energy, and so we have the usual $h/e$ Aharonov-Bohm periodicity.
Rigorous discussion of the resonant scattering will appear in Sec.\/
\ref{THexcess}, where electron interactions will be taken into
account.

\begin{figure}[th]
\centerline{\epsfig{file=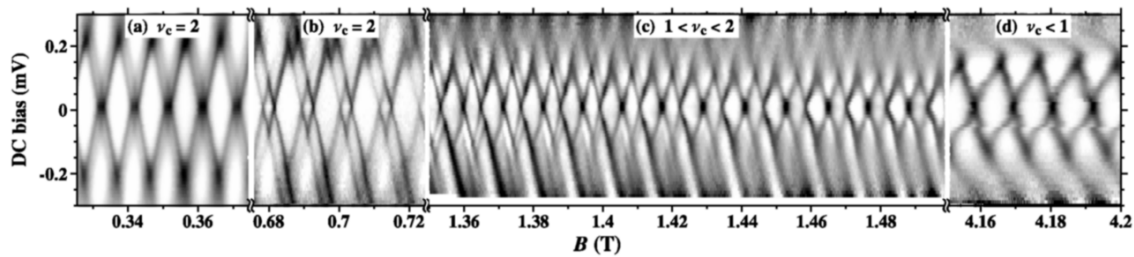,width=15cm}}
\vspace*{8pt} \caption{ Coulomb-diamond plots for the antidot. The
differential conductance of the antidot is plotted as a function of
the magnetic field $B$ and the source-drain dc bias. The same gate
voltages are used throughout. Dark regions correspond to positions
of resonance peaks [in (a) and (b)] or dips [in (c) and (d)]. The
background variation in the signal was subtracted to increase the
contrast. From Ref. \protect\cite{Kataoka_charging}. }
\label{CBdiamond}
\end{figure}

The antidot charging effect can be further analyzed from the
experimental observation \cite{Kataoka_charging} of the dependence
of the differential conductance $dG_{\rm{ad}}/dV_{\rm{sd}}$ on the
magnetic field $B$ and the source-drain dc bias $V_{\rm{sd}}$, which
is shown in Fig.~\ref{CBdiamond}. The plots show ``Coulomb
diamonds'', which have been used to investigate Coulomb
blockade in quantum dots \cite{ReviewDot,ReviewDot2}, with the
net charge being changed by the gate voltage rather than the magnetic
field. From the height of the diamonds, one can extract the energy
$\Delta E_{\rm tot} = e^2/C + \delta \epsilon$
to add one electron to the
dot, where $e^2/C$ is the charging energy of the dot and
$\delta \epsilon$
is the level spacing of the single-particle states with the
same spin. One can apply the same model
to an antidot to see whether there is any charging energy. In
Fig.~\ref{CBdiamond}(a)-(c), there are spin-split
resonances since the antidot has both spin states in those cases.
The spin-split resonances are manifested by the two neighboring
bigger and smaller diamonds, the sizes of which are expected to
correspond to $e^2/C + \delta \epsilon - E_{\rm Z}$ and $e^2/C +
E_{\rm Z}$. Here, $E_{\rm Z} = \epsilon^{\rm Z}_\downarrow -
\epsilon^{\rm Z}_\uparrow$ is the Zeeman splitting.
The average size of
$\Delta E_{\rm tot}$ for spin-up and down diamonds is $e^2/C +
\delta \epsilon/2$ and is found to be almost constant
throughout the range in
Fig.~\ref{CBdiamond}(a)-(c). Its value $\Delta E_{\rm tot} \simeq
140  \, \mu e {\rm V}$ is in good agreement with the analysis of the
temperature dependence of the Aharonov-Bohm scattering by the
antidot. Another important result of the
diamond plots is that they reveal the excitation spectrum of the
antidot as shown in the additional parallel lines to the smaller
diamonds. The energy gap between the parallel lines and the diamonds
can be interpreted as $\delta \epsilon - E_{\rm Z}$ or $E_{\rm Z}$,
depending on around which diamonds the excitation line appears. This
allows one to deduce the charging-energy contribution to $\Delta
E_{\rm tot}$.

The roughly constant value of $\Delta E_{\rm tot}$ throughout a wide
range of magnetic field can be understood as a coincidence, from the
expectation that the magnetic field dependence of $\delta \epsilon$
scales as $\sim 1/B$ for the antidot with a constant
potential slope [see Eq. (\ref{SPlevelspacing})], while $e^2/C$ can
increase for larger $B$ due to stronger magnetic confinement. The
analysis in Ref. \cite{Kataoka_charging} indicates $e^2/C \sim 100 -
120 \, \mu e {\rm V}$ at $B = 1.4 \, {\rm T}$. $\Delta E_{\rm tot}$
drops rapidly in the regimes where the antidot couples strongly with
the extended edge channels. This is because the coupling enhances
the capacitance, and thus the charging energy becomes smaller in
those regimes.

All these observations show that charging can indeed play
an important role in the antidot, depending on the relative sizes of
the single-particle and charging energies. The energy levels of the
antidot can be strongly affected by the charging energy, the
positions of the Aharonov-Bohm resonances may be determined by the
net charge around the antidot, and there is an accompanying
accumulation and relaxation of excess charge around the antidot.
The importance of this finding is that antidot Aharonov-Bohm
oscillations are not only governed by the quantization of enclosed
magnetic flux, but also by the capacitive energy of the system. This
charging model has led to the understanding of various effects in
antidots, as described in the following sections.

\subsection{$h/2e$ Aharonov-Bohm oscillations}
\label{EXPdouble}

In a quantum dot at a high magnetic field, the Aharonov-Bohm effect
can occur in combination with Coulomb charging
\cite{Taylor,Alphenaar,Sachrajda_QD_ex,Staring}. When the dot has
two spin species, {\it i.e.}, when the filling factor in the dot is
two, it was found that resonances from one spin species occur
exactly halfway between the neighboring resonances of the second
spin species. To explain this feature, a charging model of the dot
was proposed \cite{Dharma-Wardana,Sachrajda_QD_th}. In this section,
we will describe similar behavior
\cite{Ford,Sachrajda,Kataoka_double} observed in an antidot with
$\nuC = 2$. In certain ranges of magnetic field and antidot gate
voltage, the $\nuC=2$ antidot can show resonant tunneling with
period in magnetic field $\Delta B = \phi_0 / (2S) = h/(2eS)$, which
is exactly the half the Aharonov-Bohm period in Eq.
(\ref{ABperiod}).
Such observation of the $h/2e$ Aharonov-Bohm oscillations and the
spectator behavior of the antidot molecular states, which will be
reviewed in the next section, cannot be explained by the
noninteracting model, and they provided a strong motivation of
studying the antidot charging effect \cite{Kataoka_charging}
reviewed in Sec.\/ \ref{EXPcharging}.

\begin{figure}[th]
\centerline{\epsfig{file=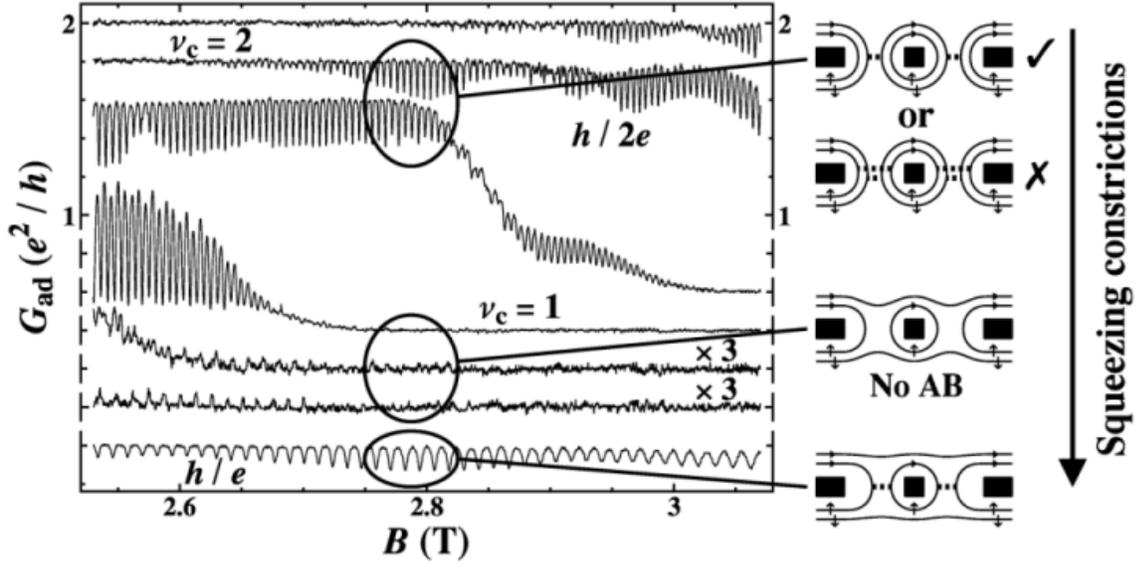,width=15cm,}} \vspace*{8pt}
\caption{ Left: $h/2e$ Aharonov-Bohm oscillations for the antidot
with $\nuC = 2$ in the regime of magnetic fields $B \sim 2.6 - 3 \,
{\rm T}$. Two curves are expanded by a factor of 3, as indicated.
Right: Schematic diagrams of antidot states and extended edge
channels (both are represented as solid lines) corresponding to the
cases of different constriction widths shown in the left panel.
Black dots indicate surface gates while the dotted lines represent
the tunneling between the antidot and the extended edge channels.
From Ref. \protect\cite{Kataoka_double}. } \label{DoubleAB_figure1}
\end{figure}

The top three curves in Fig.~\ref{DoubleAB_figure1} show typical
experimental data of the $h/2e$ Aharonov-Bohm oscillations, which
appear in the antidot with $\nuC = 2$. Since there are states of
each spin orientation encircling the antidot, one may conjecture
that the $h/2e$ oscillations occur as a result of two $h/e$
oscillations out of phase, each assigned to one spin species.
However, it was found \cite{Ford} that above certain magnetic
fields, the two sets of $h/e$ oscillations locked exactly in
antiphase with no difference in amplitude, leaving apparently pure
$h/2e$ oscillations. These $h/2e$ oscillations can be also observed
\cite{Ford} when the antidot gate voltage is swept
at a fixed magnetic field.

The experimental results in Fig.~\ref{DoubleAB_figure1} reveal an
important feature of the $h/2e$ oscillations. The data show the
Aharonov-Bohm oscillations as the constrictions on either side of
the antidot are squeezed by making the side-gate voltages more
negative. In the top three curves, the antidot filling factor is
$\nuC \sim 2$ and the oscillations have period $h/2e$, while in the
bottom curve, where $\nuC < 1$, $h/e$ oscillations appear. The fifth
and sixth curves from the top do not exhibit any Aharonov-Bohm
oscillations, except for very small peaks with period $h/e$ above
the $\nuC = 1$ plateau at the low-$B$ end.
In the case of the bottom
curve, the $h/e$ oscillations clearly come from back-scattering through the
antidot states with spin up (1--1 in the notation of Sec.\/
\ref{LOWB_AB}). As the constrictions are made wider, the $h/e$
oscillations stop as the spin-up antidot states become decoupled
from the extended edge channels that propagate through the
constrictions (fifth and sixth curves). Then, when the constrictions
are made even wider and the tunneling distance between the spin-up
antidot state and spin-up extended edge channels becomes larger,
there is no natural explanation why the back-scattering through the
spin-up states should recover. This means that for $h/2e$ Aharonov-Bohm
oscillations, tunneling is only
through antidot states with down spin, which have larger Zeeman
energy than spin-up states.

\begin{figure}[th]
\centerline{\epsfig{file=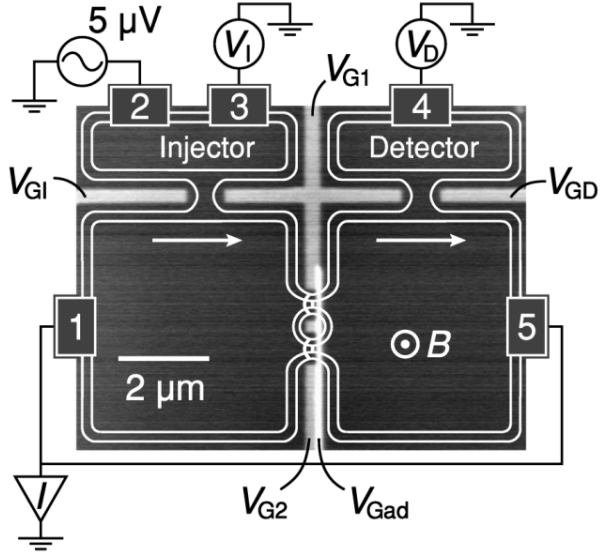,width=8cm}} \vspace*{8pt}
\caption{ Schematics of the measurement circuit for the
spin-injection and detection experiment. Metal gates directly on the
GaAs surface are seen in grey, and the second metal layer (on the
cross-linked PMMA) is seen in bright white. Edge states in the case
of $\nu_{\rm I}=\nu_{\rm D}=1$ and $\nuC=2$ are shown as
white lines. The arrows show the direction of electron flow. From
Ref. \protect\cite{Kataoka_Spin}.} \label{spininjsample}
\end{figure}

Indeed, in later experiments using selective injection and detection
of spin-resolved edge channels \cite{Kataoka_Spin}, it was confirmed
that the antidot states with up spin do not provide resonant paths
in the $h/2e$ Aharonov-Bohm oscillations. The selective injection
and detection can be used to distinguish the source of the resonance
scattering, because the equilibration length of parallel-propagating
edge channels can be extremely long \cite{Wees,Mueller} (longer than
$1 \, {\rm mm}$ depending on the conditions) especially at high
magnetic fields, when the spatial overlap between the channels is
small. In Ref. \cite{Kataoka_Spin}, as shown in
Fig.~\ref{spininjsample}, split gates are used to control the
current injection and detection of spin-resolved edge channels by
controlling the filling factors $\nu_{\rm I}$ and $\nu_{\rm D}$ of
the injector and the detector, respectively.

The effect of the selective injection and detection can be observed
in the detector nonequilibrium conductance defined as
\begin{eqnarray}
  G_{\rm D} = \frac{I}{V_{\rm D}} = \nu_{\rm I} \frac{e^2}{h} \frac{V_{\rm I}}{V_{\rm D}}.
\end{eqnarray}
Here, $I = \nu_{\rm I} (e^2/h) V_{\rm I}$ is the current flowing
through the injector constriction, and $V_{\rm I}$ and $V_{\rm D}$
are the voltages measured at the 2DEGs behind the injector and
detector constrictions, respectively. We consider the particular
case where the detector constriction is narrower than the injector
(i.e.\/ $\nu_{\rm I} \geq \nu_{\rm D}$). If there is no scattering
from the injected edge channels to other non-injected edge channels
(the injected edge channels maintain the potential $V_{\rm I}$), the
2DEG region behind the detector constriction charges up to $V_{\rm
I}$ (i.e.\/ $V_{\rm D}=V_{\rm I}$), as there is no drain contact
here and the potentials of all incoming edge channels are $V_{\rm
I}$. Therefore, $G_{\rm D}=G_{\rm I}$ is expected, where $G_{\rm I}
= I / V_{\rm I} = \nu_{\rm I} e^2/h$ is the injector two-terminal
conductance. On the other hand, if inter-edge-state scattering
occurs, the potential of the injected edge channels is lower than
$V_{\rm I}$ at the detector constriction, and therefore $V_{\rm
D}<V_{\rm I}$ and $G_{\rm D}>G_{\rm I}$. Similarly, if
back-scattering via antidot states (instead of inter-edge-state
scattering) occurs from the injected edge channels, the edge
channels lose their potential, resulting in $G_{\rm D}>G_{\rm I}$.

\begin{figure}[th]
\centerline{\epsfig{file=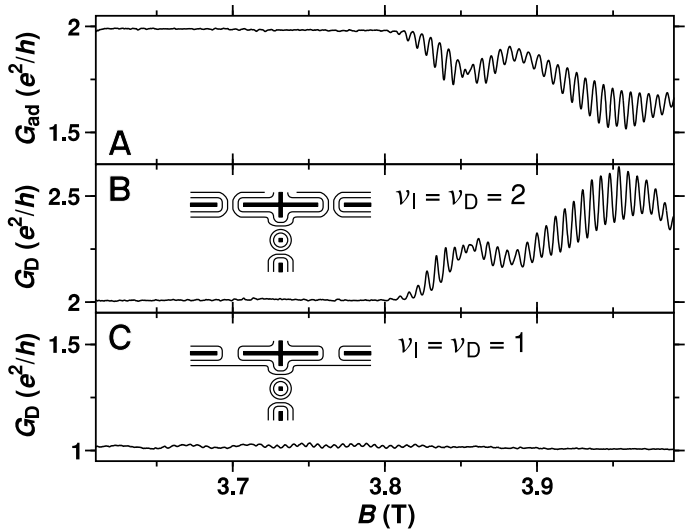,width=9cm}}
\vspace*{8pt} \caption{ (a) $G_{\rm ad}$ showing $h/2e$
Aharonov-Bohm oscillations. (b) Detector conductance $G_{\rm D}$
with $\nu_{\rm I}=\nu_{\rm D}=2$ and (c) with $\nu_{\rm I}=\nu_{\rm
D}=1$. The facts that $G_{\rm D}$ in (b) mirrors $G_{\rm ad}$ almost
perfectly and that oscillations are absent in (c) imply that only
the spin-down edge channel of the lowest Landau level is involved in
the resonant back-scattering process. From Ref.\/
\protect\cite{Kataoka_Spin}. } \label{DoubleAB_figure2}
\end{figure}

By this method, the spin states involved in antidot resonances can
be detected. Figure~\ref{DoubleAB_figure2}(a) shows $h/2e$
Aharonov-Bohm oscillations in the antidot conductance, and
Figs.~\ref{DoubleAB_figure2}(b) and (c) show the nonequilibrium
detector conductance with $\nu_{\rm I}=\nu_{\rm D}=2$ and $\nu_{\rm
I}=\nu_{\rm D}=1$, respectively.  When both spin states are injected
and detected [Fig.~\ref{DoubleAB_figure2}(b)], the shape of the
$h/2e$ oscillations that appear in $G_{\rm D}$ resembles that in
$G_{\rm ad}$ (only mirror imaged). On the other hand, when only the
spin-up edge channel is injected and detected, such oscillations are
not observed.  This provides direct evidence that, for pure $h/2e$
Aharonov-Bohm oscillations, the resonance occurs only through the
spin-down states (2--2 tunneling in the notation of Sec. \/
\ref{LOWB_AB}).

The $h/2e$ oscillations of spin-down states cannot be understood
from the noninteracting model described in Sec.\/ \ref{SPENERGY}. In
the noninteracting model, the $h/2e$ oscillations should be a simple
composition of the two $h/e$ oscillations coming from the two spin
species, and the phase shift between the two $h/e$ oscillations is
determined by the ratio between the Zeeman energy and the
single-particle level spacing (see Fig.~\ref{desp}). Since the ratio
depends on the antidot potential around the Fermi level and the
magnetic field, the noninteracting model cannot provide an
explanation of the sample-independent feature, namely the $\pi$ phase
shift. Moreover, in the absence of interactions, both spins should
participate in the resonant scattering, which is ruled out by the
experimental observation that only the spin species with the larger
Zeeman energy contributes to the resonances. Thus, it is necessary
to have a model taking electron interactions into account. The
formation of compressible rings around the antidot
\cite{Kataoka_double} was suggested to explain the $h/2e$
oscillations.  Later, a more generalized model was introduced
\cite{Sim}, considering capacitive interactions between excess
charges. We will review these models in Secs. \ref{EXPcompressible}
and \ref{THcapacitive}, respectively.

Concluding this section, we note that the $h/2e$ oscillations are
not the full story of $\nuC = 2$ antidots. We will see in Sec.
\ref{EXPKondo} that the Kondo effect \cite{Kataoka_Kondo} can also occur
in $\nuC = 2$ antidots when the magnetic field is not as strong as
in Fig.~\ref{DoubleAB_figure1}.  An explanation for the antidot
Kondo effect using the capacitive interaction model appears in Sec.\/
\ref{THKondo}.

\subsection{Spectator modes in antidot molecules}
\label{EXPmolecule}

In the previous section, we have seen that electron interactions can
strongly affect the period of Aharonov-Bohm scattering. Similar
behavior has been found \cite{Gould} experimentally in an antidot
molecule, a system of two coupled antidots. When the antidot
molecule has both the ``molecular'' orbitals circulating the whole
molecule and the ``atomic'' orbitals formed along each antidot, the
Aharonov-Bohm scattering by the molecule exhibits only the peaks or
dips with the period corresponding to the size of the atomic
orbitals, and the molecular orbitals seem not to participate in the
resonant scattering, {\it i.e.,} the molecular orbitals behave as
``spectator'' modes.  We will discuss this behavior in this section.

\begin{figure}
\centerline{\psfig{file=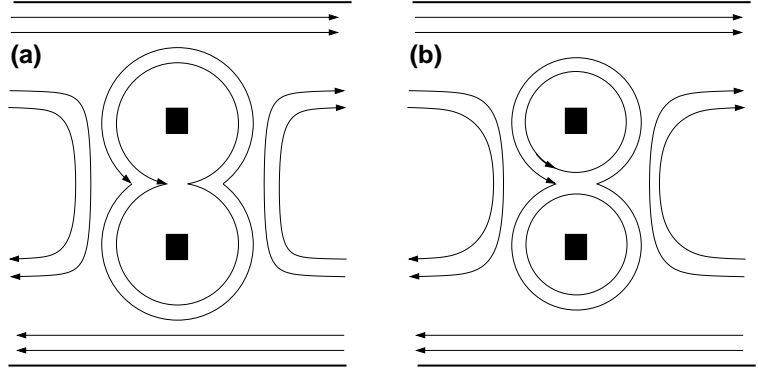,width=10cm}}
\vspace*{8pt} \caption{ Schematic diagrams of a $\nuC = 4$ antidot
molecule. The central two black squares indicate the antidot gates,
and the lines with arrows are the edge states. The antidot molecule
consists of two antidots and it has two spin-unresolved edge states
coming from the lowest two Landau levels. (a) All the states near
the Fermi level have ``molecular'' character when the two antidots
strongly couple to each other. (b) When the two antidots couple less
strongly, the outer states coming from the second lowest Landau
levels remain as molecular states while the inner states from the
lowest Landau levels become ``atomic'' states localized at one of
the antidots.} \label{ADMolecule}
\end{figure}

Figure~\ref{ADMolecule} shows a schematic view of the antidot
molecule. In Ref. \cite{Gould}, two antidots of diameter $0.2 \, \mu
{\rm m}$ are fabricated symmetrically across the width of a long
wide channel. The width of the narrow constriction between the two
antidots is $0.2 \, \mu {\rm m}$. The antidot
molecule has two spin-unresolved edge states coming from the lowest
two Landau levels, {\it i.e.}, $\nuC = 4$. By tuning two independent
antidot gate voltages, one can continuously transform the states of
the antidot molecule from those having ``molecular'' character to
those having ``atomic'' nature. For example, when the two antidots
strongly couple to each other, electrons cannot pass the
constriction between the two antidots, and all the edge states
circulate around both the antidots, having molecular character, as
shown in Fig.~\ref{ADMolecule}(a). On the other hand, when the two
antidots are tuned to couple less strongly, the outer states coming
from the second lowest Landau level keep their molecular nature
while the inner states from the lowest Landau level become atomic
states circulating around only one of the two antidots.

\begin{figure}
\centerline{\psfig{file=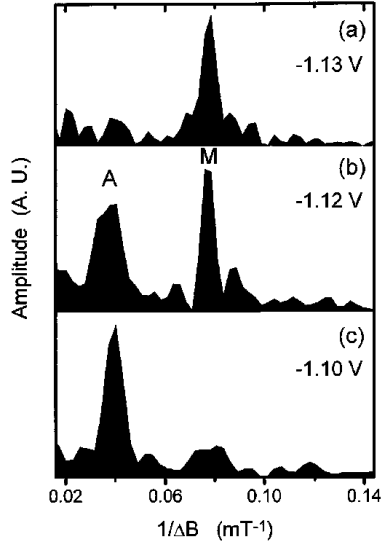,width=5cm}} \vspace*{8pt}
\caption{Fourier transforms of the Aharonov-Bohm oscillations for
the cases (a) that only molecular orbitals are formed, (b) that the
antidot gate voltage is adjusted just above the center
constriction's pinch-off gate voltage, and (c) that both molecular
and atomic orbitals are formed. Surprisingly, in (c), only the
periodicity corresponding to the atomic orbitals are shown. From
\protect\cite{Gould}.}\label{ADMolecule2}
\end{figure}

The nature of the states of the two antidots was studied by
observing the Aharonov-Bohm oscillations at fields below 1\,T. In
the case shown in Fig.~\ref{ADMolecule}(a), the period of the
Aharonov-Bohm scattering was found to be consistent with the value
estimated from the total size of the two antidots [see
Fig.~\ref{ADMolecule2}(a)]. So one can conclude that the molecular states
contribute to the Aharonov-Bohm oscillations, as expected from a
noninteracting model. However, such a simple story cannot describe
experimental observations in the case of Fig.~\ref{ADMolecule}(b),
where the antidot molecule has both the outer molecular orbitals
circulating the whole molecule and the inner atomic orbitals formed
along each antidot. One might expect from a noninteracting model
that Aharonov-Bohm scattering has two possible periods, one
corresponding to the total size of the molecule and the other to the
size of a single antidot, because the antidot molecule has both
types of states. Or else, only molecular modes might be observed if
the inner atomic modes are decoupled from the extended edge
channels. However, surprisingly, only the periodicity corresponding
to the atomic modes was found experimentally [see
Fig.~\ref{ADMolecule2}(c)]; both of the two periods exist over an
extremely narrow range of antidot gate voltage just above the
center constriction's pinch-off gate voltage [see
Fig.~\ref{ADMolecule2}(b)]. This indicates that the outer molecular
orbitals seem not to participate in the resonances.

Two different directions of understanding this ``spectator''
behavior of the outer molecular orbitals were proposed. One is to
consider disorder that would destroy the resonances coming from the
molecular orbitals \cite{Takagaki_ADM}, while the other is to
consider Coulomb interactions \cite{Gould}. We will briefly note the
possible effect of Coulomb interactions on the molecular orbitals in
Sec.\/ \ref{EXPcompressible}. More studies are required to
understand the mechanism of the spectator behavior of the molecular
orbitals, for example to confirm the observation in more
than one device.

\subsection{Line shape of Aharonov-Bohm resonances}
\label{EXPline}

In the previous two sections, we have seen possible signatures of
Coulomb interactions in the Aharonov-Bohm period. Another signature
of Coulomb interactions has been also experimentally observed in the
line shape of the resonant scattering by Maasilta and
Goldman \cite{Maasilta}. In the observation, the temperature dependence of a
single Aharonov-Bohm resonance peak was analyzed in the regime of
$\nuC = 1$ and $1/3$ to determine the coupling parameter
$\alpha_{\rm coup}$ between the antidot gate voltage and the energy
of antidot bound states at the Fermi level. The behavior of
$\alpha_{\rm coup}$ cannot be understood within the noninteracting
approach in Sec.\/ \ref{SP}. The study of the line shape has been
extended \cite{Karakurt} to question the presence of the formation
of compressible regions around the antidot at $\nuC = 2$ (see Sec.
\ref{EXPcompressible}). In this section, we will describe the study
of the line shape of a single resonance in the regime of integer
$\nuC$.

We first briefly summarize possible temperature dependences of the
line shape of a resonance in various parameter regimes of
temperature $k_{\rm B} T$, resonance level spacing $\Delta E$, and
level broadening $\Gamma \equiv (\Gamma_{\rm L} + \Gamma_{\rm R})/2$, following
Ref. \cite{Karakurt}. Here, the resonant states are located between
two (left and right) reservoirs, and their coupling energies to the
reservoirs are $\Gamma_{\rm L}$ and $\Gamma_{\rm R}$. In the Coulomb-blockade
regime with $\Gamma, \Delta E \ll k_{\rm B} T$, tunneling occurs
through many resonant states around energy $\epsilon_0$, resulting
in the line shape \cite{Averin,Shekhter} of
\begin{eqnarray}
G_T & = & G_p \frac{(\mu - \epsilon_0) / k_{\rm B} T}{\sinh [(\mu -
\epsilon_0 )/ k_{\rm B} T]}, \, G_p = \frac{e^2}{h} \frac{\rho
\Gamma_{\rm L} \Gamma_{\rm R}}{2 \Gamma}, \label{LINE_many}
\end{eqnarray}
if the density of states $\rho$ is constant.
On the other hand, for
$\Gamma, k_{\rm B} T \ll \Delta E$, only one resonant state with
energy $\epsilon_0$ contributes to the line shape
\cite{Beenakker_QD,Maasilta_FQHE}. In the case of $\Gamma \ll k_{\rm
B} T \ll \Delta E$, thermal broadening governs the line shape as
\begin{eqnarray}
G_T & = & G_p \cosh^{-2} (\frac{\mu - \epsilon_0}{2 k_{\rm B} T}),
\, G_p = \frac{e^2}{h} \frac{\pi \Gamma_{\rm L} \Gamma_{\rm R}}{4 k_{\rm B}
T \Gamma}, \label{LINE_thermal}
\end{eqnarray}
while in the case of $k_{\rm B} T \ll \Gamma \ll \Delta E$ one has
the Breit-Wigner Lorentzian line shape,
\begin{eqnarray}
G_T & = & \frac{e^2}{h} \frac{\Gamma_{\rm L} \Gamma_{\rm R}}{(\mu - \epsilon_0)^2
+ \Gamma^2}.
\label{LINE_BW}
\end{eqnarray}

\begin{figure}
\centerline{\psfig{file=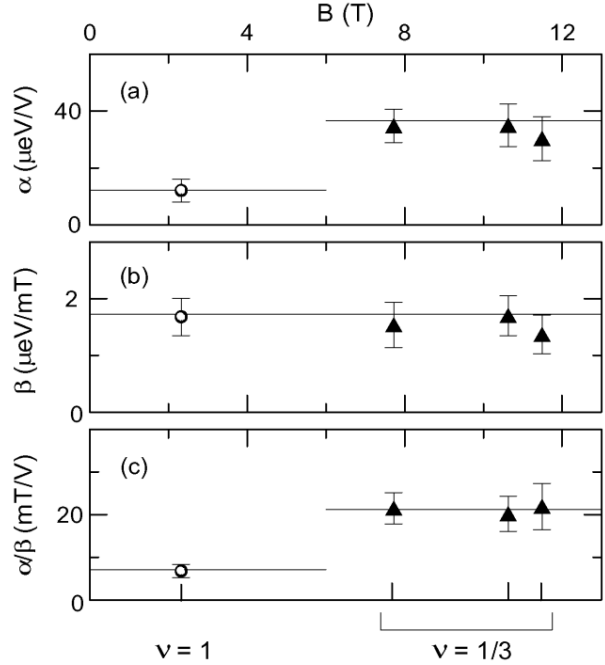,width=8cm}} \vspace*{8pt}
\caption{ (a) The coupling parameters $\alpha_{\rm coup}$, (b)
$\beta_{\rm coup}$, and (c) their ratio are shown as a function of
magnetic field. Both the cases of $\nuC = 1$ and $\nuC = 1/3$ are
investigated. From Ref \protect\cite{Maasilta}.}
\label{LINESHAPE_fig}
\end{figure}

In the work \cite{Maasilta} by Maasilta and Goldman, a single
resonance peak was analyzed as a function of the magnetic field $B$
or of the voltage $V_{\rm BG}$ on a large gate on the back of the
sample. At all measurement temperatures, all the resonant
Aharonov-Bohm peaks were found to fit the thermally broadened
Fermi-Dirac line shape in Eq. (\ref{LINE_thermal}) very well. The
line shape of $G_T \propto \cosh^{-2} [ \alpha_{\rm coup} (V_{\rm
BG}^m - V_{\rm BG})/(2 k_{\rm B} T) ]$ or $G_T \propto \cosh^{-2} [
\beta_{\rm coup} (B^m - B)/(2 k_{\rm B} T) ]$ was used, depending on
whether $V_{\rm BG}$ or $B$ varies. Here, the coupling parameters
are introduced as
\begin{eqnarray}
\alpha_{\rm coup} & \equiv & | d(E_m - \mu) / d V_{\rm BG}|,
\label{alpha_coup} \\
\beta_{\rm coup} & \equiv & | d(E_m - \mu) / d B|,
\end{eqnarray}
$E_m$ is the energy of the $m$-th resonance state, and $V_{\rm
BG}^m$ and $B^m$ are the positions of the $m$-th peak in gate
voltage and magnetic field, respectively.

As shown in Fig. \ref{LINESHAPE_fig}, the coupling parameter was
found to have a constant value of $\alpha_{{\rm coup}, \nuC =1} = 12
\pm 4 \, \mu e {\rm V}/{\rm V}$ in the regime of $\nuC = 1$,
whatever the field at which this was chosen to occur. Surprisingly,
this value is very similar to the value of $\alpha_{{\rm coup}, B=0}
= d \mu / dV_{\rm BG}$ obtained in the model of two-dimensional
electrons at zero magnetic field. Moreover, it has also a simple
relationship with the value observed in the $\nuC = 1/3$ regime as
$\alpha_{\rm coup, \nuC = 1/3} \simeq 3 \alpha_{\rm coup, \nuC =
1}$. As a result, the addition energy $\Delta E_{\rm tot}$, which
can be obtained from the coupling parameters and the Aharonov-Bohm
period in gate voltage ($\Delta V_{\rm BG}$) or in magnetic field
($\Delta B$) via $\Delta E_{\rm tot} = \alpha_{{\rm coup}, \nuC =1}
\Delta V_{\rm BG} = \beta_{{\rm coup}, \nuC =1} \Delta B$, has an
almost constant value for a wide range of $B$. This constant value
of $\Delta E_{\rm tot}$ cannot be understood in a noninteracting
model, in which one may expect $\Delta E_{\rm tot} \sim 1/B$ [see
Eq. (\ref{SPlevelspacing})].  Maasilta and Goldman suggested that
the constant $\Delta E_{\rm tot}$ arises from electron interactions
in the self-consistent electrostatics of the quantum Hall edges.
Interestingly, a similar result was found in the antidot charge
detection experiment \cite{Kataoka_charging} as explained in Sec.\/
\ref{EXPcharging}. In Ref. \cite{Kataoka_charging}, the constant
$\Delta E_{\rm tot}$ was explained as coincidental cancellation in
the $B$ dependence of the charging energy and the single-particle
level spacing. The observation of constant $\Delta E_{\rm tot}$
indicates that Coulomb interactions are important in the structure
of antidot states.

\subsection{Compressible regions around antidots}
\label{EXPcompressible}

Along the extended edges of quantum Hall systems, strips of
compressible regions, where the in-plane electric field can be
screened, are considered to form where Landau levels intersect the
Fermi level \cite{Beenakker,Chang,Chklovskii1,Chklovskii2}.
This has been confirmed
by experiments in bulk 2DEGs (for example, see Ref. \cite{Wei}). One
may therefore question whether similar edge reconstruction causes
the formation of compressible regions around an antidot, since an
antidot provides an edge region of a quantum Hall system although
the length of the antidot edge is finite. In this section, we will
discuss whether the formation of compressible regions around an
antidot is compatible with existing experimental observations, and
introduce recent experimental efforts
\cite{Kataoka_double,Karakurt,Kataoka_comment,Goldman_reply,Michael}
of searching for compressible regions around an antidot.

\begin{figure}
\centerline{\epsfig{file=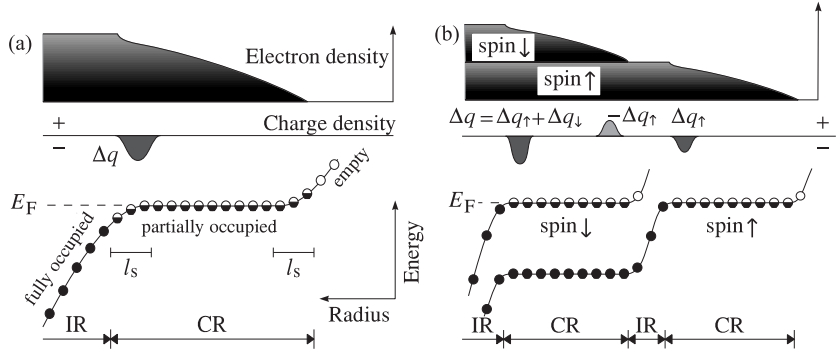,width=11cm}}
\vspace*{8pt} \caption{ Schematic diagrams of electron densities and
energy-level structures around an antidot in the cases of (a) $\nuC
= 1$ and (b) $\nuC = 2$ where compressible regions are assumed to be
formed. In (b), the two spin branches of the lowest Landau levels
form two compressible regions (CRs) separated by one incompressible
region (IR). The outer compressible region tries to screen excess
charge $\Delta q_\uparrow$ of the inner spin-up compressible region,
therefore, $-\Delta q_{\uparrow}$ piles up at the inner edge of the
outer compressible region in the case of perfect screening. As a
result, the net charge $\Delta q = \Delta q_\uparrow + \Delta
q_\downarrow$ built up at the outer edge of the outer compressible
region is the sum of the charging of both spins.
Modified from Ref. \protect\cite{Kataoka_double}. } \label{CRModel}
\end{figure}

Compressible regions formed along extended quantum Hall edges
consist of partially filled states, so there is a continuous density of states
around the Fermi level. One might consider that in the case of
antidots this feature is incompatible with the experimental
observation of discrete Aharonov-Bohm conductance resonances, because the
discrete resonances require states to pass the Fermi energy one by one,
{\it i.e.,} a discrete density of states. However, the introduction
of the charging effect discussed in Sec. \ref{EXPcharging} removes
the requirement of a discrete density of states for observing
discrete resonances---resonant tunneling only occurs once per
period, when the net charge around the antidot is $\pm e/2$, so that
there is no energy cost for tunneling. Then the screening property
of compressible regions can give the key to a heuristic picture for
understanding electron interactions in antidots, as we will see
below in a compressible-ring model.

The model with compressible rings around a
$\nuC = 2$ antidot was suggested \cite{Kataoka_double} to explain
the $h/2e$ Aharonov-Bohm oscillations. In this model, two
compressible rings, {\it i.e.,} spin-up and down rings, are assumed
to form around the antidot, although it is not clear
\cite{Kataoka_double,Karakurt,Kataoka_comment,Goldman_reply,Ihnatsenka,Michael}
whether such compressible
regions can be formed for the size of experimentally fabricated
antidots; we will return to this issue later in this section. In
both the compressible rings, a net excess charge accumulates as the
magnetic field increases, as shown in Fig.~\ref{CRModel}(b). The
incompressible ring formed between the two compressible rings acts as
an insulating barrier, so that the two compressible regions form a
capacitor. This capacitor mediates the interactions between the
excess charges accumulated at the inner ring and those at the outer
ring. Note that this interaction depends on the screening nature of
the compressible rings (at least the outer one) and the
environment of the antidot. Thus in the outer edge of the outer
compressible ring, the total excess charge, coming from the two
compressible rings, can pile up, and its relaxation period satisfies
$\Delta B = \phi_0 / (2S) = h/(2eS)$, resulting in $h/2e$
oscillations that arise from tunneling through the spin-down states
only (``2--2'' tunneling). This is in good agreement with the
experimental observations shown in Figs.~\ref{DoubleAB_figure1} and
\ref{DoubleAB_figure2}.

Another way of looking at this model is to consider the outer
compressible region as a conducting cylinder (to the extent that
electric field lines from the inner compressible ring terminate on
the outer one rather than extending to the gates ten times further
away). Increasing the magnetic field ``pumps'' charge into the
region bounded by the (insulating) incompressible strip outside this
cylinder, at a rate of $e$ per $h/e$ of flux increase, for
\emph{each} of the spin-up and down Landau levels. By Gauss's Law,
it is irrelevant what happens inside the conducting cylinder. The
extra charge must appear on the outer edge of the cylinder, so it
accumulates at a rate of $e$ per $h/2e$. Thus every time the flux
increases by $h/2e$, an electron can leave, and so there are
tunneling resonances, as for conventional Coulomb-blockade, with
period $h/2e$, and hence ``double-frequency'' oscillations are
observed.
In Fig.~\ref{CRModel}(b), $\Delta q_\downarrow$ is the change in
charge in the outer compressible region, but it is interesting to
realize that this is not purely caused by extra spin-down charge. In
fact, though the charge in the spin-up Landau level mainly goes into
the inner ring (as $\Delta q_\uparrow$), the remainder serves to
increase the electron density in that spin's incompressible states
in the region of the other spin's compressible region (filled
circles in that region in Fig.~\ref{CRModel}(b)).

We note, firstly, that this model really only requires a
compressible outer ring, so the inner one may be incompressible (as
found in recent calculations \cite{Ihnatsenka}, see Sec.\/
\ref{THcompressible}) and secondly, that the model was later
generalized \cite{Sim} into a capacitive interaction model without
the assumption of compressible regions as such (see Sec.\/
\ref{THcapacitive}).

This compressible-ring model might also give a clue to understanding
the spectator modes observed in antidot molecules discussed in
Sec.\/ \ref{EXPmolecule}. Excess charges accumulated in the inner
atomic orbital region can be screened by the outer molecular
orbitals. Then, the total excess charge piled up in the outer
orbital region comes from both the outer and the inner regions, and
the relaxation of the total excess charge does not necessarily have
the resonance period corresponding to the total size of the antidot
molecule, though it seems unlikely that it would have
the required (atomic) periodicity.
Also, the excitation spectra observed in Fig.~\ref{CBdiamond}
imply that compressible regions do not form at low
fields below about 2~T.
More studies are required to understand the mechanism of the
spectator behavior.

\begin{figure}
\centerline{\epsfig{file=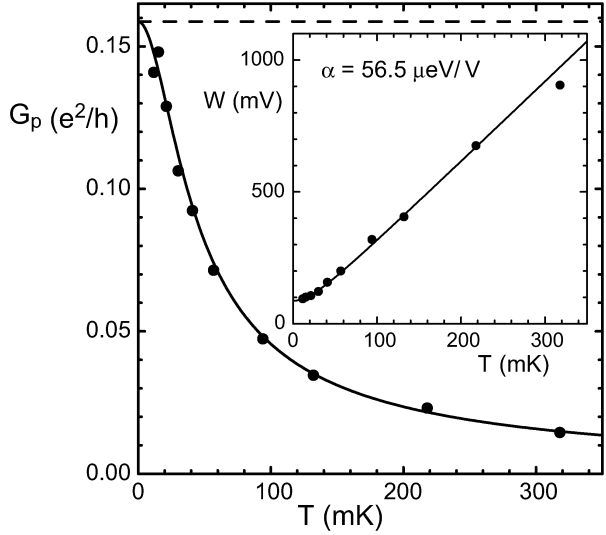,width=8cm}} \vspace*{8pt}
\caption{ Peak tunneling conductance $G_p$ as a function of
temperature $T$. $G_p$ is obtained by fitting line shapes of
measured peaks to $G_T = G_p \cosh^{-2} [(V_{G,{\rm AD}0} -
V_{G,{\rm AD}})/W]$, which comes from Eq. (\ref{LINE_thermal}). For
temperatures $35 \, {\rm mK} < T < 320 \, {\rm mK}$, $G_p \propto
1/T$ (solid curve). Inset: The dependence of the width parameter
[see Eq. (\ref{alpha_coup})]. For $T > 35 \, {\rm mK}$, $W \propto
T$. From Ref. \protect\cite{Karakurt}. } \label{GpTdependence}
\end{figure}

Although the compressible-ring model may be useful in understanding
certain experimental observations, it has been controversial whether
compressible regions really form around antidots, or if so under
what conditions.  Karakurt and coworkers studied the line
shape of resonant peaks in the regime of $\nuC = 2$ in order to
understand the antidot state structure \cite{Karakurt}. The radius
of the measured antidot was estimated as $0.34 \, \mu{\rm m}$ from
the Aharonov-Bohm period, and the range of magnetic
field investigated was around $B \sim 1 \, {\rm T}$
in the $\nuC = 2$ regime. To
see any possible signature of the formation of compressible regions
around a $\nuC = 2$ antidot, the measured line shape was fitted by
the two curves in Eqs. (\ref{LINE_many}) and (\ref{LINE_thermal}).
Over the range of temperature, $35 \, {\rm mK} < T < 320 \,
{\rm mK}$, $G_p$ was found to follow $G_p \sim 1/T$ (see
Fig.~\ref{GpTdependence}), indicating that Eq. (\ref{LINE_thermal})
provides a better fit. This means that the resonant tunneling
mechanism through the antidot is different from the simple resonance
model in the regime of $\Gamma, \Delta E \ll k_{\rm B} T$, where a
$T$-independent $G_p$ is expected. Based on this observation and on
the fact that there should exist many partially filled
single-particle states in compressible
regions \cite{Beenakker,Chang,Chklovskii1,Chklovskii2},
Karakurt and his coworkers concluded that
there is no compressible region around the studied antidot.

However, this conclusion was questioned
\cite{Kataoka_comment,Goldman_reply}. The assumption made by
Karakurt \emph{et al}.\/ was that the simple resonance model with a
large number of states within $k_{\rm B} T$ should describe the
temperature dependence of tunneling via a compressible region. The
$T$-independent $G_p$ of the line shape in Eq. (\ref{LINE_many}) is
derived based on the assumption that the density of states is
constant over a range wider than $k_{\rm B} T$. This is unlikely to be the
case in compressible regions, where the density of states is
expected to have a peak at the Fermi level. Calculations \cite{Lier}
suggest that the width of the peak in the density of states is of
the order of $k_{\rm B} T$ and is therefore temperature-dependent.
In such a case, Eq. (\ref{LINE_many}) is not applicable, and the
$1/T$ dependence measured by Karakurt \emph{et al}.\/ is expected for
$G_p$, assuming that the width of the compressible region, and hence
the number of states through which tunneling occurs, are independent
of $T$.

In addition, one may easily expect the formation of compressible
regions to be strongly dependent on magnetic field, antidot size,
electron density, etc. Therefore, the absence of compressible
regions in one experimental regime may not mean their absence in
another. Compressible regions were invoked by Kataoka \emph{et
al}.\/ in order to explain ``double-frequency'' oscillations
\cite{Ford,Sachrajda,Kataoka_double,Kataoka_Spin} at $B \sim 3$~T,
three times higher than the range investigated in Ref.
\cite{Karakurt}.

%Karakurt and his coworkers
%measured
%$h/2e$ oscillations at relatively low magnetic field of 1 T,
%whereas in other experiments higher magnetic fields $\sim 3$ T
%were required to observe pure $h/2e$ oscillations
%\cite{Ford,Sachrajda,Kataoka_double,Kataoka_Spin}.

Michael \emph{et al}.\/ investigated the single-particle energy
spacing $\delta \epsilon$ and the charging energy, both deduced
from the excitation spectrum, as a function of $B$ ($B < 2$~T)
\cite{Michael}. It was found that $\delta \epsilon$ decreased
faster than the expected $1/B$ dependence. No excitation spectrum
has been observed in the ``pure double-frequency'' regime at
higher fields, where the compressible model describes the system
well \cite{Kataoka_double}. This may be evidence that the antidot
potential flattens, forming compressible regions, as the magnetic
field increases.  However, in the experiments, the antidot
potential is not circularly symmetric because of side
constrictions. The bulging of the states in the constrictions can
also lead to a similar deviation from the $1/B$ behavior
of $\delta \epsilon$ as the edge states are reflected from the
constriction \cite{Bassett}, though whether this fully explains
the behavior of $\delta \epsilon$ is not yet clear;
the data in Ref. \cite{Michael} were mainly taken on resonant
transmission peaks above the $\nuC = 2$ plateau, where the effect of
the edge-state reflection from the constriction is expected to be weak.
Complete understanding of this regime still remains elusive.

In order to investigate the possibility of antidot compressible
regions from the theoretical side, Ihnatsenka and Zozoulenko
recently reported \cite{Ihnatsenka}, based on spin density-functional theory,
that the formation mechanism of the compressible
regions around an antidot is very different from the case of
extended edges \cite{Chklovskii1,Chklovskii2}, and that the exchange
interaction is very important in the case of $\nuC = 2$. This
theoretical work will be discussed in Sec. \ref{THcompressible}.
Further systematic experimental and theoretical studies on the
formation of compressible regions around an antidot are required,
and such studies should be valuable in understanding local disorder
regions in the integer quantum Hall regime.

\subsection{Kondo-like zero-bias anomaly}
\label{EXPKondo}

\begin{figure}[th]
\centerline{\psfig{file=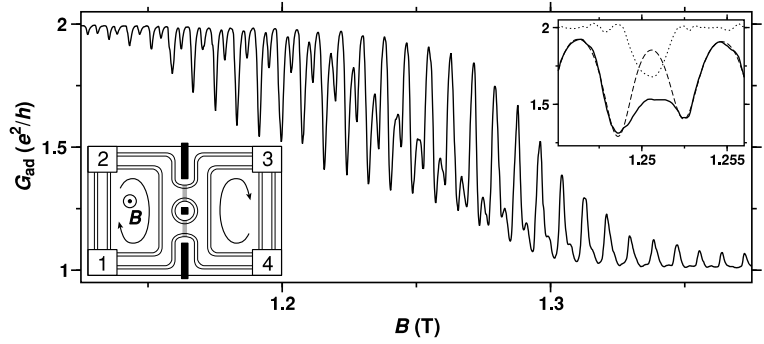,width=10cm}}
\vspace*{8pt} \caption{ Antidot conductance $G_{\rm ad}$ at a
lattice temperature of $25 \, {\rm mK}$ as a function of the
magnetic field $B$ in the transition region between the $\nuC = 2$
and 1 plateaus. Top-right inset: The fitting of a pair of dips
around $B = 1.25 \, {\rm T}$ by using four dips proportional to the
Fermi-function derivative. The solid curve is the experimental
result, while the dashed line is the fit. The dotted line shows the
difference between the experimental curve and the fit (offset by
$2e^2/h$). Bottom-left inset: Schematic view of the measured antidot
geometry, which is identical to the setup in Fig.~\ref{geometry}.
The numbered rectangles on the corners represent Ohmic contacts.
From Ref. \protect\cite{Kataoka_Kondo}. } \label{Kondo_figure1}
\end{figure}

The Kondo effect is one of the most well-studied many-body
phenomena. It arises from the interactions between a localized
electron spin and many free electrons \cite{Kondo,Hewson}. Recently,
there has been renewed interest in the effect, as it was predicted
\cite{Ng,Glazman} and observed
\cite{Goldhaber-Gordon,Cronenwett,van_der_Wiel,Goldhaber-Gordon2} in
quantum dots. The localized spin arises naturally in a quantum dot,
when the dot is in the Coulomb-blockade regime and contains an odd
number of electrons. The experimental tunability of the parameters
of quantum dots enables the study of the Kondo effect in great
detail. In Sec.\/ \ref{EXPcharging}, we saw some similarities
between a quantum dot and an antidot in the integer quantum Hall
regime, such as Coulomb blockade. However, despite the similarities,
one may not expect the Kondo effect in antidots because of the large
Zeeman splitting and hence the lack of apparent spin degeneracy. In
fact, in a $\nuC = 2$ antidot, Kondo-like behavior in the
conductance of an antidot has been observed experimentally, with
no Zeeman splitting as a function of DC bias. We note that
Kondo behavior in strong magnetic fields has also been observed in a
quantum dot \cite{Keller}, and attributed to an unpaired
electron at the edge of the dot that arises periodically as
electrons are redistributed between the center and the edge, but
this does show the expected Zeeman splitting in DC bias. Therefore,
the mechanism behind the antidot Kondo effect must be different from
the cases of quantum dots. In this section, we will review the
experimental observation of the Kondo-like behavior, while the
theoretical model for the effect will be discussed in Sec.
\ref{THKondo}.

Figure~\ref{Kondo_figure1} shows typical behavior of the conductance
$G_{\rm ad}$ of an antidot (see also Fig.~\ref{geometry}) in the
transition regime between $\nuC = 2$ and $\nuC = 1$. The effective
antidot radius is estimated from the measured Aharonov-Bohm period
as $r_{\rm AD} \sim 0.36 - 0.40 \, \mu {\rm m}$, depending on the
antidot gate voltage, and the bulk filling factor is $\nu_{\rm bulk}
\simeq 10$.  At low magnetic field $B$ where $G_{\rm ad}$ is close
to the $\nuC = 2$ quantum Hall plateau value of $2e^2 / h$, one can
see that there are two conductance dips in one Aharonov-Bohm period
$\Delta B = h/(eS)$. As $B$ increases, the coupling between the
antidot and the extended edge channels becomes stronger, since the
edge states move towards the center of the antidot constriction. As
a result, the amplitude of the dips increases. At the same time, two
neighboring dips become paired and eventually unrecognizable as two
independent dips about $1.3 \, {\rm T}$ as $G_{\rm ad}$ approaches
the $\nuC =1 $ plateau value of $e^2/h$. Very similar features were
observed between the $\nuC = 2$ and $\nuC = 1$ plateaus with
different gate voltages and with different magnetic fields from 0.8
to 1.5 ${\rm T}$. At magnetic fields above $3 \, {\rm T}$, the
pairing of the dips disappears, and instead pure $h/2e$
Aharonov-Bohm oscillations appear.

The pairing was analyzed by using a fit based on the Fermi-function
derivative. As shown in the inset of Fig.~\ref{Kondo_figure1}, such
a simple fit cannot reproduce the observed data. The fit shows that
there seems to be another dip in the intra-pair gap, resulting in a
large discrepancy in that region. This extra dip cannot be explained
by the noninteracting model in Sec. \ref{SP}, in which there are
only two dips within the $B$ range of one Aharonov-Bohm period.

\begin{figure}[th]
\centerline{\psfig{file=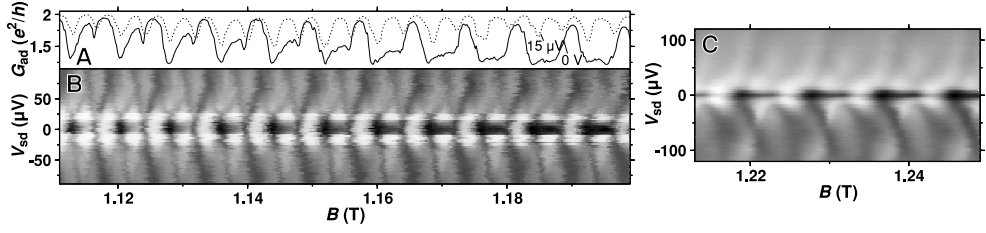,width=13cm}} \vspace*{8pt}
\caption{ The dependence of $G_{\rm ad}$ on source-drain bias
voltage $V_{\rm sd}$. (a) $G_{\rm ad}$ vs $B$ at $25 \, {\rm mK}$
with $V_{\rm sd} = 0 \, {\rm V}$ (solid line) and with $V_{\rm sd} =
15 \, \mu {\rm V}$ (dashed). (b) Gray-scale plot of the differential
conductance $G_{\rm ad}$ against $B$ and $V_{\rm ad}$ (white:
$2e^2/h$, black: $1.2 e^2/h$). The horizontal dark lines along
$V_{\rm ad} = 0$ show the extra dip (see text). (c) Similar
gray-scale plot to (b) in the regime of stronger coupling between
the antidot and the extended edge channels (white: $1.8 e^2/h$,
black: $1.1 e^2/h$). From Ref. \protect\cite{Kataoka_Kondo}. }
\label{Kondo_figure2}
\end{figure}

\begin{figure}[th]
\centerline{\psfig{file=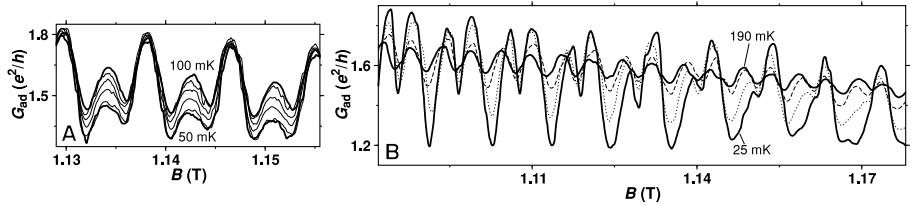,width=12cm}} \vspace*{8pt}
\caption{ (a) Temperature dependence of $G_{\rm ad}$ vs $B$ with
small increments of temperature $\sim 10 \, {\rm mK}$. (b) The same
as (a), but representative traces from a wider range of $T$. From
Ref. \protect\cite{Kataoka_Kondo}. } \label{Kondo_figure3}
\end{figure}

The extra dip has the following features. Firstly, it is absent in a
magnetic field larger than $3 \, {\rm T}$. Secondly, when a small
source-drain dc bias $V_{\rm sd}$, even as small as 15 $\mu$V, is
applied, the extra dip vanishes, leaving two well-defined dips, as
shown in Fig.~\ref{Kondo_figure2}(a). The Coulomb-diamond structures
of dark lines arise at large bias $V_{\rm sd}$ in
Fig.~\ref{Kondo_figure2}(b), indicating the two well-defined dips.
Note that from the height of the diamond, the charging energy can be
estimated to be $\sim 60 \, \mu e {\rm V}$. Thirdly, the extra dip
becomes stronger as $B$ increases [see Fig.~\ref{Kondo_figure2}(b)
and \ref{Kondo_figure2}(c)]. We note that the coupling between the
antidot and the extended edge channels increases in stronger $B$.
Fourthly, the extra dip becomes suppressed as the temperature $T$
increases, as shown in Fig.~\ref{Kondo_figure3}. At $T \sim 190 \,
{\rm mK}$, the extra dip is almost absent. The amplitude of the
extra dip decreases monotonically as $T$ increases.

The above bias and temperature dependences of the extra dip are
strikingly similar to those of the Kondo effect in quantum dots
(although  the zero-bias anomaly persists over too small a range of
temperatures for a logarithmic temperature dependence, which is the
hallmark of the Kondo effect, to be discernible). Moreover, stronger
coupling normally enhances the Kondo features, which is also in good
agreement with the above dependence of the extra dip on the
coupling. The observation of the antidot Kondo effect may not be
surprising, based on the similarities between quantum dots and
antidots. On the other hand, a counter-argument would be that the
Kondo effect cannot occur, because the antidot has considerable
Zeeman splitting of $E_{\rm Z} = 30 \, \mu e {\rm V}$ at $B = 1.2 \,
{\rm T}$. The width of the observed zero-bias anomaly of
$\sim 20 \, \mu e {\rm V}$ implies that any splitting in bias can be
at most about one-third of the expected Zeeman splitting at that
field. We will see in Sec. \ref{THKondo}, based on a capacitive
interaction model,
that the $\nuC = 2$ antidot can indeed show this
antidot Kondo effect.

\section{Theoretical studies on electron interactions in
antidots} \label{TH}

The experimental observations discussed in the previous sections
indicate that electron-electron interactions may play an
important role in an antidot in the integer quantum Hall regime. To
theoretically treat electron interactions in a general situation,
one can either construct a phenomenological model Hamiltonian or
numerically solve a full many-body Hamiltonian using some
approximations. The former way is powerful in the sense that it can
provide simple understanding and basic physics of the system of
interest, but it relies on a few phenomenological parameters. On the
other hand, the latter can provide the justification of the former
way and predict many-body ground states, but it requires intensive
numerical calculations.
The two approaches are complementary to each other. The
effects of electron interactions in an integer quantum Hall antidot
have been studied in both ways.

The first theoretical treatment of electron interactions in an
integer quantum Hall antidot was reported only recently \cite{Sim}.
It provides a phenomenological model for a $\nuC = 2$ antidot, based
on capacitive interactions of excess charges localized around the
antidot. This model is similar to the capacitive-interaction model
for a quantum dot \cite{ReviewDot,ReviewDot2,Glazman_CI}, and it may
explain the key features of the nontrivial Aharonov-Bohm
oscillations accompanied by the antidot charging effect, the $h/2e$
Aharonov-Bohm oscillations, and the antidot Kondo effect discussed
in Secs. \ref{EXPcharging}, \ref{EXPdouble}, and \ref{EXPKondo},
respectively. This phenomenological model has been tested
\cite{Sim,Hwang} by a Hartree-Fock numerical study. The Hartree-Fock
calculation has shown that a maximum-density droplet of holes can be
the ground state of an antidot in some parameter regimes, and that
the transitions of the droplet ground states can be well described
by the capacitive-interaction model. A numerical calculation based
on spin density-functional theory has also been performed
\cite{Ihnatsenka} very recently, to investigate the formation of
compressible regions around an antidot. This study shows that the
widths of the compressible regions can be narrower than those of
compressible regions formed along the extended edges of a quantum
Hall system, and
finds that
exchange interactions can suppress the formation of some of the possible
compressible regions.

In this section, we review the above theoretical works on electron
interactions in an integer quantum Hall antidot. It includes the
total Hamiltonian of the antidot (Sec.\/ \ref{THHamiltonian}), the
origin of excess charges (Sec.\/ \ref{THexcess}), the
capacitive-interaction model for a $\nuC = 2$ antidot and its
description of the $h/2e$ oscillations (Sec.\/ \ref{THcapacitive}),
an explanation of the antidot Kondo effect based on the
capacitive-interaction model (Sec.\/ \ref{THKondo}), the
Hartree-Fock study and the prediction of a hole maximum-density
droplet (Sec.\/ \ref{THHF}), and the spin density-functional study
of compressible regions around antidots (Sec.\/
\ref{THcompressible}).

\subsection{Total Hamiltonian}
\label{THHamiltonian}

We consider the antidot system shown in Fig.~\ref{geometry}. A
strong perpendicular magnetic field ($B > 1$ T) is applied so that
the system is in the integer quantum Hall regime with bulk filling
factor $\nu_{\rm bulk}$. The antidot local filling factor is chosen
to be $\nuC = 1$ or 2 ($\nuC \le \nu_{\rm bulk}$) for comparison
with experiments. In this case, antidot states come only from the
two spin-split branches of the lowest Landau level; we
assume that the Landau-level mixing from the higher Landau levels
may be neglected
due to the strong magnetic field even in the presence of electron
interactions. It is assumed that the localized antidot states and
the extended edge channels are tunnel-coupled, and that the period
of the Aharonov-Bohm oscillations is much smaller than the strength
of the applied magnetic field, $\Delta B = h / (eS) \ll B$. This
condition is achieved when the effective antidot area $S$ is large
enough or when $B$ is sufficiently strong. This is a key assumption
of the capacitive-interaction model for an antidot, as will be
discussed in Secs. \ref{THexcess} and \ref{THcapacitive}, because it
allows one to have a sufficiently slowly varying antidot potential
$V_{\rm AD}(r)$ within the scale of the magnetic length $l_B$.

The total Hamiltonian of the system can be written as
\begin{eqnarray}
H_{\rm tot} = H_\mathrm{AD} + H_{\rm edge} + H_{\rm tun}.
\label{TOTALH}
\end{eqnarray}
The Hamiltonian $H_\mathrm{AD}$ for the antidot states is
\begin{eqnarray}
H_\mathrm{AD} = \sum_{m\sigma} \epsilon_{m \sigma}
c^\dagger_{m\sigma} c_{m\sigma} + \frac{1}{2} \sum_{m m' n n' \sigma
\sigma'} W_{m m' n n'} c_{m \sigma}^\dagger c_{m' \sigma'}^\dagger
c_{n' \sigma'} c_{n \sigma}, \label{HAD}
\end{eqnarray}
where $c_{m \sigma}^\dagger$ creates an electron with spin $\sigma$
in the localized single-particle state enclosing $m$ magnetic flux
quanta, $\epsilon_{m \sigma}$ is the single-particle energy, and $W$
is the Coulomb interaction. The single-particle energy $\epsilon_{m
\sigma} = \hbar \omega_c/2 + \epsilon^{\rm Z}_\sigma + V_{\rm
AD}(m)$ comes from the lowest Landau-level energy $\hbar
\omega_c/2$, Zeeman energy $\epsilon^{\rm Z}_\sigma$, and the
antidot potential energy $V_{\rm AD}$. The antidot potential energy
is governed by the antidot gate voltage and includes the positive
background term preserving total charge neutrality. The Hamiltonian
$H_{\rm edge}$ of the extended edge channels is written as
\begin{eqnarray}
H_{\rm edge} = \sum_{ik\sigma} \epsilon_{ik\sigma}
c^\dagger_{ik\sigma} c_{ik\sigma},
\end{eqnarray}
where $c^\dagger_{ik\sigma}$ creates an electron in the extended
edge state with momentum $k$, energy $\epsilon_{ik\sigma}$, spin
$\sigma$, and Landau-level index $i$. Note that electron
interactions of the extended edge channels are neglected in this
regime of integer filling factor $\nuC$. The extended edge channels
can tunnel-couple to antidot states through incompressible quantum
Hall regions in between with coupling strength tuned by the side and
antidot gate voltages. The tunneling Hamiltonian $H_{\rm tun}$ is
written as
\begin{eqnarray}
H_{\rm tun} = \sum_{ikm\sigma} V^i_{km\sigma} c^\dagger_{ik\sigma}
c_{m\sigma} + \textrm{H.c.}
\end{eqnarray}
We only consider the extended edge channels coming from the first
and the second Landau levels ($i \in \{\aaa, \aaa', \bbb, \bbb' \}$) [see
Fig.~\ref{geometry}], since the extended edge channels of higher
Landau levels have smaller and negligible tunneling amplitudes
$V^i_{km\sigma}$ to antidot states.

\subsection{Derivation of excess charge}
\label{THexcess}

In the capacitive-interaction model for antidots, the antidot
Hamiltonian $H_{\rm AD}$ in Eq. (\ref{HAD}) can be simply described
by the capacitive coupling of excess charges formed around the
antidot. In this section, we provide the origin of the excess charge
and derive its dependence on magnetic field $B$ for a large-size
antidot.

We first sketch the origin and typical behavior of the excess
charge. A many-body state of an antidot can be constructed by
single-particle antidot states discussed in Sec.\/ \ref{SP}, which
are phase-coherent closed orbits encircling the antidot. The $m$-th
single-particle state encloses an integer number $m$ of magnetic
flux quanta, $m = BS_m / \phi_0$, where $S_m$ is the area enclosed.
As $B$ adiabatically increases by $\delta B (< \Delta B)$, each
single-particle state moves with respect to the antidot potential
$V_{\rm AD}(r)$ by $\Delta r_m \sim - \sqrt{S_m} \delta B / B$,
adjusting the enclosed area in order to keep the flux $BS_m$
constant. Thus the position of electron density of the many-body
state moves towards the center of the antidot. Since the positive
background charge does not move at all at the same time, this
displacement of electron density results in charge imbalance around
the antidot, {\it i.e.}, continuous accumulation of local excess
charge (see Fig.~\ref{EXCESS}) {\cite{Ford,Kataoka_charging}. In
this way, the variation of magnetic field $B$ modifies the excess
charge of the antidot, and thus it acts like a gate voltage for a
quantum dot.

The accumulated excess charge relaxes via resonant
tunneling. The relaxation is a discrete event because of the
insulating incompressible regions between the excess charge and
extended edge channels, and it can be observed as a resonant peak or
dip in tunneling conductance $G_{\rm ad}$, depending on the nature
of its coupling to the extended edge channels, as discussed in Secs.
\ref{SPTUNNELING} and \ref{EXPsetup}. Note that in the fractional
quantum Hall regimes, the relaxation may be accompanied by
fractional charge tunneling \cite{FQHE}.

Below, we derive the magnetic-field dependence of the excess charge
$\delta q_\sigma$. For simplicity, we first consider a $\nuC=1$
antidot. The extension of the derivation to the case of $\nuC=2$ is
straightforward, as we will see later.

To see the behavior of the excess charge as a function of magnetic
field $B$, one may start with a constant-interaction model. In the
constant-interaction model for a quantum dot
\cite{ReviewDot,ReviewDot2,Glazman_CI}, electron interactions are
described by an effective capacitive interaction. It works well in
the case where the dot is almost isolated, so that the number of
electrons within the dot can be treated as a good quantum number.
The effective capacitive interaction is modeled by a set of
capacitances, which are treated as constant within a narrow range of
gate voltage applied to the dot. One can adopt a similar strategy
for the antidot when it weakly couples to extended edge channels
only via tunneling through the incompressible regions in between.
For a given magnetic field $B_0$ and a fixed antidot gate voltage, a
part of the antidot Hamiltonian $H_{\rm AD}$ in Eq. (\ref{HAD}),
which includes the antidot potential $V_{\rm AD}(m)$ and the
electron-electron interaction $W_{mm'nn'}$ terms, can be
approximately described by an effective capacitive interaction,
\begin{eqnarray}
E_\mathrm{AD}^{\nuC = 1}(B_0) = \frac{(N^{\rm AD} e -
Q_G(B_0))^2}{2C}. \label{effH_nu1}
\end{eqnarray}
Here, $C$ is an effective antidot capacitance, $N^{\rm AD}$ is the
total number of electrons occupying the single-particle antidot
states,
%and effectively participating in electron-electron interactions,
and $Q_G$ is the positive ``antidot gate charge''
governed by the antidot gate voltage.  The spin index $\sigma$ is
dropped in this case of $\nuC=1$.  As an antidot is an open
system, $N^{\rm AD}$ may not be a well-defined quantity. However, a
strict definition of $N^{\rm AD}$ is not necessary here, as one can see
below.  The
important property of $N^{\rm AD}$ is that it can vary only by an
integer number due to charge discreteness.  On the other hand, the
antidot gate charge $Q_G$ can vary continuously. This leads to
Coulomb-blockade oscillations as a function of the antidot gate
voltage. We note that
the contribution from the single-particle energies (due
to a rising antidot potential), the single-particle level spacing,
can be absorbed into the capacitive
energy in Eq. (\ref{effH_nu1}) \cite{Geller2}, either if $N^{\rm
AD}$ is large or if the single-particle energy levels can be
approximated to follow a linear function of angular momentum $m$
with a constant level spacing, as described below.

As $B$ varies from $B_0$ by $\delta B$ ($< \Delta B \ll B_0$), the
position $r_m$ of the single-particle state enclosing $m$ magnetic
flux quanta is shifted by $\delta r_m$ with respect to the antidot
potential $V_{\rm AD}(r)$, resulting in the energy variation of
the state $\delta \epsilon_m = dV_{\rm AD}(r)/dr|_{r=r_m} \delta
r_m$, since the amount of flux enclosed by each single-particle
state is preserved. This preservation, $\delta (B \pi r_m^2) = 0$,
gives $\delta r_m = - (r_m / 2B_0) \delta B$, so $\delta
\epsilon_m$ can be approximated as a linear function of $\delta B$
when $V_{\rm AD}(r)$ varies slowly on the scale of the magnetic
length $l_B(\equiv\sqrt{\hbar}/eB)$ and is locally linear in $r$.
The energy shift coming from the variations of the cyclotron and
Zeeman energies is also linear in $\delta B$, and thus can be
absorbed into $\delta \epsilon_m$. By introducing the average
value $\overline{\delta \epsilon_m}$ of $\delta \epsilon_m$ for
occupied single-particle states, one can find the shift of total
single-particle energy $\delta E (\delta B) = N^{\rm AD}
\overline{\delta \epsilon_m}$. The shift can be rewritten in terms
of a function $Q_B (\delta B) \equiv - C \overline{\delta
\epsilon_m} (\delta B) / e$ as
\begin{eqnarray}
\delta E (\delta B) =  - e N^{\rm AD} Q_B(\delta B) / C.
\end{eqnarray}
Notice that $Q_B(\delta B)$ is a linear function of $\delta B$
when $dV_{\rm AD}(r)/dr$ can be assumed to be constant as in a
large-size antidot. By summing this single-particle energy shift
and the capacitive interaction in Eq. (\ref{effH_nu1}), one can
arrive at the dependence of the total energy of the antidot on
$\delta B$:
\begin{eqnarray}
E_\mathrm{AD}^{\nuC=1}(B_0 + \delta B) & = & \frac{(N^{\rm AD} e -
Q_G(B_0) - Q_B(\delta B))^2}{2C} + E_\mathrm{AD,0}^{\nuC=1}
\nonumber \\
& = & \frac{\delta q^2}{2C} + E_\mathrm{AD,0}^{\nuC=1},
\label{effH_nu1_2}
\end{eqnarray}
where $E_\mathrm{AD,0}^{\nuC=1}$ is a term independent of $N^{\rm
AD}$ ( the dependence of $E_\mathrm{AD,0}^{\nuC=1}$ on $N^{\rm
AD}$ may be ignored under the assumption of large $N^{\rm AD}$ or
constant $\delta \epsilon_m$). From this expression, one can
interpret $Q_B$ as an ``antidot magnetic-gate charge'', and
$\delta q$ as an excess charge given by
\begin{eqnarray}
\delta q = N^{\rm AD} e - Q_G(B_0) - Q_B(\delta B).
\label{excess_charge_nu1}
\end{eqnarray}
This expression shows that the excess charge $\delta q$ varies
with magnetic field. The magnetic-field dependence of $\delta q$
can be also derived \cite{Geller,Geller2} in a field-theoretical
language with the action $\delta \mathcal{L} = \vec{j} \cdot
\vec{A}$, where $\vec{j}$ is the current density along the antidot
circumference and $A$ is the vector potential.
The control of the excess charge near an antidot by using
magnetic field is reminiscent of the charge
control by a gate voltage in a quantum dot.

By using the form of the effective energy in Eq.
(\ref{effH_nu1_2}), one can study the Coulomb blockade by a
large-size $\nuC=1$ antidot. The antidot magnetic-gate charge
$Q_{B}(\delta B)$ has a real value, similarly to $Q_G$, thus one
has Coulomb-blockade oscillations of $N^{\rm AD}$ or $\delta q$ as
a function of magnetic field. Relaxation of $\delta q$ is allowed
whenever
\begin{eqnarray}
E^{\nuC=1}_\mathrm{AD}(\delta q \pm e)  =
E^{\nuC=1}_\mathrm{AD}(\delta q), \label{resonance_condition_nu1}
\end{eqnarray}
which is satisfied by $\delta q = \pm e/2$. We remark that
the energy in Eq. (\ref{effH_nu1_2}) and the condition
(\ref{resonance_condition_nu1}) can provide a good description for
the charging and relaxation behavior of $\nuC = 1$ antidots (see
Sec. \ref{EXPcharging}) studied experimentally in Ref.
\cite{Kataoka_charging}. Note that the energy in Eq. (\ref{effH_nu1_2})
does not include the contribution from neutral excitations
\cite{Geller2}, which have the energy gap of the single-particle
level spacing in the noninteracting limit. Therefore, the condition
(\ref{resonance_condition_nu1}) is valid at bias and temperature
lower than the neutral-excitation energy.

To understand the periodic nature of $\delta q(B)$, we may imagine
the simple situation where the electron density around an antidot is
only shifted without any charge redistribution. This situation can
be achieved only when an incompressible region or a
maximum-density droplet of holes \cite{Hwang} (see Sec.\/
\ref{THHF}) is formed around an antidot. In this case, one can
estimate $\delta q = - 2\pi e n_{\rm e} r \delta r$ for the change $\delta
B \equiv B - B_0$ ($<\Delta B \ll B$), where $r$ is the effective
antidot radius, $\delta r$ is the shift caused by $\delta B$, and
$n_{\rm e}$ is the electron density. Since $\delta (B \pi r^2) = 0$ and
the density can be expressed as $n_{\rm e} = eB / h$ in the $\nuC = 1$
regime, one can find a simple but intuitive dependence of $\delta q$
on magnetic field $B$,
\begin{eqnarray}
\delta q(B) \sim - \Delta Q_B(B) \sim - e (B - B_0) / \Delta B.
\label{excess_charge_nu1_2}
\end{eqnarray}
Therefore, the amount of accumulation of $\delta q$ in one
Aharonov-Bohm period $\Delta B$ is one electron charge $e$ in the
$\nuC = 1$ regime. This accumulation is combined with the
relaxation, giving rising to the expected result that $\delta q$ has
the same periodicity as the Aharonov-Bohm oscillations. When one
takes the redistribution of electron density into account, $\delta
q$ may be approximated to be a linear function of $B$ for a
large-size antidot with potential varying smoothly over several
magnetic lengths. In this case, the accumulation and relaxation of
$\delta q$ are expected still to be periodic.
%$\delta q$ is no longer a linear function of $B$,
This periodic nature of the
$B$-dependent evolution of the excess charge is consistent with the
experimental findings in the $\nuC = 1$ regime
\cite{Kataoka_charging} (see Sec. \ref{EXPcharging}).

The extension of the above derivation to the $\nuC=2$ case is
straightforward. In this case, one can introduce spin-dependent
excess charges, $\delta q_\uparrow$ and $\delta q_\downarrow$,
coming from the spin-split branches of the lowest Landau level, and
a two-by-two capacitance matrix (with spin index $\sigma$) is
required to describe the capacitive interactions between the excess
charges; the detailed form of the model Hamiltonian will be
introduced in Eq. (\ref{effH}) in Sec.\/ \ref{THcapacitive}. By
generalizing Eq. (\ref{excess_charge_nu1}), one can use an
expression for $\delta q_\sigma$
\begin{equation}
\delta q_\sigma = eN_\sigma^{\rm AD} - Q_{G\sigma} - Q_{B\sigma}(B),
\label{excess_charge}
\end{equation}
where $N_\sigma^{\rm AD}$, $Q_{G\sigma}$, and $Q_{B\sigma}(B)$ are
the spin-dependent occupation number, antidot gate charge, and
antidot magnetic-gate charge, respectively. When the antidot
potential varies linearly within the scale of magnetic length $l_B$,
one may use an approximate linear form,
\begin{equation}
Q_{B\sigma}(B) + Q_{G\sigma} = e ( a_\sigma B + b_\sigma ),
\label{excess_charge2}
\end{equation}
between two adjacent relaxation events of $(\delta q_\uparrow,
\delta q_\downarrow)$ within a sufficiently small range of magnetic
field, as in the above $\nuC = 1$ case. Here, $a_\sigma e $ is the
rate of excess charge accumulation with increasing $B$, $a_\sigma >
0$, and $b_\sigma e$ originates from the positive background charge.
From the discussion around Eq. (\ref{excess_charge_nu1_2}), one has
$a_\sigma \simeq 1/ \Delta B_\sigma = S_\sigma / \phi_0$, where
$S_\sigma$ is the effective area enclosed by $\delta q_\sigma$. When
$S \simeq S_\sigma \gg S_\uparrow - S_\downarrow$, one has
$a_\uparrow \simeq a_\downarrow \simeq S / \phi_0$. Note that a
recent Hartree-Fock numerical calculation \cite{Hwang} has shown the
behavior of excess charge to be consistent with the linear form in
Eq. (\ref{excess_charge2}), which will be discussed in
Sec.~\ref{THHF}.

In the $\nuC = 2$ case, the evolution of excess charges, $(\delta
q_\uparrow, \delta q_\downarrow)$, as a function of magnetic field
is not as simple as in the $\nuC = 1$ case, because of the
interactions between $\delta q_\uparrow$ and $\delta q_\downarrow$.
As will be shown in the next two sections, the capacitive
interaction between $\delta q_\uparrow$ and $\delta q_\downarrow$
leads to a characteristic evolution of $(\delta q_\uparrow, \delta
q_\downarrow)$ which can explain the $h/2e$ Aharonov-Bohm
oscillations and the Kondo effect discussed in Secs. \ref{EXPdouble}
and \ref{EXPKondo}.

\subsection{Capacitive interaction model for $\nuC=2$ antidots}
\label{THcapacitive} In this section,
we introduce the capacitive interaction model Hamiltonian for a
$\nuC = 2$ antidot, and discuss the accumulation and relaxation of
excess charges $(\delta q_\uparrow, \delta q_\downarrow)$ as a
function of magnetic field $B$ \cite{Sim}. Various sequences of
single-particle normal resonant relaxations and Kondo-type resonant
relaxations can appear in the evolution of $(\delta q_\uparrow,
\delta q_\downarrow)$, depending on the interaction strength between
$\delta q_\uparrow$ and $\delta q_\downarrow$ and the spin-dependent
effective antidot area $S_\sigma$.

In the $\nuC = 2$ case, the localized excess charges $\delta
q_\uparrow$ and $\delta q_\downarrow$ are spatially separated from
each other and from extended edge channels by insulating
incompressible regions. This feature allows us to describe the
antidot Hamiltonian (\ref{HAD}) by using capacitive coupling of
excess charges. By generalizing the model Hamiltonian for the $\nuC
= 1$ antidot in Eq. (\ref{effH_nu1_2}), one can write the total
energy $E^{\nuC = 2}_\mathrm{AD} (B)$ of a $\nuC = 2$ antidot as
\begin{eqnarray}
E^{\nuC = 2}_\mathrm{AD}(\delta q_\uparrow(B), \delta
q_\downarrow(B)) = \frac{1}{2} \sum_{\sigma \sigma'} \delta q_\sigma
\left(C^{-1}\right)_{\sigma \sigma'} \delta q_{\sigma'}.
\label{effH}
\end{eqnarray}
Here, $\delta q_\sigma$ has the form in Eq. (\ref{excess_charge})
and the effective capacitance matrix $C_{\sigma \sigma'}$ is written
as
\begin{eqnarray}
C^{-1} = \left( \begin{array}{cc}
(C^{-1})_{\uparrow \uparrow} &  (C^{-1})_{\uparrow \downarrow} \\
(C^{-1})_{\downarrow \uparrow} & (C^{-1})_{\downarrow \downarrow}
\end{array} \right) =
\frac{1}{C_{\uparrow \uparrow} C_{\downarrow \downarrow} -
C_{\uparrow \downarrow}^2} \left( \begin{array}{cc}
C_{\downarrow \downarrow} & - C_{\uparrow \downarrow} \\
- C_{\uparrow \downarrow} & C_{\uparrow \uparrow}
\end{array} \right).
\end{eqnarray}
$C_{\sigma \sigma'}$ is a classical electrostatic quantity if
alternating compressible and incompressible regions
\cite{Chklovskii1,Chklovskii2} are formed around the antidot, while
it is a phenomenological parameter in other cases. One has
$C_{\uparrow \uparrow} < C_{\downarrow \downarrow}$, since $\delta
q_\downarrow$ is located further away from the center of the antidot
than $\delta
q_\uparrow$ due to its higher Zeeman energy. In addition,
$C_{\uparrow \downarrow} < 0$ and $|C_{\uparrow \downarrow}|$ is the
smallest among $C_{\sigma \sigma'}$, as it is a mutual capacitance.
Note that the cyclotron energy and the Zeeman energy are counted in
the definitions of $\delta q_\sigma$ and $C$, as they also cause the
displacement and the separation of each $\delta q_\sigma$.

For large antidots with $B \gg \Delta B$, $\delta q_\sigma (B)$ in
Eqs. (\ref{excess_charge}, \ref{excess_charge2}) and $E_\mathrm{AD}$
in Eq. (\ref{effH}) can provide a good model describing the Coulomb
blockade and charge relaxations of a $\nuC = 2$ antidot, as in the
capacitive interaction model for the $\nuC =1 $ case (see Sec.\/
\ref{THexcess}). Moreover, since $C_{\sigma \sigma'}$ varies only
slightly within one Aharonov-Bohm period, one can take it as
constant; in this case, the dependence of $E^{\nuC = 2}_\mathrm{AD}$
on $B$ is governed only by $\delta q_\sigma (B)$. The validity of
the capacitive interaction model in Eqs. (\ref{excess_charge},
\ref{excess_charge2}, \ref{effH}) has been tested \cite{Hwang} using
a Hartree-Fock approach (see Sec.~\ref{THHF}). This test shows that
the transition of a hole maximum-density-droplet ground state of an
antidot can be described well by the model.

When antidot states weakly couple to extended edge channels, Coulomb
blockade prohibits the relaxation of $(\delta q_\uparrow, \delta
q_\downarrow)$ unless one of the following conditions is satisfied.
First, a normal resonance occurs whenever $E_\mathrm{AD}^{\nuC = 2}$
is invariant under single-electron tunneling [see Eq.
(\ref{resonance_condition_nu1}) for $\nuC = 1$], i.e.,
\begin{eqnarray}
E^{\nuC = 2}_\mathrm{AD}(\delta q_\uparrow, \delta q_\downarrow) & =
& E^{\nuC = 2}_\mathrm{AD}(\delta q_\uparrow \pm e, \delta
q_\downarrow), \label{resonance_condition_up}
\end{eqnarray}
or
\begin{eqnarray}
E^{\nuC = 2}_\mathrm{AD}(\delta q_\uparrow, \delta q_\downarrow) & =
& E^{\nuC = 2}_\mathrm{AD}(\delta q_\uparrow, \delta q_\downarrow
\pm e). \label{resonance_condition_down}
\end{eqnarray}
Another process
can occur via spin-flip cotunneling, in
which an electron tunnels into an antidot state and another electron
with the opposite spin tunnels out via a virtual state or vice
versa. The condition for this is
\begin{equation}
E^{\nuC = 2}_\mathrm{AD}(\delta q_\uparrow, \delta q_\downarrow) =
E^{\nuC = 2}_\mathrm{AD}(\delta q_\uparrow \pm e, \delta
q_\downarrow \mp e). \label{resonance_condition_Kondo}
\end{equation}
Among many available virtual states, the one with the lowest energy
(thus with the biggest contribution) is either $(\delta q_\uparrow
\pm e, \delta q_\downarrow)$ or $(\delta q_\uparrow, \delta
q_\downarrow \mp e)$.
The cotunneling process, combined with other spin-nonflip cotunneling,
can give rise to a Kondo resonance, which will be described in Sec. \/
\ref{THKondo}. We call Eq. (\ref{resonance_condition_Kondo})
as Kondo resonance condition hereafter.

By combining the resonant relaxation conditions and the
accumulations of excess charge one can study the evolution of
$(\delta q_\uparrow, \delta q_\downarrow)$ as a function of magnetic
field $B$. As discussed before in Sec.\/ \ref{THexcess}, the
accumulation of $\delta q_\sigma$ may be approximated as a linear
function of
$\delta q_\sigma = - e(a_\sigma B  + b_\sigma )$
[see Eq.
(\ref{excess_charge2})] between two adjacent relaxation events,
assuming that the antidot potential $V_{\rm AD}(r)$ varies linearly within
the scale of magnetic length $l_B$ and $\Delta B \ll B$. The two
important parameters governing the evolution of excess charges are
the ratio of accumulation rates $a_\uparrow / a_\downarrow$ and the
ratio of capacitances $\alpha \equiv |C_{\uparrow
\downarrow}|/C_{\uparrow \uparrow}$, as we will see below.

\begin{figure}[th]
\centerline{\psfig{file=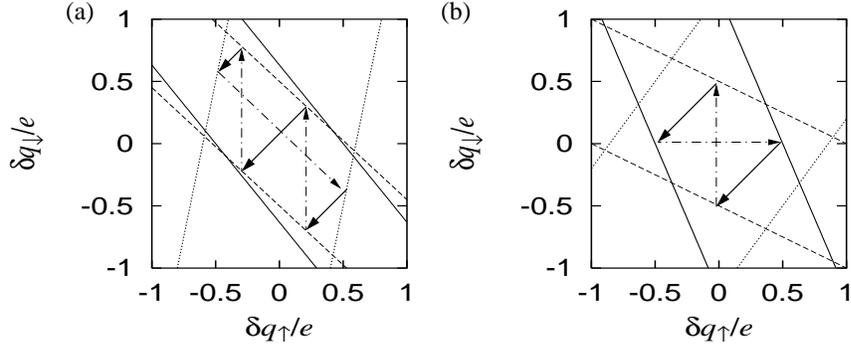,width=12cm}} \vspace*{8pt}
\caption{ Evolution of the excess charges ($\delta q_\uparrow$,
$\delta q_\downarrow$) as a function of magnetic field $B$ in the
case of $S_\uparrow = S_\downarrow$. The solid-arrow trajectories
indicate the direction of the evolution with increasing $B$. When
one of resonance conditions is met, ($\delta q_\uparrow$, $\delta
q_\downarrow$) jumps following horizontal (normal spin-up electron
tunneling), vertical (normal spin-down electron tunneling), or
diagonal dashed-dotted (Kondo resonance) arrows. Solid, dashed, and
dotted lines indicate resonance conditions of Eqs.\
(\ref{resonance_condition_up}), (\ref{resonance_condition_down}),
and (\ref{resonance_condition_Kondo}), respectively, and they
constitute the boundaries of a hexagonal cell, in which ($\delta
q_\uparrow$, $\delta q_\downarrow$) evolves.
Two different parameters are used: (a) $\alpha \equiv |C_{\uparrow
\downarrow}|/C_{\uparrow \uparrow} = 0.95$ and (b) $0.5$. For both
the cases, $C_{\downarrow\downarrow}/C_{\uparrow\uparrow}=1.2$ is
chosen. From Ref. \protect\cite{Sim}. } \label{trajectory}
\end{figure}

To follow the evolution of excess charges, it is useful to draw the
trajectories of evolution in a two-dimensional plane of ($\delta
q_\uparrow$,$\delta q_\downarrow$), as shown in
Fig.~\ref{trajectory}. As the accumulation of ($\delta
q_\uparrow$,$\delta q_\downarrow$) is determined by the rates
$a_\uparrow e$ and $a_\downarrow e$ [see Eq.
(\ref{excess_charge2})], the evolution trajectory follows a line
parallel to $\delta q_\uparrow = (a_\uparrow / a_\downarrow) \delta
q_\downarrow$ with increasing $B$, until it jumps at one of the
resonance conditions. Therefore, the evolution trajectory of
$(\delta q_\uparrow, \delta q_\downarrow)$ is confined in a
hexagonal cell, the boundaries of which are defined by the three
resonance conditions in Eqs.
(\ref{resonance_condition_up}, \ref{resonance_condition_down},
\ref{resonance_condition_Kondo}).
The trajectories (solid arrows in Fig.~\ref{trajectory}) parallel to
the line of $\delta q_\uparrow = (a_\uparrow / a_\downarrow) \delta
q_\downarrow$ mean the accumulation of excess charges, while their
jumps (following dashed-dotted arrows) between the boundaries of the
cell indicate one of the three possible resonant relaxations. The
horizontal and vertical jumps correspond to the normal resonant
tunneling of single electron with spin up and that of spin down
[Eqs. (\ref{resonance_condition_up}) and
(\ref{resonance_condition_down})], respectively, while the diagonal
jumps occur whenever the Kondo resonance condition [Eq.
(\ref{resonance_condition_Kondo})] is satisfied.

Now one can see the dependence of the evolution trajectories, or of
the sequence of resonant tunneling events, on the ratios $a_\uparrow
/ a_\downarrow$ and $\alpha \equiv |C_{\uparrow
\downarrow}|/C_{\uparrow \uparrow}$. We first consider the case of
$a_\uparrow \simeq a_\downarrow$, which can appear when the spatial
separation of $\delta q_\uparrow$ and $\delta q_\downarrow$ is small
compared with their average radius, {\it i.e.}, when $S_\downarrow -
S_\uparrow \ll S_\downarrow$. In this case, as $B$ increases, the
ground-state value of ($\delta q_\uparrow$,$\delta q_\downarrow$)
evolves parallel to the line of $\delta q_\uparrow = \delta
q_\downarrow$. By using this fact, one can show geometrically very
easily that the evolution trajectory $(\delta q_\uparrow, \delta
q_\downarrow)$ is always closed in the hexagonal cell and periodic
with period $\Delta B = \phi_0 /S $ in this case of $a_\uparrow
\simeq a_\downarrow$.

In the case of $a_\uparrow \simeq a_\downarrow$, one can further
notice that there are two possible types of evolution trajectory
when $C_{\uparrow \uparrow} < C_{\downarrow \downarrow}$, depending
on the parameter $\alpha \equiv |C_{\uparrow \downarrow}|/C_{\uparrow
\uparrow}$ and the initial value of $(\delta q_\uparrow, \delta
q_\downarrow)$ at a starting value of magnetic field. The first type
(i) consists of two consecutive tunneling events of spin-down
electrons as well as one intermediate Kondo resonance in one
Aharonov-Bohm period $\Delta B$ [see Fig.~\ref{trajectory}(a) and
Fig.~\ref{sequence2}], while the other type (ii) is composed of
alternating tunneling events of spin-up and down electrons with an
arbitrary phase difference [Fig.~\ref{trajectory}(b) and
Fig.~\ref{sequence2}]. In type (i), the tunneling of a spin-up
electron, which is assumed to have smaller Zeeman energy, is {\em
always} Coulomb blockaded, while in type (ii), the sequence of
resonant tunneling events shows two {\em independent} periodic
events (each with Aharonov-Bohm period $\Delta B$) of the normal
spin-up and down resonances. If $\alpha$ has the maximum value of
$\alpha = 1$, all trajectories are of type (i) regardless of the
initial value of $\delta q_\sigma$. In the other extreme case of
$\alpha=0$ (minimum), i.e., $C_{\uparrow \downarrow}=0$, the two
spin states are completely uncorrelated and all trajectories are of
type (ii). For $0<\alpha<1$, both types are allowed depending on
the initial values of the $\delta q_\sigma$. For larger $\alpha$, more
trajectories follow type (i). One may notice that the type (i)
trajectory itself depends on $\alpha$. In type (i), the
difference in magnetic field between the two consecutive tunneling
events of spin-down electrons within one Aharonov-Bohm period
$\Delta B$ is exactly $\Delta B / 2$ for $\alpha = 1$ (maximum)
while for smaller $\alpha$ it starts to deviate
from $\Delta B / 2$ to $\Delta B / (1 + \alpha)$ or to
$\Delta B \alpha / (1 + \alpha)$.

The experimental observations for $\nuC = 2$ antidots
\cite{Ford,Sachrajda,Kataoka_double,Kataoka_Kondo,Kataoka_Spin} are
consistent with the type (i) trajectory at $\alpha \simeq 1$. In
high magnetic fields around $3$~T, they show the $h/2e$
Aharonov-Bohm oscillations, i.e., evenly spaced resonance dips of
antidot conductance that appear two times per $\Delta B$
\cite{Ford,Sachrajda,Kataoka_double}. In lower magnetic fields
around $1.2$~T \cite{Kataoka_Kondo}, another dip related to the
Kondo resonance can appear approximately once every $\Delta B$ at
low temperatures below the Kondo temperature $T_{\rm K}$ (see Sec.
\ref{THKondo}), while the Kondo-related dips disappear and the
$h/2e$ Aharonov-Bohm oscillations emerge at high temperature.
%the $h/2e$ oscillations in this low field regime are imperfect,
%which may be understood as the deviation from the type (i) trajectories
%due to the fact that the single-particle energy difference,
%$\epsilon_{sp} - \EF$, ignored in Eqs.
%(\ref{resonance_condition_up},\ref{resonance_condition_down},\ref{resonance_condition_Kondo}) can play a role as
%the single-particle level spacing
%may not be much smaller than the charging energy \cite{Kataoka_Kondo}.
Moreover, a spin-resolved measurement \cite{Kataoka_Spin} shows that
only the electrons with spin down can contribute to the $h/2e$
oscillations. Therefore, based on this agreement, one may argue
that the experiments are in the parameter regime of $\alpha \simeq
1$, where the mutual Coulomb interactions between $\delta
q_\uparrow$ and $\delta q_\downarrow$ are as strong as those
characterized by the self capacitances $C_{\uparrow \uparrow}$ or
$C_{\downarrow \downarrow}$.

\begin{figure}[th]
\centerline{\psfig{file=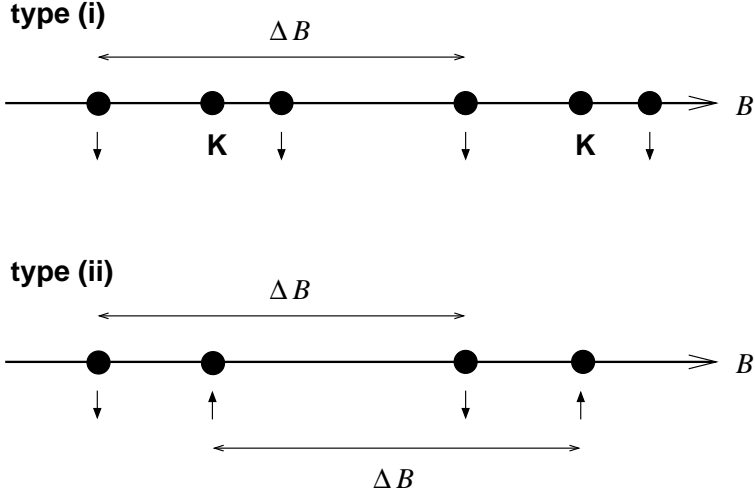,width=10cm}} \vspace*{8pt}
\caption{ Two types of possible sequence of relaxation events
(filled circles) of $(\delta q_\uparrow,\delta q_\downarrow)$ as a
function of magnetic field $B$ in the case of $a_\uparrow \simeq
a_\downarrow$. The normal resonances of electrons with spin up,
those with spin down, and Kondo resonances are marked by up arrow, down
arrow, and the letter ``K'', respectively. In the type (i)
sequence, a pair of spin-down normal resonances periodically appears
with Aharonov-Bohm period $\Delta B$, while Kondo resonances occur
once within the pair in one $\Delta B$. The position of the Kondo
resonances depends on the initial point of the evolution of $(\delta
q_\uparrow, \delta q_\downarrow)$, and there exist no spin-up normal
resonances for type (i). In the type (ii) sequence, spin-up and
down normal resonances occur alternately, while there is no Kondo
resonance. Each normal resonance appears periodically with period
$\Delta B$. } \label{sequence2}
\end{figure}

When the areas enclosed by $\delta q_\uparrow$ and $\delta
q_\downarrow$ are different, $S_\downarrow \ne S_\uparrow$, the
accumulation rate $a_\sigma$ of excess charges is spin-dependent so
that the ground-state value of ($\delta q_\uparrow$,$\delta
q_\downarrow$) evolves parallel to the line of $\delta q_\uparrow =
(a_\uparrow / a_\downarrow) \delta q_\downarrow$ with increasing
$B$. The resulting trajectories depend on $\alpha$ as in the above
case of $S_\downarrow = S_\uparrow$. For $\alpha = 1$, as in type (i), all
trajectories of ($\delta q_\uparrow$,$\delta q_\downarrow$) have
resonant tunneling of spin-down electrons as well as Kondo
tunneling, while they cannot have resonant tunneling of spin-up
electrons due to Coulomb blockade. However, in
contrast to type (i) resonances, they are not periodic, although each of
their spin-down resonant-tunneling and Kondo tunneling events has
its own periodicity ($\Delta B_\downarrow$ and $\Delta B_{\rm K}$,
respectively). The relation between $\Delta B_\downarrow$ and
$\Delta B_{\rm K}$ is found to be
$\Delta B_{\rm K} / \Delta B_\downarrow = 1 + a_\downarrow / a_\uparrow$.
For $0 < \alpha < 1$, the trajectory
does not have a simple structure. All of spin-up resonant tunneling,
spin-down resonant tunneling, and Kondo tunneling can appear in this
case. For $\alpha = 0$, all trajectories consist of two independent
periodic spin-up and down resonances (with period $\Delta B_\uparrow
= \phi_0/S_\uparrow$ and $\Delta B_\downarrow =
\phi_0/S_\downarrow$, respectively), similar to type (ii), as
easily expected.

\subsection{Antidot Kondo effect}
\label{THKondo}

In this section, we describe in detail how the Kondo effect
can arise
in the capacitive interaction model $E^{\nuC = 2}_\mathrm{AD}$ [Eq.
(\ref{effH})] for a $\nuC = 2$ antidot. We first introduce the
mapping of $E^{\nuC = 2}_\mathrm{AD}$ to a Kondo Hamiltonian,
discuss an effective Zeeman energy originated from the
spin-dependent tunneling strength, which might suppress the Kondo
resonance, and finally analyze the experimental observation of
antidot conductance in the Kondo regime.

We first show the capacitive interaction model can be mapped into a
Kondo impurity Hamiltonian, as in the cases of quantum dots
\cite{Glazman_CI}. In the vicinity of a Kondo resonance condition
[Eq. (\ref{resonance_condition_Kondo})], only the two lowest
[$(\delta q_\uparrow,\delta q_\downarrow)$ and $(\delta q_\uparrow +
e,\delta q_\downarrow -e)$] and the next two excited states
[$(\delta q_\uparrow+e,\delta q_\downarrow)$ and $(\delta
q_\uparrow,\delta q_\downarrow-e)$] among possible antidot states
are important when the charging energy is strong enough. Thus one
can ignore all the other excited states, which may affect the Kondo
effect only slightly as in multilevel quantum dots
\cite{Inoshita_multilevel,Boese}. Using these four lowest states one
can map $E^{\nuC = 2}_\mathrm{AD}$ into the Anderson impurity model,
given by a truncated Hamiltonian
\begin{eqnarray}
H_{\rm imp}
= \sum_\sigma \epsilon_\sigma d_\sigma^\dagger d_\sigma +
U d_\uparrow^\dagger d_\downarrow^\dagger d_\downarrow d_\uparrow,
\nonumber
\end{eqnarray}
where $d_\sigma^\dagger$ creates an electron in the impurity site.
The two lowest states constitute the two singly occupied impurity
states, while the next two excited states are the empty and doubly
occupied states. Defining $E_\mathrm{empty} \equiv E^{\nuC =
2}_\mathrm{AD}(\delta q_\uparrow+e,\delta q_\downarrow)$, one can
identify the energy levels of the Anderson impurity Hamiltonian as
\begin{eqnarray}
\epsilon_\uparrow & = & E^{\nuC = 2}_\mathrm{AD}(\delta
q_\uparrow,\delta q_\downarrow) -
E_\mathrm{empty}, \nonumber \\
\epsilon_\downarrow & = & E^{\nuC = 2}_\mathrm{AD}(\delta
q_\uparrow+e,\delta q_\downarrow-e) -
E_\mathrm{empty}, \nonumber \\
U + \epsilon_\uparrow + \epsilon_\downarrow & = & E^{\nuC =
2}_\mathrm{AD}(\delta q_\uparrow,\delta q_\downarrow-e) -
E_\mathrm{empty}. \nonumber
\end{eqnarray}
One can find that the difference between $\epsilon_\uparrow$ and
$\epsilon_\downarrow$ is zero only at the point where the Kondo
resonance condition (\ref{resonance_condition_Kondo}) is satisfied.
As a result, the Anderson impurity site has an {\em effective}
Zeeman energy of $\Delta \epsilon \equiv \epsilon_\uparrow -
\epsilon_\downarrow$ in the vicinity of the Kondo resonance points.
Note that the real Zeeman energy splitting $E_{\rm Z}$ is irrelevant
to the antidot Kondo effect, as the Zeeman energy can be absorbed in
the definition of the excess charges, as discussed before. Now one
can replace the total Hamiltonian $H_{\rm tot}$ in Eq.
(\ref{TOTALH}) with that of the Anderson impurity coupled to the
extended edge channels by amplitude $V^i_{k\sigma}$:
\begin{eqnarray}
H_{\rm K} =
H_{\rm imp}
+ H_{\rm edge} +
\sum_{ik\sigma} V^i_{k\sigma}
c^\dagger_{ik\sigma} d_\sigma + \textrm{H.c.}
\label{Anderson}
\end{eqnarray}
The $k$ dependence of $V^i_{k\sigma}$ is usually ignored,
for simplicity.
%In Fig. \ref{sequence}, we show the correspondence
%of spin-flip cotunneling processes
%between the $\nuC = 2$ antidot and usual quantum dots with
%odd number occupation.
%The combinations of the spin-flip cotunneling
%and the spin-nonflip cotunneling (not shown in Fig. \ref{sequence})
%are the origin of the antidot Kondo effect.

%\begin{figure}[th]
%\centerline{\psfig{file=sequence.eps,width=11cm}}
%\vspace*{8pt}
%\caption{
%This figure shows a clear analogy of
%spin-flip cotunneling processes between
%(a) antidot states
%and (b) quantum dot states with single electron occupation.
%The antidot states are defined as
%$A_{\uparrow} \equiv (\delta q_{\uparrow}, \delta q_{\downarrow})$,
%$A_{\downarrow} \equiv (\delta q_{\uparrow} + e,
%\delta q_{\downarrow} - e)$,
%$C_1^{+} \equiv (\delta q_{\uparrow} + e, \delta q_{\downarrow})$,
%and
%$D_1^{-} \equiv (\delta q_{\uparrow}, \delta q_{\downarrow} - e)$,
%while the quantum dot states is denoted as
%$(n_{\uparrow},n_{\downarrow})$, where
%$n_{\sigma}$ is the number of occupied electrons with spin $\sigma$.
%The addition of an electron with the spin marked by the arrows from
%leads to dot is represented by solid arrows, while
%the substraction from dot to leads by dashed ones.
%}
%\label{sequence}
%\end{figure}

To see whether the Kondo effect can appear in real experimental
situations, one needs to estimate energy scales from the experimental
data (Ref. \cite{Kataoka_Kondo}). In the experiment of Ref.
\cite{Kataoka_Kondo}, the $h/2e$ oscillations were recovered at high
temperature (see Fig.~\ref{Kondo_figure3}), where the Kondo effect
was suppressed. This indicates $\alpha \simeq 1$ in the
experimental situation, so one may focus on the type (i)
trajectory with $\alpha = 1$. Since only spin-down electrons cause
the normal resonances in the type (i) trajectories, one may
rewrite Eq. (\ref{effH}) as
\begin{eqnarray}
E^{\nuC = 2}_\mathrm{AD}(\delta q_\uparrow, \delta q_\downarrow) =
\frac{(\delta q_\downarrow + \alpha \delta q_\uparrow)^2}{2C_{\rm
out}} + \frac{\delta q_\uparrow^2}{2C_{\rm in}},
\end{eqnarray}
where
$C_{\rm out} = C_{\downarrow \downarrow} - \alpha |C_{\uparrow \downarrow}|$
and $C_{\rm in} = C_{\uparrow \uparrow}$.
From this expression, one finds that
$e^2/C_{\rm out}$ corresponds to the charging energy ($\sim 60$ $\mu$eV)
measured in Ref. \cite{Kataoka_Kondo},
from which
one can estimate
$- \epsilon_\sigma$ and $U + \epsilon_\sigma$
based on energy relations of the Anderson impurity \cite{Hewson}
$- \epsilon_\sigma$, $U + \epsilon_\sigma \sim e^2/(2C_{\rm out})$.
To estimate further $e^2/C_{\rm in}$, one can use the following fact.
In the experiment of Ref. \cite{Kataoka_Kondo},
a Kondo dip appears
only within a certain finite window $\delta B$ of
magnetic field in each Aharonov-Bohm period.
The finite window may be related to the effective Zeeman energy
$\Delta \epsilon$, which can split and suppress the Kondo resonance
\cite{Meir2}.
For $\alpha = 1$,
the Kondo resonance occurs at
$\delta q_\uparrow = \pm e/2$
(where $\Delta \epsilon = 0$),
and in its vicinity where
$\delta q_\uparrow = \pm (1/2 - p)e$,
one finds
$\Delta \epsilon = p e^2/C_{\rm in}$.
From this,
$e^2/C_{\rm in}$ is roughly estimated to be of the order of
10 $\mu$eV using the
experimental results \cite{Kataoka_Kondo} of
$\Delta \epsilon \lesssim$ 10 $\mu$eV
(the energy scale of the zero-bias anomaly),
the width $\delta B$ of a Kondo dip $\simeq \Delta B / 4$,
and $p \simeq \delta B / \Delta B$.
By using the estimated energy scales and from
a numerical renormalization-group calculation
\cite{Wilson_NRG, Costi_NRG, Hewson},
the Kondo temperature was
found \cite{Sim} to be $T_{\rm K} \sim 1~\mu$eV,
which is similar to the energy scale of
the zero-bias anomaly of the antidot Kondo effect
\cite{Kataoka_Kondo}.

One interesting point of the antidot setup in Fig.~\ref{geometry}
is that
the resonance width
$\Gamma_\sigma(E) \equiv \sum_i \Gamma^i_\sigma(E)$
has spin dependence
since
the distance between the antidot and the extended edge
channel $\aaa$ (see Fig.~\ref{geometry})
is spin dependent, {\it i.e.},
$V^{i=\aaa(\aaa')}_{k\uparrow} \neq V^{i=\aaa(\aaa')}_{k\downarrow}$,
where $\Gamma^i_\sigma(E) = 2 \pi \sum_k
|V^i_{k\sigma}|^2 \delta (E - \epsilon_{ik\sigma})$.
The spin-dependent $\Gamma_\sigma$'s renormalize
the effective Zeeman energy as
$\Delta \tilde{\epsilon} (\delta q_\uparrow, \delta q_\downarrow)$,
%$\Delta \epsilon (\delta q_\uparrow, \delta q_\downarrow)
%\to \Delta \tilde{\epsilon} (\delta q_\uparrow, \delta q_\downarrow)$,
as in
quantum dots coupled to
ferromagnetic leads \cite{Lopez,Martinek}.
This changes the Kondo resonance
condition to
\begin{eqnarray}
E_\mathrm{AD}(\delta q_\uparrow \pm e, \delta q_\downarrow \mp e) =
E_\mathrm{AD}(\delta q_\uparrow, \delta q_\downarrow)
+ \Delta \tilde{\epsilon},
\label{redefined}
\end{eqnarray}
instead of Eq. (\ref{resonance_condition_Kondo}).
When Eq. (\ref{redefined}) is satisfied,
there is no renormalized effective Zeeman splitting so that
one can estimate \cite{Haldane}
$T_{\rm K} \sim (\sqrt{\Gamma U}/2) \exp(\pi \epsilon (U+\epsilon)/\Gamma U)$,
where $\Gamma = (\Gamma_\uparrow + \Gamma_\downarrow)/2$.

It is worth pointing out that, at high magnetic fields $B \sim
3$~T, the Kondo dips have not been found experimentally
\cite{Kataoka_double}. In this regime, the Kondo dips may be
suppressed, leaving $h/2e$ oscillations, since the spin-up antidot
states are almost decoupled from all extended edge channels.

\begin{figure}[th]
\centerline{\psfig{file=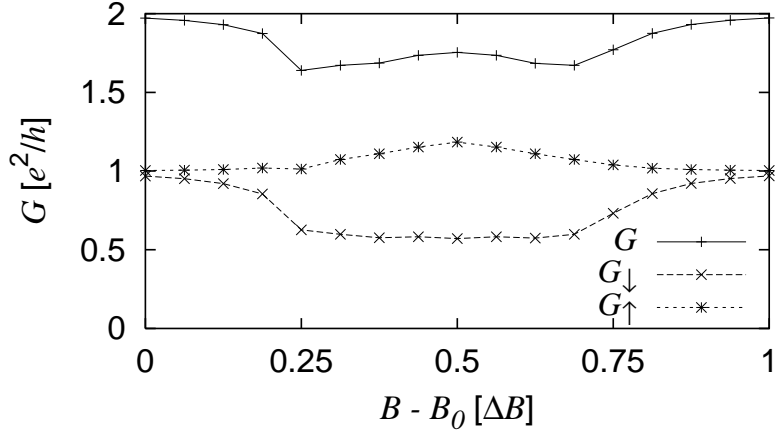,width=11cm}} \vspace*{8pt}
\caption{ Calculated antidot conductance $G_{\rm ad}(B)$ and its
spin-dependent component $G_\sigma(B)$ around a magnetic field $B_0$
($ \sim 1$ T). $G_{\rm ad}(B)$ matches well the experimental data in
the inset of Fig.\/ \ref{Kondo_figure1}. This plot of $G_{\rm
ad}(B)$ is obtained by solving the Hamiltonian $H_{\rm K}$ in Eq.
(\ref{Anderson}) using the numerical renormalization-group method,
and by applying the result into Eq. (\ref{transmission}). In the
calculation, the dependence of the coupling $V^i_{k \sigma}$ on the
channel index $i$ and on $\sigma$ is fully taken into account; the
dependence of $V^i_{k \sigma}$ on $k$ is ignored and the extended
edge channels are approximated to have infinite bandwidth, to see
the low-energy limit. From Ref. \protect\cite{Sim}. }
\label{NRGconductance}
\end{figure}

Now we describe the line shape of the antidot Kondo effect (see the
inset of Fig.~\ref{Kondo_figure1}). In Ref. \cite{Sim}, the line
shape was calculated by using the zero-temperature expression of the
antidot conductance $G_{\rm ad}$ (as well as that of its spin
component $T_\sigma$) derived in Sec.\/ \ref{SPTUNNELING} [see Eqs.
(\ref{transmission}) and (\ref{conductance})]. The parameters used
in the conductance expression, the level broadening
$\Gamma^i_\sigma$'s, can be estimated from Ref. \cite{Kataoka_Kondo}
in the following way: since the experimental result of the antidot
conductance $G_{\rm ad}(B)$ shows dips below $2e^2/h$ [see
Fig.~\ref{Kondo_figure1}] and the excess charges follow a type (i)
trajectory, one can expect that the spin-down electrons favor
{\em backward} scattering, {\it i.e.,} $\Gamma^{i=\aaa,\aaa'}_\downarrow >
\Gamma^{i=\bbb,\bbb'}_\downarrow$ and $T_\downarrow < 0$. On the other
hand, the experimental result of the $\Delta B$ periodic peaks at
stronger magnetic fields near the plateau of $G_{\rm ad} \simeq
e^2/h$ may be interpreted as normal resonances of spin-up electrons.
Combined with the fact that $\Gamma^{i=\aaa,\aaa'}_\sigma$ is smaller for
weaker $B$, such an interpretation indicates that the spin-up electrons
favor {\em forward} scattering, {\it i.e.,}
$\Gamma^{i=\aaa,\aaa'}_\uparrow < \Gamma^{i=\bbb,\bbb'}_\uparrow$ and
$T_\uparrow > 0$ (opposite to the case for spin down), for weaker $B$
where $G_{\rm ad} > e^2/h$. The type (i) trajectory with $\alpha =
1$, the above features of the level broadenings, and the calculation
of $\sin^2 \theta_\sigma$ based on a numerical renormalization group
approach allow one to get the antidot conductance $G_{\rm ad}(B)$
for one Aharonov-Bohm period, as shown in Fig.~\ref{NRGconductance}.
The result of $G_{\rm ad}(B)$ is in qualitatively good agreement
with the experimental data in Fig.~\ref{Kondo_figure1}.

The spin-dependent component $G_\sigma$ in Fig.~\ref{NRGconductance}
can be understood as follows.
$G_\uparrow$ has only one Kondo peak,
while $G_\downarrow$ shows a mixture of two normal
and one intermediate Kondo resonance dips.
As a result,
at the center of the Kondo resonance,
even if the unitary Kondo limit is reached,
$G_{\rm ad}$ can be greater than $e^2/h$.
The center and the width of the Kondo resonance
are governed by Eq. (\ref{redefined}) and $e^2/C_{\rm in}$, respectively.
This spin-dependent behavior
of $G_\sigma$,
which results from
the type (i) trajectory of excess charges and
the scattering nature of
multiple extended edge channels by the antidot
({\it i.e.}, the sign of $T_\sigma$),
is an interesting feature of the antidot Kondo effect.

%\subsection{Hartree-Fock calculations on antidot ground states}
\subsection{Hole maximum-density-droplet antidot ground state}
\label{THHF}

In the previous section, the properties of antidots in the integer
quantum Hall regime have been described based on a capacitive
interaction model. To have a more solid understanding, it is
necessary to go beyond such a phenomenological model by using
numerical methods. However, the antidot system is an open-geometry
problem and thus may require heavy calculations to treat its
interaction effects. The difficulty was avoided in a recent
Hartree-Fock study \cite{Hwang} by using a particle-hole
transformation. The transformation converts the antidot into a
quantum dot of holes. This indicates that one can study an antidot
by applying the numerical methods used for a large-size quantum dot.

In the Hartree-Fock calculation in Ref.~\cite{Hwang}, the properties
of the ground state of an antidot containing about 300 holes have
been investigated. The main findings of the work are that in a
certain parameter range, the ground state of the antidot can be a
hole maximum-density droplet, the state without edge reconstruction,
and that the transition of the ground states as a function of the
magnetic field can be well described by the evolution of the excess
charges derived in the capacitive interaction model in Secs.
\ref{THexcess} and \ref{THcapacitive}. This study will be described
in this section.

By using a particle-hole transformation
\cite{Sim_HF,Johnson_PHtransform} of the type $c_{m,\sigma}
\rightarrow h_{m,\sigma}^\dagger$ and $c_{m,\sigma}^\dagger
\rightarrow h_{m,\sigma}$, one can map the $\nuC = 2$ antidot system
described by the Hamiltonian (\ref{HAD}) into a system confining a
finite number of holes, for which one can find the eigenstates by
using diagonalization.
Note that only the lowest Landau levels with the two spin
branches are considered for particles and then the particle-hole
transformation is performed; the effect of the unoccupied higher
Landau levels can be ignored as they provide spatially uniform
density of holes and therefore constant energy shift.
The transformed total Hamiltonian of holes
can have the form \cite{Sim_HF,Hwang}
\begin{eqnarray}
H &=& \sum_{m \sigma}^{m_{\rm c}} \varepsilon_{m \sigma}^{\rm h}
h_{m \sigma}^\dagger h_{m \sigma}
+ \frac{1}{2} \sum_{m m' n n' \sigma \sigma'}^{m_{\rm c}}
\langle m m' |W|n n' \rangle
h_{n' \sigma'}^\dagger h_{n \sigma}^\dagger
h_{m \sigma} h_{m' \sigma'} + E_{\rm T}, \
\nonumber \\
E_T & = & 2 \sum_m^{m_{\rm c}} V_{\rm AD}(m)
- 2 \sum_m^{m_{\rm c}} W_m^{\rm H} - \sum_m^{m_{\rm c}} W_m^{\rm X},
\label{holeHam} \\
\varepsilon^{\rm h}_{m \sigma} & = & - V_{\rm AD}(m) +  W_m^{\rm X} -
\epsilon^{\rm Z}_\sigma, \label{single_hole_energy}
\end{eqnarray}
where $\varepsilon^{\rm h}_{m \sigma}$ is the effective
single-particle energy of a hole with angular momentum
$m$, $W_m^{\rm H} = \sum_{m'}^{m_{\rm c}} \langle m m' |W| m m'
\rangle$, and $W_m^{\rm X} = \sum_{m'}^{m_{\rm c}} \langle m m'
|W| m' m \rangle$. Here, the angular-momentum wave functions
(\ref{LLsymmetricgauge}) of the lowest Landau level are used as the
single-particle basis functions, and $m_{\rm c}$ is a
cutoff value of angular momentum (chosen sufficiently large). The
first term $V_{\rm AD}(m) = \langle m | V_{\rm AD}(r) | m \rangle$
of $\varepsilon^{\rm h}_{m \sigma}$ in Eq.
(\ref{single_hole_energy}) means the hole confinement energy coming
from the antidot potential $V_{\rm AD}$, while the second term
represents the change in exchange energy $W^{\rm X}$
when an electron with angular
momentum $m$ disappears; the constant Landau energy $\hbar \omega_c
/ 2$ is included in $V_{\rm AD}(m)$. A Hartree term $W^{\rm H}$
is absent in $\varepsilon_m$, since it is canceled by the
contribution of the positive background charges. The Zeeman term of
$\varepsilon^{\rm h}_{m \sigma}$ has the opposite sign to
the corresponding term in Eq. (\ref{HAD}). The final constant term
$E_{\rm T}$ of the Hamiltonian $H$ is the total energy of
an electron system in which electrons are occupied from $m=1$ to
$m_c$ states. Note that when the cutoff $m_c$ is much larger than
the total number of holes, $W_m^X$ in Eq. (\ref{single_hole_energy})
can be treated as a constant \cite{Yang2} so that
$\varepsilon^{\rm h}_{m \sigma}$ is mainly governed by
$V_{\rm AD}(m)$. In Ref. \cite{Hwang}, a bell-shape antidot
potential is chosen,
\begin{eqnarray}
V_{\rm AD}(r) = \left\{ \begin{array}{ll}  \frac{1}{2}\hbar\omega_c-\frac{1}{2}m^{*}\Omega^2r^2, &  r<r_{\rm t} \\
\frac{1}{2} \hbar \omega_c + \frac{p}{r^2} + q, & r_{\rm t} <r<r_{\rm s} \\
\textrm{constant}, &    r>r_{\rm s},
\end{array} \right.
\label{bellshape1}
\end{eqnarray}
from which one can easily get the analytic expression for $V_{\rm AD}(m)$.
Notice that
in the interval $ r<r_{\rm t}$ the bell-shape potential is
inverse parabolic, while in
$ r_{\rm t}<r<r_{\rm s}$ the curvature changes sign.
Here, $p$ is a parameter determining the curvature in $ r_{\rm t} < r <
r_{\rm s}$,
while $q$ is chosen for $V_{\rm AD}(r)$ to be continuous at $r = r_{\rm t}$.

%\vspace*{1cm}
\begin{figure}[th]
\vspace*{1cm}
\centerline{\psfig{file=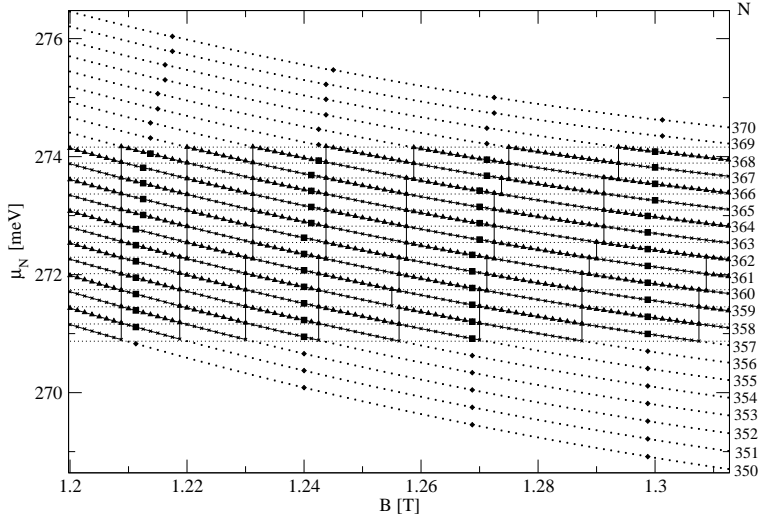,width=10cm}}
\vspace*{8pt}
\caption{
Chemical potential, $\mu_{N^{\rm h}} \equiv E_{N^{\rm h}+1} - E_{N^{\rm h}}$,
vs magnetic field $B$ for a bell-shaped antidot potential.
Twenty-one different values of
$N^{\rm h} = N^{\rm h}_\uparrow + N^{\rm h}_\downarrow \in [350,370]$ are used.
The horizontal dotted line represents the {\em hole} Fermi energy.
Filled squares represent the spin-flip Kondo transitions
$| N^{\rm h}_\uparrow, N^{\rm h}_\downarrow \rangle \to
| N^{\rm h}_\uparrow + 1, N^{\rm h}_\downarrow - 1 \rangle$
as $B$ increases,
while vertical jumps show the spin-down resonances
 $| N^{\rm h}_\uparrow, N^{\rm h}_\downarrow \rangle \to
| N^{\rm h}_\uparrow, N^{\rm h}_\downarrow + 1 \rangle$.
Following the topmost zigzag solid line, for example, the hole
maximum-density droplet ground state
evolves as
$|N^{\rm h}_{\uparrow}, N^{\rm h}_{\downarrow}
\rangle = |128,233 \rangle
\rightarrow |128,234 \rangle
\Rightarrow |129,233 \rangle
\rightarrow |129,234 \rangle
\rightarrow |129,235 \rangle$
$\Rightarrow |130,234 \rangle
\rightarrow |130,235 \rangle
\rightarrow |130,236 \rangle
\Rightarrow |131,235 \rangle
\rightarrow |131,236 \rangle
\rightarrow |131,237 \rangle
\Rightarrow |132,236 \rangle$
with increasing $B$.
Similarly, in the fourth zigzag solid line,
$|N^{\rm h}_{\uparrow}, N^{\rm h}_{\downarrow} \rangle$
evolves as
$|127,231 \rangle
\rightarrow |127,232 \rangle
\Rightarrow |128,231 \rangle
\rightarrow |128,232 \rangle
\rightarrow |128,233 \rangle
\Rightarrow |129,232 \rangle
\rightarrow |129,233 \rangle$
$\rightarrow |129,234 \rangle
\Rightarrow |130,233 \rangle$.
Here $\rightarrow$ and $\Rightarrow $ indicate, respectively,
the spin-down normal resonance and spin-flip Kondo transitions.
The parameters of this antidot can be found in Ref. \protect\cite{Hwang}.
From Ref. \protect\cite{Hwang}.
}
\label{HFdata}
\end{figure}

The above transformation maps
the antidot problem into
a numerically tractable finite-size hole system,
so that one can study the ground-state properties of an antidot.
The properties of the ground state are determined by
the competition between the hole confinement potential $(-V_{\rm AD})$ and
the Coulomb energy $(E^{\rm H} + E^{\rm X})$.
The confinement potential favors a smaller value of total angular
momentum of the state.
As a result,
in the limit of strong hole confinement $-V_{\rm AD}(r)$,
the ground state is a maximum-density droplet of holes,
which has a single-Slater-determinant form,
\begin{eqnarray}
| N^{\rm h}_\uparrow, N^{\rm h}_\downarrow \rangle =
h_{{N^{\rm h}_\downarrow -1},\downarrow}^\dagger \cdots h_{0 \downarrow}^\dagger
h_{{N^{\rm h}_\uparrow -1},\uparrow}^\dagger \cdots h_{0 \uparrow}^\dagger
| 0 \rangle.
\end{eqnarray}
The total number of holes is $N^{\rm h} = N^{\rm h}_\uparrow +
N^{\rm h}_\downarrow$,
and $N^{\rm h}_\downarrow$ is equal to or larger than $N^{\rm h}_\uparrow$
due to the Zeeman energy.
The spatial splitting between spin-up and down edges
(or $N_{\downarrow}-N_{\uparrow}$) of the droplet
depends on the relative strength of the confinement energy and
electron-electron interactions;
for a given $N$,
smaller ($N^{\rm h}_{\downarrow}-N^{\rm h}_{\uparrow}$)
is favored by stronger confinement, since the droplet size is
determined by $N^{\rm h}_\downarrow$.
In Ref. \cite{Hwang}, it was found that
in certain ranges of system parameters
the maximum-density droplet of holes is the ground state of
the bell-shaped antidot containing about 300 holes.
For weaker confinement, the droplet is not the
ground state any more and edge reconstruction such as
the formation of compressible regions may occur.
Note that an electronic maximum-density droplet has been
studied theoretically \cite{Yang2,Yang1} and experimentally
\cite{OKlein,Oosterkamp}.

Figure~\ref{HFdata} shows the transitions between the hole-droplet
ground states of the antidot studied in Ref. \cite{Hwang} as a
function of $B$. The numerical results can be compared with the
prediction of the capacitive interaction model in Sec.
\ref{THcapacitive}. In the parameter regime studied in
Fig.~\ref{HFdata}, the transitions of the ground state follow the
type (i) process with $\alpha = 1$ (see Fig.~\ref{sequence2}).
Moreover, the transitions are found \cite{Hwang} to be equivalent to
those determined by the evolution of the excess charges $\delta
q_\sigma (B) / e = N^{\rm AD}_\sigma - a_\sigma B - b_\sigma$ in
Eqs. (\ref{excess_charge},\ref{excess_charge2}) and by the
transition conditions
(\ref{resonance_condition_up},\ref{resonance_condition_down},\ref{resonance_condition_Kondo});
the parameters of the evolution can be uniquely determined as
$a_\uparrow = 34.78$ and $a_\downarrow = 54.11$. This comparison
shows that the capacitive-interaction model may work very well for
the droplet states.

The droplet studied in Fig.~\ref{HFdata} has the following
properties. First, the electron interactions between spin-up and
down edges of the droplet are very strong, as indicated by the
result of the comparison, $\alpha = 1$. Second, since it follows the
type (i) process, it can give rise to the $h/2e$ Aharonov-Bohm
oscillations. Third, the studied droplet has spin-dependent areas
$S_\uparrow \neq S_\downarrow$, as $N^{\rm h}_\uparrow \simeq 130$
and $N^{\rm h}_\downarrow \simeq 235$. In this case, only the
spin-down electrons may be allowed to tunnel between the droplet and
the extended edge channels. Therefore, as discussed in
Sec.~\ref{THKondo}, the Kondo effect will be suppressed in this
case. To see the Kondo effect, we need $S_\uparrow \simeq
S_\downarrow$, which can be obtained for a stronger antidot
confinement potential than that studied in Fig.~\ref{HFdata}.

Concluding this section, we note that the Hartree-Fock
approach discussed here is valid only when the ground state of the
antidot is the hole maximum-density-droplet state. In the next
section, we will review a numerical study based on
density-functional theory in which the formation of compressible regions
around an antidot, another interesting and important type of
antidot ground state, can be studied.

\subsection{Density-functional studies on the compressibility of
antidot states}
\label{THcompressible}

As discussed in Sec.\/ \ref{EXPcompressible}, it is interesting to
study whether compressible regions can be formed around an antidot.
Recently, Ihnatsenka and Zozoulenko \cite{Ihnatsenka} studied the
formation of compressible regions around an antidot numerically by
using spin-density-functional theory. They found that screening is
less effective in a finite-length antidot edge than in an extended edge
with infinite length. This numerical approach will be described in
this section.

\begin{figure}[th]
\centerline{\psfig{file=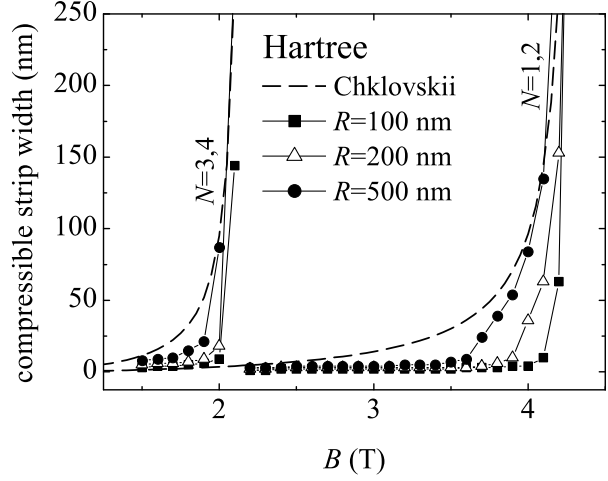,width=8cm}}
\vspace*{8pt} \caption{ Width of the compressible regions around
an antidot formed by an antidot gate with various radii $R$ as a
function of magnetic field, calculated within the Hartree
approximation at temperature $T = 1$~K for the case of spinless
electrons. It is compared with the width of compressible regions
formed along extended edge channels calculated by Chklovskii and his
coworkers \protect\cite{Chklovskii1,Chklovskii2}. $N=1$,2 and 3,4
refer to the subband number corresponding to the two lowest
(spin-degenerate) edge states. From Ref. \protect\cite{Ihnatsenka}.
} \label{SDFT1}
\end{figure}

In the work by Ihnatsenka and Zozoulenko \cite{Ihnatsenka}, a form
of antidot potential (induced from a circular gate of radius
100--500~nm) obtained by Davies \cite{Davies} was used, and Hartree
interactions, exchange correlations, and the effect of Zeeman
splitting were taken into account within the framework of
density-functional theory in the local spin-density approximation
\cite{Parr}. Compressible regions are defined as the spatial windows
in which effective single-particle energy $E$, obtained by solving
self-consistently a Schr\"{o}dinger equation within the
density-functional theory, satisfies $|E - \EF| < 2 \pi k_{\rm B}
T$. The features of the compressible regions were studied separately
for the spinless and spinful cases in magnetic fields of 1--5~T
($\nuC = 2$ is achieved around $B = 4$ T).

For the spinless case, where the exchange correlations are not taken
into account, the calculation shows that the width of the compressible
region formed around an antidot becomes narrower, deviating from
the value in the extended-edge cases
\cite{Chklovskii1,Chklovskii2},
as the radius of the antidot becomes smaller (see Fig.~\ref{SDFT1}).
This finding can be understood as follows.
Compared with the extended edge cases, the Hartree contribution
is stronger in the antidot due to finite-size effects.
The stronger Hartree potential repels electrons from the antidot
edge, and thus screening properties become weaker,
resulting in a steeper effective antidot potential preventing
the formation of compressible regions.

\begin{figure}[th]
\centerline{\psfig{file=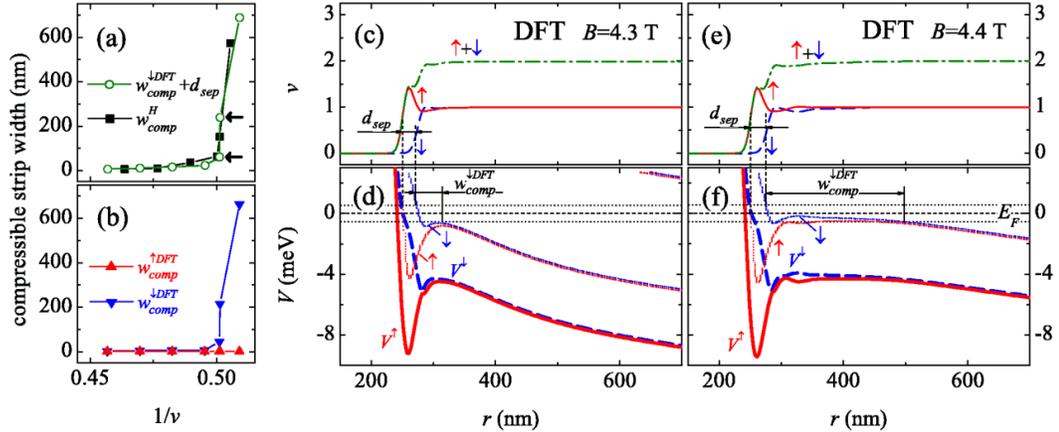,width=14cm}}
\vspace*{8pt} \caption{ (a) Width of the compressible regions formed
around an antidot for spinless electrons as a function of $1/\nu$,
where $\nu$ is the antidot filling factor. The result of a Hartree
approximation (filled squares) is almost identical to that calculated
using density-functional theory. (b) Width of the compressible
regions in the spinful case, calculated using density-functional
theory. In the parameter regime studied in this figure, the spin-up
compressible region is not formed. (c) and (e) Electron density
profiles for different magnetic fields, and (d) and (f) the
corresponding magnetosubband structures. From Ref.
\protect\cite{Ihnatsenka}.} \label{SDFT2}
\end{figure}

For the spinful case, the compressible regions around the antidot
show more distinct deviation from the extended edge cases. The
contribution from exchange interactions suppresses the formation of
the compressible region of spin-up electrons, which is assumed to
have lower Zeeman energy than spin down. In the parameter ranges
studied in Ref. \cite{Ihnatsenka} it was found that the spin-up
compressible region is not formed [see Fig.~\ref{SDFT2}(b)].
Moreover, the spin-up and down states at the Fermi level are
spatially separated from each other [Figs.~\ref{SDFT2}(c)-(f)].
These results indicate that the edge structure around the antidot
could be surprisingly different from that along extended edges due
to finite-size effects.

The two numerical studies in this and the previous sections provide
useful information on electronic structure around antidots.
However, they serve just as a starting point, as they explore
only narrow parameter ranges (antidot size, magnetic field, electron density,
the smoothness of antidot potential, etc).
More systematic numerical studies are required to have the phase
diagram which shows when the hole maximum-density droplet,
the formation of compressible regions, or any other possible
edge reconstruction is the ground state of an antidot.
Such studies will be very valuable in order to
understand electron interactions in antidots or local disorder
regions in the integer quantum Hall regime.

\section{Summary and perspectives}
\label{SUMMARY}

We have reviewed experimental and theoretical works devoted to
electron interactions in an antidot in the integer quantum Hall regime.
To conclude this review,
we emphasize
the usefulness
of antidots in future studies.
An antidot can be created
as an experimentally tunable system like a quantum dot.
Therefore,
understanding antidot systems will open up a systematic way of
studying the nature of electron states in a quantum Hall system,
such as localized chiral edge states, quasiparticle tunneling,
electron-electron interactions, compressible regions and screening
of electric (Hall) fields, etc.
For example, it may provide a useful way of studying
local disordered regions in quantum Hall systems,
which can be modeled by a combination of antidots and quantum dots
\cite{Zhitenev,Finkelstein,Ilani,Cobden}.

Although, as discussed in this review, many features of
antidots have been understood in the past decade,
there are still many open questions.
Some examples are listed below.
\begin{itemize}
\item
The phase diagram of the ground state (the hole maximum-density
droplet, the formation of compressible regions, or any other
possible edge reconstruction) of an antidot has not been studied.
Especially, the formation of compressible regions around an antidot
is still controversial
\cite{Karakurt,Kataoka_comment,Goldman_reply,Ihnatsenka} (see Secs.
\ref{EXPcompressible} and \ref{THcompressible}).
\item It is not understood how the change occurs from a
low-magnetic-field regime, with spin-split pairs of states,
which can be described by a noninteracting model (see Fig.~\ref{LowB}
and its description in Sec. \/ \ref{EXPsetup}),
to higher-field regimes, where interaction effects such as
exchange, charging, and Kondo physics emerge, with separate regions
of states for the two spins. The transitions may be sudden or gradual.
\item In the regime of
strong coupling between antidot states and extended edge channels,
where the antidot is not isolated,
there is no proper theoretical model describing interaction effects.
\item A full description of the spectator mode in antidot molecules
is required (see Sec.\/ \ref{EXPmolecule}).
\item Interaction effects in an antidot with filling factor $\nuC$
larger than 2 still remain to be investigated in detail
experimentally and theoretically.
\end{itemize}
Future studies in these directions are highly desirable.

In addition, we very briefly note the connection between antidot
studies in the integer and fractional quantum Hall regimes. Some of
the most interesting aspects of the latter regime are the fractional
charge and statistics of quasiparticle excitations. As an antidot
can provide an experimentally controllable finite edge of the
fractional quantum Hall state, it has been used for studying the
fractional excitations. Below, we summarize the previous works on
antidots in the fractional regime and discuss possible connection to
the studies of the integer regime as well as possible future
directions.
\begin{itemize}
\item The initial aim of studying an antidot in the fractional
regime was to detect the fractional charge of quasiparticles
\cite{Goldman,Franklin,Franklin2,Goldman01,Goldman_FQHE,Goldman96,Goldman97}
and to measure the energy structure \cite{Maasilta}. The ratio of
the Aharonov-Bohm periods as a function of the external magnetic
field and a back-gate voltage is related to the change in charge
around the antidot in sweeping from one conductance peak to the
next, i.e.\/ in moving from one resonance to its neighbor. This
charge has been found to be $\sim e/3$ for $\nuC = 1/3$
\cite{Goldman}. It was originally pointed out that such antidot
studies of periodicities \cite{Goldman,Franklin,Franklin2} are made
in equilibrium (on a timescale much longer than the lifetime of
quasiparticles), so the system could rearrange during each $h/e$ or
$e/3$ period on that timescale by methods other than the tunneling
of a quasiparticle \cite{Franklin,Franklin2}. However, the resonant
tunneling that gives rise to a conductance peak is itself measuring
the passage of individual particles [in general, at zero bias,
resonant tunneling of single charged particles occurs when the
many-particle states with and without that charge are degenerate].
This implies that, in conjunction with the above measurement of the
difference in trapped charge of these two states \cite{Goldman}, the
particles do have this measured fractional charge, $e/3$, as
expected for $\nu=1/3$ quasiparticles. Similar measurements imply
that the fractional charge of quasiparticles in the 2/5 state is
$e/5$ \cite{Goldman96}. Shot-noise measurements
\cite{Picciotto,Saminadayar}, performed out of equilibrium using a
quantum point contact, have also measured fractional charge.
\item In the fractional regime, the antidot state can be described
by a chiral Luttinger liquid. Its Hamiltonian has a similar form
to the Hamiltonian of the integer regime; see Eq.
(\ref{effH_nu1_2}) of Sec.~\ref{THexcess}. It has been suggested
\cite{Geller,Geller2} that an antidot can be used to distinguish
experimentally between Luttinger and Fermi liquids, as the
Aharonov-Bohm oscillations can show the crossover between them.
The crossover remains to be tested experimentally
\cite{Maasilta_FQHE}. It has been also investigated
\cite{Braggio}, based on the chiral Luttinger-liquid theory, that
the statistics of quasiparticle tunneling in an antidot system can
reveal its features in high moments of the tunneling current.
These theoretical works suggest that the antidot can provide an
experimental system for studying a finite-size Luttinger liquid.
\item We have seen that the many-body ground state of an antidot
and the reconstruction of the antidot edge are important issues in
the integer regime (see Secs.~\ref{EXPcompressible} and
\ref{THcompressible}). The edge reconstruction is also important
in the fractional regime as it is determined by the interplay of
electron-electron interaction in the fractional regime, the
single-particle electrostatic edge potential (or antidot
potential), and effects of finite temperature. A recent numerical
prediction \cite{Wan} on edge reconstruction in the fractional
regime has been linked to experimental results \cite{Ye} on the
microwave conductivity of two-dimensional electron systems with an
array of antidots. More systematic studies on the edge
reconstruction are required to understand excitations of an
antidot in both the integer and fractional regimes. \item An
antidot molecule has been also studied \cite{Maasilta_molecule} in
the fractional regime, focusing on coherent tunneling between two
antidot bound states (for the integer regime, see
Sec.~\ref{EXPmolecule}). In the antidot studied, the coherent
tunneling rate between the two bound states was found to be an
order of magnitude higher than the phase-breaking rate. Coulomb
blockade of an antidot molecule has been theoretically studied
very recently as well \cite{AverinNesteroff}. These works may be a
starting point for implementing quantum computation using antidots
\cite{Averin_QC,Camino}. \item It has been suggested
\cite{Bonderson,Stern} to use antidots for detecting non-Abelian
braiding statistics of the Pfaffian state \cite{Moore,Greiter} in
the $\nu = 5/2$ regime. An experimental test of this proposal is
still needed, and such a test would provide a way of constructing
topologically protected qubits based on a non-Abelian fractional
state.
\end{itemize}

Finally, we would like to point out that antidots can provide
various powerful experimental tools. For example, they can be used
to study departures from the Onsager relations in nonlinear
mesoscopic transport \cite{Sanchez}, and can constitute a spin
injector or spin filter \cite{Zozoulenko}, which may be useful for
spintronics using quantum-Hall edge channels.

\section*{Acknowledgements}

We are grateful to L. C. Bassett for critical reading and
useful discussions. We also thank V. J. Goldman, A. S. Sachrajda, and
I. V. Zozoulenko for providing their figures and for valuable
comments. We are supported by a Korean Research Foundation Grant
funded by the Korean Government (KRF-2006-331-C00118,
KRF-2005-070-C00055), and by the UK EPSRC.

\end{document}